\documentclass[a4paper,11pt]{article}
\pdfoutput=1

\usepackage{jcappub}  
\usepackage{amsmath}  
\usepackage{amsfonts} 
\usepackage{amssymb}  
\usepackage{mathrsfs} 
\usepackage[version=3]{mhchem}		
\usepackage{graphicx} 
\usepackage{epstopdf} 
\usepackage{hhline}
\usepackage{subfig}

\usepackage{soul} 

\DeclareMathOperator{\erf}{erf}

\newcommand{\cm}{~\mathrm{cm}} 
\newcommand{\fm}{~\mathrm{fm}} 
\newcommand{\erg}{~\mathrm{erg}} 
\newcommand{\Gy}{~\mathrm{Gyr}} 
\newcommand{\g}{~\mathrm{g}} 
\newcommand{\GeV}{~\mathrm{GeV}} 
\newcommand{\kms}{~\mathrm{km}~\mathrm{s}^{-1}}
\newcommand{\ecm}{~e\operatorname{-}\mathrm{cm}}

\newcommand{\s}{~\mathrm{s}} 

\newcommand{\ED}{\mathrm{ED}}
\newcommand{\MD}{\mathrm{MD}}
\newcommand{\AN}{\mathrm{AN}}

\begin{document}



	\title{Effect of electromagnetic dipole dark matter on energy transport in the solar interior}
	\author[a,b]{Ben Geytenbeek,}
	\author[a]{Soumya Rao,}
	\author[c]{Pat Scott,}
	\author[d]{Aldo Serenelli,}
	\author[c]{Aaron C. Vincent,}
	\author[a]{Martin White}
	\author[a,1]{and Anthony G. Williams\note{ORCID: \url{http://orcid.org/0000-0002-1472-1592}}}
	\affiliation[a]{ARC Centre of Excellence for Particle Physics at the Terascale \& CSSM, Department of Physics, University of Adelaide, Adelaide, South Australia 5005, Australia}
	\affiliation[b]{Cavendish Laboratory, J.J. Thomson Avenue, Cambridge CB3 0HE, United Kingdom}
	\affiliation[c]{Department of Physics, Imperial College London, Blackett Laboratory,
	Prince Consort Road, London SW7 2AZ, United Kingdom}
	\affiliation[d]{Institut de Ci\`{e}ncies de l'Espai (ICE-CSIC/IEEC), Campus UAB, Carrer de Can Magrans s/n, 08193 Cerdanyola del Vall\`{e}s Spain}
	
	\emailAdd{bg364@cam.ac.uk}
	\emailAdd{soumya.rao@ncbj.gov.pl}
	\emailAdd{p.scott@imperial.ac.uk}
	\emailAdd{aldos@ice.csic.es}
	\emailAdd{aaron.vincent@imperial.ac.uk}
	\emailAdd{martin.white@adelaide.edu.au}
	\emailAdd{anthony.williams@adelaide.edu.au} 
	
	\abstract{In recent years, a revised set of solar abundances has led to a discrepancy in the sound-speed profile between helioseismology and theoretical solar models. Conventional solutions require additional mechanisms for energy transport within the Sun. Vincent et al. have recently suggested that dark matter with a momentum or velocity dependent cross section could provide a solution. In this work, we consider three models of dark matter with such cross sections and their effect on the stellar structure. In particular, the three models incorporate dark matter particles interacting through an electromagnetic dipole moment: an electric dipole, a magnetic dipole or an anapole. Each model is implemented in the \texttt{DarkStec} stellar evolution program, which incorporates the effects of dark matter capture and heat transport within the solar interior. We show that dark matter with an anapole moment of $\sim1\GeV^{-2}$ or magnetic dipole moment of $\sim10^{-3}\mu_p$ can improve the sound-speed profile, small frequency separations and convective zone radius with respect to the Standard Solar Model. However, the required dipole moments are strongly excluded by direct detection experiments. 
	}
	
	
	\maketitle

	

	\IfFileExists{Introduction.tex}{
		\section{Introduction}

	The problem of dark matter in the universe has led to a major theoretical and observational effort to explain its origin, spanning particle physics, astrophysics and cosmology. The current most popular paradigm is that of a non-relativistic particle with a mass on the order of a $\GeV$, which interacts weakly with Standard Model particles, motivated by the relic abundance of dark matter. However, the search for the particles themselves has proven more elusive. Much of the focus has been on terrestrial direct detection experiments. Some experiments reported some an excess of events such as DAMA~\cite{bernabei08}, CoGeNT~\cite{cogent11}, CRESST-II~\cite{angloher16} and CDMS II~\cite{cdms13}. Others have provided strong exclusion limits, including XENON10~\cite{xenon11a}, XENON100~\cite{xenon12}, COUPP~\cite{coupp12}, SIMPLE~\cite{simple12}, LUX~\cite{lux16}, CDMSlite~\cite{supercdms16} and PandaX-II~\cite{pandaxii16}. To help alleviate the ongoing conflict between these results, it is often illuminating to consider alternate, indirect approaches to detecting dark matter. In particular, the present work will consider using the Sun as a laboratory for testing dark matter physics~\cite{turckchieze12a}. 
		The sheer size of the Sun means that, with high precision measurements, very small effects can become observable. 
			
		Much of the theoretical framework for solar dark matter was developed during an explosion of work in the late 1980s~\cite{steigman78,faulkner85,spergel85,krauss85,krauss86,griest87,raby87,raby88,giudice89,roulet89} where it was proposed to provide additional energy transport in the Sun to solve the solar neutrino problem~\cite{davis68,bahcall68}. In particular, Gould and Raffelt~\cite{gould87a,gould87b,gould90a,gould90b,gould90c} developed much of the theoretical framework for solar dark matter studies which remains standard today, including a prescription for the capture rate, evaporation rate, and energy transport.
	
		As the solar neutrino problem involved the observation of a lower than expected flux of neutrinos, an obvious solution was the possibility that the core temperature of the Sun was cooler than predicted by solar models~\cite{littleton72}. A cooler core slows the fusion of helium isotopes to $\ce{^7Be}$ and $\ce{^8B}$, resulting in a lower flux of their associated neutrinos. Introducing interactions with dark matter provided a mechanism for heat to be transported out of the core of the Sun~\cite{spergel85}. Even though the solar neutrino problem was later solved with the discovery of neutrino oscillations~\cite{superkamiokande98,sno01}, the context of the crisis is necessary to explain the historical development of solar dark matter studies.
		
%
		
		
		Much of the modern description of the solar interior arises from helioseismology, the study of pressure wave propagation through the Sun.
%
%
		In particular, it is possible to predict the speed of the propagation of the waves as a function of the depth of the Sun. As the waves are acoustic, such a quantity is termed the sound-speed profile of the Sun. 
		Importantly, the speed of sound in the Sun is readily predicted independently  by theoretical solar models. Early attempts at fitting the data to models seemed to be successful at best~\cite{gough96b,grevesse98,bertello00a,bertello00b,garcia01}.
		
%
		However, updated spectroscopy of the Sun led to a downwards revision of the solar metal content~\cite{allendeprieto01,allendeprieto02,asplund04,asplund05b,scott06,melendez08,scott09c,asplund09,scott15a,scott15b,grevesse15}. The result was crucial for helioseismological fits, as the propagation of acoustic waves is affected not only by the temperature, but the composition of the Sun itself~\cite{basu04,bahcall05b,bahcall06,yang07,basu08}. The reduction in metallicity meant that the Standard Solar Models could no longer reproduce the results of helioseismological data within error~\cite{asplund05a}. Further revisions have led to a slight remediation of the errors~\cite{asplund09,serenelli09,serenelli11}, but a significant discrepancy remains~\cite{serenelli16}. Other observables also cause tension with the Standard Solar Model, in particular the radius of the convection zone $R_{\mathrm{cz}}$~\cite{basu04} and the surface helium abundance $Y_s$~\cite{serenelli09,serenelli10,serenelli11}. 
%
%
	
		Dark matter is a proposed solution to alleviate the tension that has arisen in the sound-speed profile of the Sun. The theory is conceptually simple, though has proven difficult to implement correctly in practice. Suppose that dark matter is present in the Sun, and it interacts sufficiently weakly so as not to disrupt other observables. Collisions between dark matter and atomic nuclei can provide a means of energy transfer, subject to two limiting cases. If the interaction strength is sufficiently strong, dark matter particles are in local thermal equilibrium with the surrounding material, providing additional heat transport via conduction~\cite{gould90a}. If the interaction is too strong, then the larger number of collisions means that the energy can only be transported a short distance. If the interaction strength is weaker, then the mean-free path can be sufficiently large to provide energy transport across different regions of the solar interior~\cite{gould90c}.
			
		Early investigations relating dark matter to helioseismology focussed on constant cross section spin-independent models of heavy mass ($30\GeV-100\GeV$). For the selected mass range, there was little to no modification to the sound speed profile~\cite{bottino02}. Results were shown to be dependent on the strength of the interaction. Effects start to become larger as the cross section approaches the transition between local and non-local transport, where the total luminosity transported approaches a maximum ~\cite{lopes02b,lopes02a}. 
			
		More recent developments have led to a resurgence of interest in solar dark matter in the literature. Models with lighter dark matter particles ($1\GeV-10\GeV$) have been shown to induce drastic changes to the solar profile~\cite{cumberbatch10,taoso10}. While their introduction reduces the tension near the border of the radiative and convective zones, the energy transport in the centre of the Sun is considerable. The difference between models and observations near the core increases drastically, implying a core that is much colder for the models than expected; a result heavily constrained by the measurements of neutrino fluxes. Inserting dark matter with a constant, spin-independent cross section into the Sun did not appear to resolve the problems with the sound-speed profile. 
			
		A series of recent works by Vincent et. al.~\cite{vincent14,vincent15a,vincent15b,vincent16} appears to have reconciled the sound-speed profile at the edge of the radiative zone without significantly disrupting the core of the Sun. Their solution was to use dark matter with a generalised momentum or velocity dependent cross section. By doing so, the mean cross section becomes a function of the radius of the Sun.
		However, ref.~\cite{vincent15b,vincent16} only considers general momentum or velocity dependence, not specific dark matter models. This work seeks to look further by considering three different models of momentum or velocity dependent dark matter, and investigating the resultant behaviour in the Sun. The three models considered involve the dark matter particle interacting electromagnetically, through interactions via a dipole moment. Each of these contains unique momentum and velocity dependent properties, and may provide a potential solution to the problems arising from helioseismology. Firstly, each model will be introduced, highlighting their properties, behaviour and current experimental bounds. Then, we develop the theoretical ingredients for modelling the effect of electromagnetic dipole dark matter within the star. Two of the three models require development of non-trivial combinations of functional forms considered in earlier works. Finally, full simulations for the Sun are conducted with electromagnetic dipole dark matter using the \texttt{DarkStec} package~\cite{vincent15b}.

		\section{Electromagnetic dipole dark matter}
\label{sec:dipolemoments}

	We consider three models of electromagnetic dipole dark matter: electric dipole (ED), magnetic dipole (MD) and anapole (AN) dark matter. In each model, the dark matter particles posess their eponymous electromagnetic form factor and the principal interaction with baryonic matter is via the electromagnetic force.
	All of the models have been developed extensively in the literature in varying contexts \cite{pospelov00,sigurdson04,masso09,banks10,cho10,fitzpatrick10,barger11,heo11,barger12,delnobile12,fortin12,ho13,delnobile14,cabralrosetti14,gao14,gresham14,cabralrosetti16,mohanty15}.	
	Each has unique momentum and/or velocity dependence in the cross section for a dark matter particle scattering from a nucleus. The momentum dependence arises principally because the interaction is mediated by a massless photon, rather than a massive boson, and so the denominator of the propagator is $q^2$ instead of $q^2+m^2\simeq m^2$. Further momentum-dependent terms originate due to the magnetic form factor of the nucleus, and the operator describing the interaction. One model, the magnetic dipole, has been investigated before in the context of solar physics~\cite{lopes14}. However, the formalism used does not correctly account for the energy transport at the transition between local thermal equilibrium and Knusden transport, where dark matter effects are strongest~\cite{vincent15b,vincent16}. The effect of electric dipole or anapole dark matter in the Sun has not been examined before.
	
	Each of the dipole moments may be described by a parity-violating interaction term in the Lagrangian of the theory. For the electric dipole the interaction Lagrangian is:
	\begin{align}
		\label{eq:EDlagrangian}
		\mathcal{L}_{\ED} =&~ -\frac{i}{2}\mathcal{D}\overline{\chi}\sigma^{\mu\nu}\gamma^5\chi F_{\mu\nu} ;
	\intertext{whereas for the magnetic dipole:}
		\label{eq:MDlagrangian}
		\mathcal{L}_{\MD} =&~ \frac{1}{2}\mu_\chi \overline{\chi}\sigma^{\mu\nu}\chi F_{\mu\nu} ;
	\intertext{and for the anapole:}
		\label{eq:anapolelagrangian}
		\mathcal{L}_{\AN} =&~ \frac{g}{2\Lambda^2}\overline{\chi}\gamma^\mu \gamma^5 \chi \partial^\nu  F_{\mu\nu} ,
	\end{align}
	where $\chi$ is a spinor describing the dark matter particle, $F_{\mu\nu}$ is the electromagnetic field strength tensor, and $\mathcal{D}$, $\mathcal{\mu_\chi}$ and $\frac{g}{\Lambda^2}$ are the electric, magnetic and anapole moments of the dark matter particle respectively~\cite{masso09,delnobile14}. In the non-relativistic limit, the interaction term reduces to a coupling between the particles' spin and an electric field, magnetic field or electromagnetic current respectively~\cite{barger11,ho13}. 
	
	For each of these models, it is possible to construct the differential cross section for dipole moment dark matter scattering from a nucleus via photon exchange. For the electric dipole,
	\begin{equation}
		\label{eq:edmcs}
		\left(\frac{d\sigma}{dE_R}\right)_{\mathrm{ED}} = \frac{Z^2 e^2 \mathcal{D}^2}{4\pi E_R v^2}|F_E(E_R)|^2 ;
	\end{equation}	
	for the magnetic dipole, 
	\begin{equation}
		\label{eq:mdmcs}
		\left(\frac{d\sigma}{dE_R}\right)_{\mathrm{MD}} = \frac{e^2 \mu_\chi^2}{4\pi v^2}\left[Z^2\left(\frac{v^2}{E_R} - \frac{1}{2m_N} - \frac{1}{m_\chi}\right)|F_E(E_R)|^2 + \frac{I_N + 1}{3I_N} \frac{\mu_N^2}{\mu_p^2}\frac{m_N}{m_p^2}|F_M(E_R)|^2\right] ;
	\end{equation}	
	and for the anapole,
	\begin{equation}
		\label{eq:admcs}
		\left(\frac{d\sigma}{dE_R}\right)_{\mathrm{AD}} = \frac{e^2g^2 m_N}{2\pi v^2 \Lambda^4}\left[Z^2\left(v^2 - E_R \frac{m_N}{2M_{\chi,N}^2}\right)|F_E(E_R)|^2 + E_R\frac{I_N + 1}{3I_N} \frac{\mu_N^2}{\mu_p^2}\frac{m_N}{m_p^2}|F_M(E_R)|^2\right] ;
	\end{equation}
	where $Ze$ is the charge of the nucleus, $v$ is the incoming velocity of the dark matter particle in the lab frame, $m_N$ is the nucleus mass, $m_\chi$ is the dark matter mass, $m_p$ is the proton mass, $M_{\chi,N}$ is the dark matter-nucleus reduced mass, $I_N$ is the nucleus spin, $\mu_N$ is the nucleus magnetic moment, $\mu_p$ is the nuclear magneton and $E_R = \frac{q^2}{2m_N}$ is the recoil energy for momentum transfer $q$ and $e^2 = 4\pi\alpha$~\cite{barger11,delnobile12,delnobile14}. The expressions also contain electric and magnetic form factors $F_E$ and $F_M$, which parametrise the distributions of charge and current in the nucleus. For the present investigation, the form factors are assumed to be Gaussian~\cite{helm56}, following ref.~\cite{vincent15b,vincent16}
	\begin{equation}
		|F_{E,M}(E_R)|^2 = \exp\left(-\frac{E_R}{E_0}\right) \quad,
	\end{equation}
	where $E_0 = \frac{3}{2m_N r^2}$ for nuclear radius $r$, determined empirically as $r = (0.3 + 0.89 A^{\frac{1}{3}})\fm$ for mass number $A$. It forms a simplified parametrisation of the electric charge and magnetic current distributions within the nucleus.
	Although it has recently been shown~\cite{catena15} that modifications to the form factor may affect the capture rate of dark matter particles, it is worth noting in advance that for most of the regions of interest, the capture rate is saturated, making the result independent of the form factor. For the transport of dark matter particles within the Sun, it is assumed that the collisions are of sufficiently low velocity such that the particles can effectively be treated as point particles, as justified by ref.~\cite{vincent14,vincent15b,vincent16}.
	
	The permissible range for electromagnetic dipole dark matter in the dipole moment-mass parameter space has been narrowed significantly through experimental searches. In particular, dark matter direct detection experiments are often the most effective at providing constraints~\cite{pospelov00,sigurdson04,masso09,banks10,heo11,fitzpatrick10,delnobile12,cho10,delnobile14,gresham14,barger11,ho13,fortin12}. Whereas earlier investigations suggested that electromagnetic dipole dark matter may provide an explanation for the DAMA~\cite{bernabei08} annual modulation~\cite[e.g.][]{masso09}, more recent data from LUX~\cite{lux14} appears to rule it out as a solution~\cite[e.g.][]{gresham14}. Further constraints can be derived from collider~\cite{fortin12,barger12,gao14,cabralrosetti14,cabralrosetti16} and beam dump experiments~\cite{mohanty15}. The compilation of all of the experimental bounds provides order of magnitude limits on the dipole moments as $\mathcal{D} < 10^{-21}\ecm$~\cite{masso09}, $\mu_\chi < 10^{-4}\mu_p$~\cite{delnobile12,delnobile14} and $\frac{g}{\Lambda^2} < 10^{-4} \GeV^{-2}$~\cite{delnobile14}. Note however, that the bounds on low mass $(<5\GeV)$ dark matter particles may be more relaxed, due to the finite threshold of direct detection experiments.

		\section{Theory of dark matter in the Sun}
\label{sec:solartheory}

	Given a particular model of dark matter, the next step is to construct a theoretical formalism to account for the effect of the particles on the structure of the Sun. There are two quantities to consider, assuming that the ratio of dark matter particles to baryons is small and the dark matter is asymmetric: the population of particles in the Sun, and the energy transport due to those particles. The energy transport describes the changes in temperature between different regions of the Sun. However, the amount of energy transported depends on the number of particles within the Sun; more particles means more energy transport. Both are dependent on the properties of the dark matter model being investigated, and so both are considered in turn.

	\subsection{Population of dark matter particles in the Sun}
	\label{sec:solardmpopulation}

		The number of dark matter particles $N$ in the Sun is governed by a simple differential equation
		\begin{equation}
			\label{eq:dndt}
			\frac{dN}{dt} = C(t) - A(t) - E(t),
		\end{equation}
		where $C(t)$ is the rate at which the Sun captures dark matter particles, $A(t)$ is the rate at which dark matter particles annihilate, and $E(t)$ is the evaporation rate of dark matter particles~\cite{vincent15b}. The equation reflects the three modes by which the population of dark matter particles may be changed. The most prominent mode is the capture of particles from the galactic halo. The dark matter in the galaxy is assumed to exist as distinct particles moving within some velocity distribution. As the particles in the distribution approach the Sun, they experience a gravitational attraction focusing them towards the solar disk. The particles may then scatter from a nucleus inside the Sun. If the particle loses enough energy it is considered gravitationally bound. A full, mathematical description is presented below.
		
		The second process that affects the population of dark matter particles is annihilation. 
		If the dark matter particle is self-conjugate (for example, a Majorana fermion), it will self-annihilate. The self-annihilation will significantly reduce the population of particles.
		For the electric and magnetic dipole models, we avoid this complication automatically, since the interactions in eqs.~({\ref{eq:EDlagrangian}}) and ({\ref{eq:MDlagrangian}}) vanish exactly for Majorana fermions. Indeed, the anapole is the only allowed electromagnetic form factor allowed for Majorana fermions.
		
		Even if the dark matter is not self-conjugate, any significant proportion of antiparticles will suppress the total dark matter population in the Sun, and thus reduce the impact of conduction. 
		To maximise the effect of dark matter in our model, we assume that it is asymmetric~\cite{kaplan09,petraki13,blennow15,vincent15a,vincent15b,vincent16}. That is, self-annihilations are suppressed such that $A(t) \simeq 0$.
		The assumption is not strictly required, because if dark matter and its antiparticle have different cross sections, then symmetric dark matter may still be captured in large quantities~\cite{blennow15}.
		
		Finally, the population of dark matter particles may be depleted through evaporation. Evaporation occurs when a particle scatters to acquire a velocity greater than the local escape velocity of the Sun. Such an occurrence is relatively rare, and requires a significant increase in momentum. The typical velocity for a dark matter particle in the Sun is on the order of $100\kms$, whereas the escape velocity can approach $1400 \kms$~\cite{vincent15b}. The evaporation rate has been calculated analytically in refs.~\cite{gould87a,gould90b}. If the mass of the dark matter particle is greater than the threshold `evaporation mass', then the particle loss will be largely suppressed. 
		Preliminary results from an upcoming work~\cite{busoni16}suggest that, for momentum dependent dark matter analogous to the electric dipole model, the evaporation mass is at most $\sim 4 \GeV$. For simulations with greater dark matter masses, we can safely ignore the evaporation terms in eq.~(\ref{eq:dndt}). However, the magnitude of any modification to the solar structure by dark matter for masses less than $\sim 4\GeV$ is expected to be reduced for most of the models considered here. We tentatively include models at $m_\chi = 3 \GeV$, with the provision that the results should be treated with caution. The full calculation of the evaporation rates are expected to be presented in future works.

		The capture rate of dark matter particles is the only component of eq.~(\ref{eq:dndt}) that requires explicit evaluation. The expression for the capture rate was calculated for constant-cross section dark matter by ref.~\cite{gould87b}, and is straightforward to generalise to dark matter with momentum and velocity dependence, as for each of the electromagnetic dipole models~\cite{vincent15b,vincent16}.
		
		Consider dark matter particles moving in the galactic halo with velocity $u$. The probability that the particle has a velocity in the range $u+du$ is given by $f(u)du$ for some probability distribution $f(u)$. Typically, $f(u)$ is taken to be a Maxwell-Boltzmann distribution. In the solar rest frame, which moves with velocity $u_\odot = 220 \kms$ with respect to the galactic rest frame, the distribution is given by~\cite{scott09b}
		\begin{equation}
			\label{eq:boltzmann}
			f(u) = \left(\frac{3}{2}\right)^{\frac{3}{2}} \frac{4}{\sqrt{\pi}}\frac{\rho_\chi}{m_\chi}\frac{u^2}{u_0^3}\exp\left(-\frac{3(u_\odot^2+u^2)}{2u_0^2}\right)\frac{\sinh\left(3uu_\odot/u_0^2\right)}{3uu_\odot/u_0^2} ,
		\end{equation}	
		where $m_\chi$ is the mass of the dark matter particle, $\rho_\chi = 0.38 \GeV/cm^3$ is the local dark matter density and $u_0 = 270 \kms$ is taken as the velocity dispersion.

		The capture rate of dark matter particles with differential cross section $\frac{d\sigma}{dE_R}$ is given by~\cite{gould87b}	
		\begin{equation}
			\label{eq:capturerate}
			C=4\pi\int_0^{R_\odot}r^2 dr \int_0^\infty du \frac{f(u)}{u}w(r)\Omega(w),
		\end{equation}
		where $w(r) = \sqrt{u^2 + v_{\mathrm{esc}}^2(r,t)}$ is the velocity of the particle as it falls into the gravitational potential well of the Sun for escape velocity $v_{\mathrm{esc}}(r,t)$. The local rate of particle capture $\Omega(w)$ is given by
		\begin{equation}
			\label{eq:localcapturerate}
			\Omega(w) = w(r) \sum_i n_i(r,t) \int_{\frac{m_\chi u^2}{2}}^{\frac{m_\chi w^2 \mu}{2\mu_+^2}} \frac{d\sigma_i}{dE_R} dE_R
		\end{equation}	
		with $\mu= \frac{m_\chi}{m_N}$ and $\mu_\pm = \frac{\mu\pm1}{2}$, and the sum is over the nuclear species $i$ with number density $n_i$ at radius $r$. 
		The lower bound on the integration is a requirement that the velocity of the dark matter particle after the collision be smaller than the local escape velocity of the Sun. The upper bound is a kinematic constraint which describes the maximum possible recoil energy. The differential cross section $\frac{d\sigma}{dE_R}$ is integrated numerically for each of the cross sections discussed in eqs.~(\ref{eq:edmcs}-\ref{eq:admcs}).
		
%
%
%
%
%

		There is one final element of the capture rate calculation. There exists a maximum possible capture rate that corresponds to the case where the total cross section is equal to the cross sectional area of the Sun ($\sigma = \pi R_\odot^2$). In this limit, all particles that collide with the Sun from the halo are captured, and the Sun becomes opaque to dark matter~\cite{taoso10}. The saturation limit is independent of the model chosen, and can therefore be calculated analytically~\cite{vincent15b}:
		\begin{equation}
		\label{eq:saturation} 
			C_{\mathrm{max}} = \frac{1}{3} \pi\frac{\rho_\chi}{m_\chi}R_\odot^2(t)\left( e^{-\frac{3}{2}\frac{u_\odot^2}{u_0^2}}\sqrt{\frac{6}{\pi}}u_0 + \frac{6G_N M_\odot + R_\odot(u_0^2 + 3u_\odot^2)}{R_\odot u_\odot} \erf\left[\sqrt{\frac{3}{2}} \frac{u_\odot}{u_0}\right]\right).
		\end{equation}
		The total capture rate is then the lesser of eqs.~(\ref{eq:capturerate}) and (\ref{eq:saturation}).

		The cross sections for the electric, magnetic and anapole models, given in eqs. (\ref{eq:edmcs}), (\ref{eq:mdmcs}) and (\ref{eq:admcs}) respectively, have a dependence on the recoil energy and the incoming velocity. The capture rates for each model are then calculated by inserting these cross sections into eq.~(\ref{eq:localcapturerate}).  
		
		Because the capture rate is dependent on the elemental density $n$ at each time-step and radius, a fair comparison of the capture rates requires a complete model of the Sun. 
		\begin{figure}[t]
			\centering
				\IfFileExists{../plots/SICapRate.eps}{
					\subfloat{\includegraphics[width=0.48\textwidth]{../plots/SICapRate.eps}}
					\subfloat{\includegraphics[width=0.48\textwidth]{../plots/EDCapRate.eps}}
				}{
					\subfloat{\includegraphics[width=0.48\textwidth]{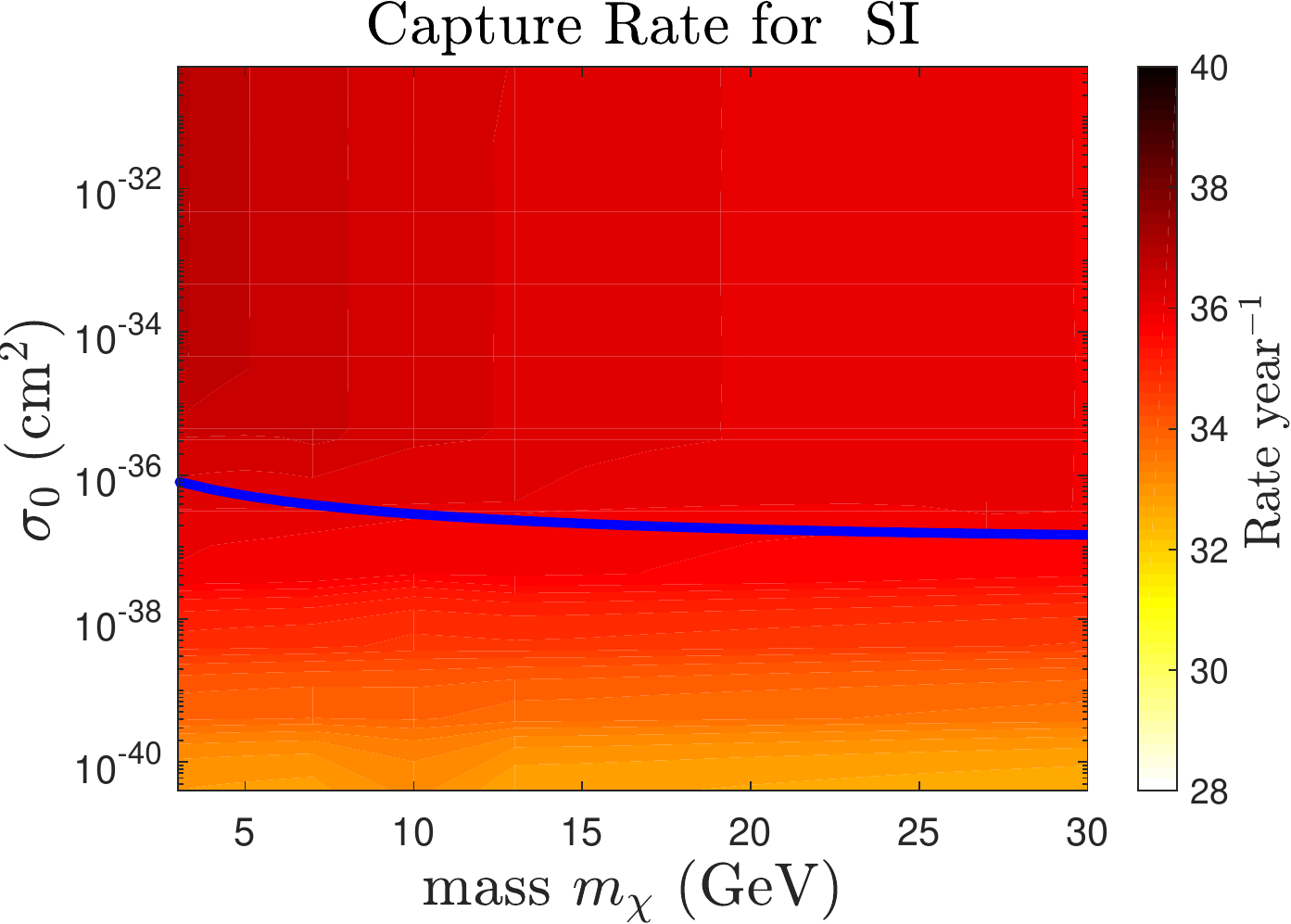}}
					\subfloat{\includegraphics[width=0.48\textwidth]{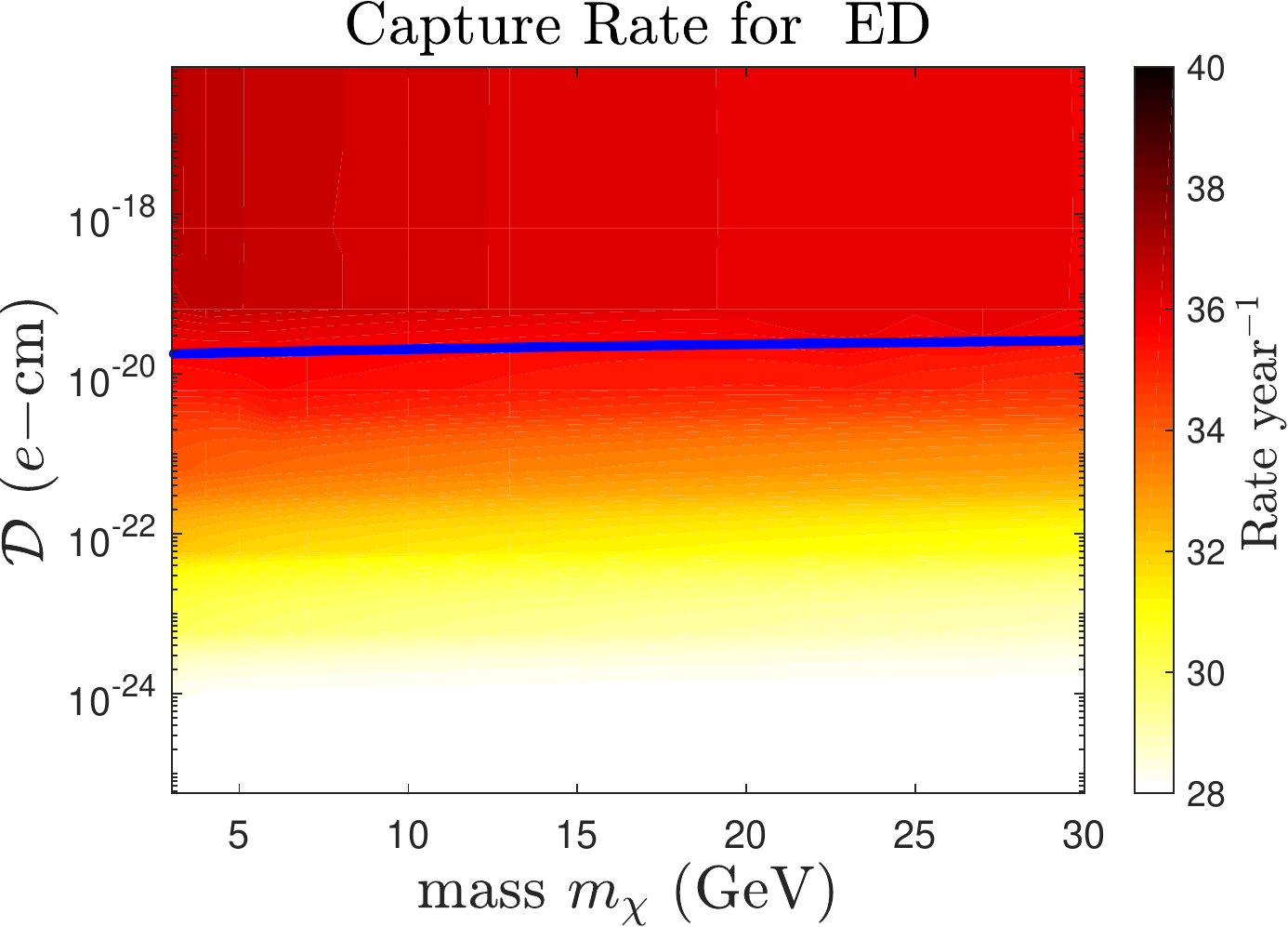}}
				}
				\IfFileExists{../plots/MDCapRate.eps}{
					\subfloat{\includegraphics[width=0.48\textwidth]{../plots/MDCapRate.eps}}
					\subfloat{\includegraphics[width=0.48\textwidth]{../plots/ANCapRate.eps}}
				}{
					\subfloat{\includegraphics[width=0.48\textwidth]{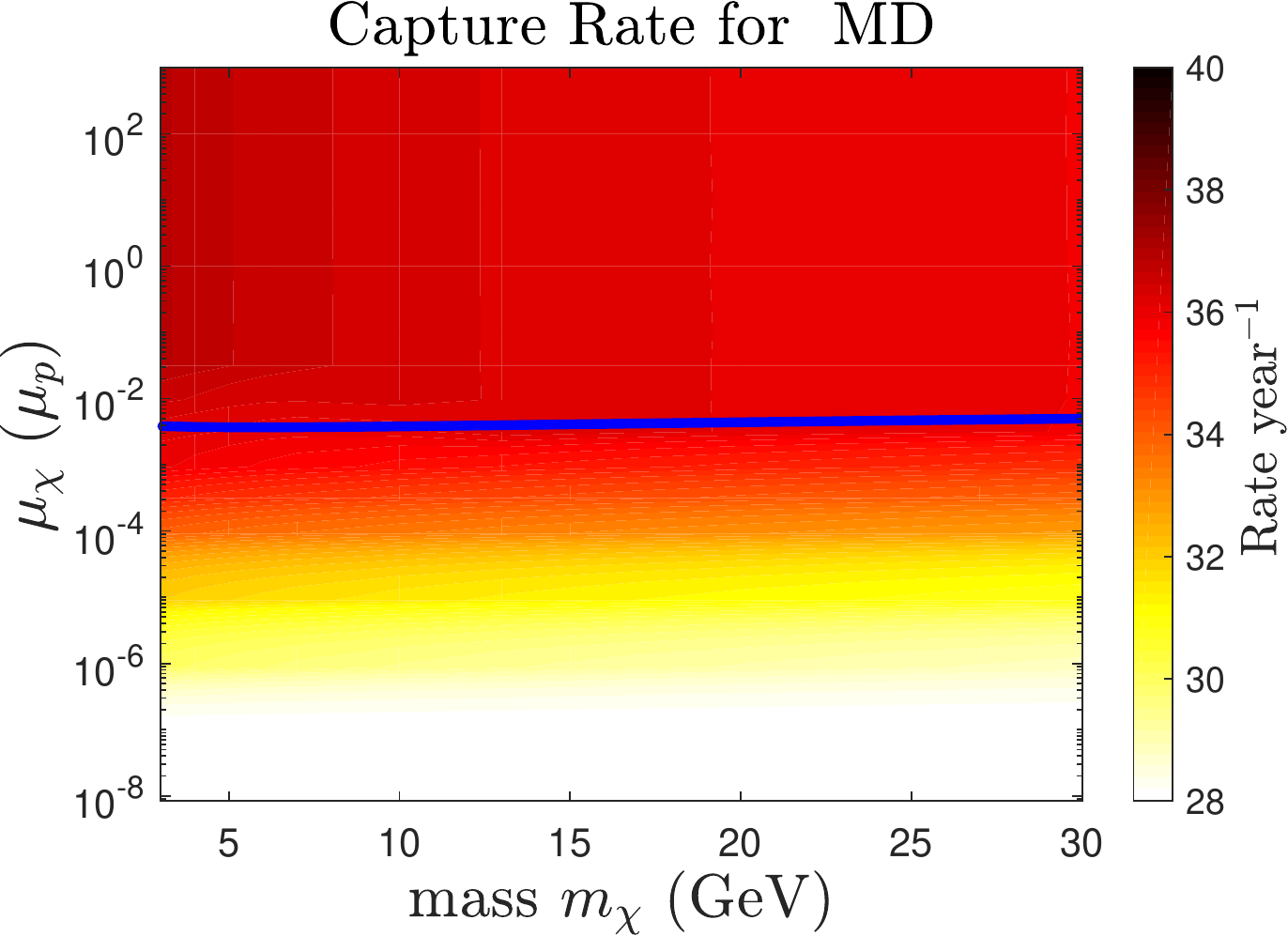}}
					\subfloat{\includegraphics[width=0.48\textwidth]{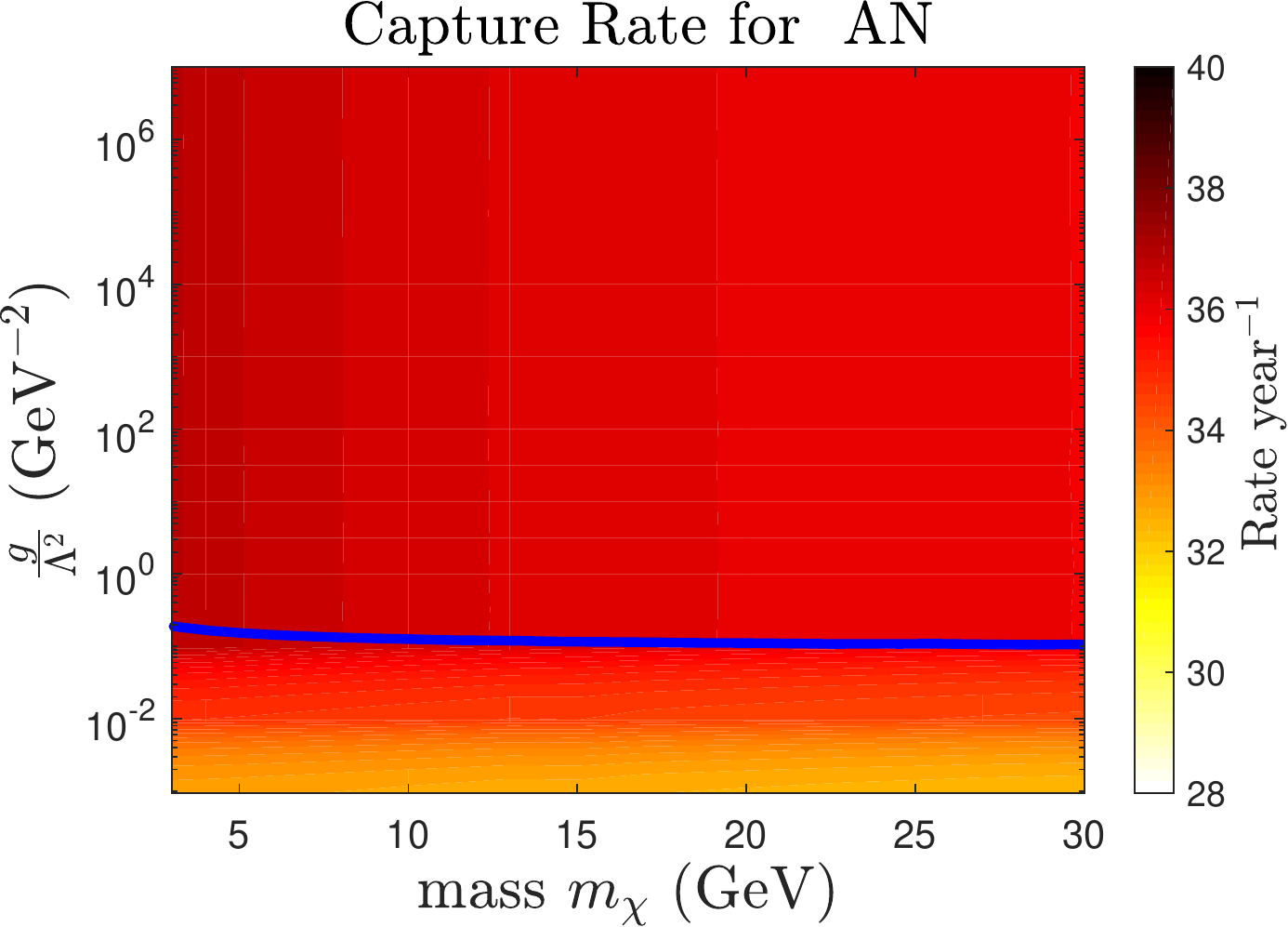}}
				}	
			
%
			\caption{Capture rate for spin independent dark matter (top left), electric dipole dark matter (top right), magnetic dipole dark matter (bottom left) and anapole dark matter (bottom right). The colour axis shows the base-10 logarithm of the capture rate in units of $\mathrm{year}^{-1}$. The point at which saturation first occurs is shown as a blue line.}
			\label{fig:capture}
		\end{figure}		
		Figure~\ref{fig:capture} compares solar-age capture rates of the AGSS09ph~\cite{asplund09,serenelli09} standard solar model (SSM) for different parameter combinations in the three electromagnetic models and one constant, spin-independent model.
		Models with a large capture rate appear red or black, and models with a small capture rate appear yellow or white. The cut-off from eq.~(\ref{eq:saturation}) is shown as a blue line, found by a few steps of a Newton-Raphson iteration. Models above the blue line are saturated, models below are not. The effect of the geometric cut-off is clear; for each model the capture rate increases significantly with the coupling strength until it reaches the cut-off (which is dependent on the mass of the particle), where it plateaus to a constant value. Above the cut-off, it is a weak function of the mass of the particle. 
		
		
	\subsection{Energy transport}
	\label{sec:transport}
	
		The dark matter density and luminosity as a function of radius within the Sun provide a description of the transport of energy due to dark matter particles. These two variables describe the full effect on solar observables and are therefore critical to calculate. The theoretical framework for the transport equations is extensively developed in refs.~\cite{gould90a,vincent14,vincent16}.
		After much calculation, the distribution of dark matter in the Sun in the limit of local thermal equilibrium is is given by~\cite{gould90a}
		\begin{equation}
			\label{eq:nchi}
				n_\chi(r) = n_\chi(0) \left[\frac{T(r)}{T(0)}\right]^{-\frac{3}{2}} \exp\left[-\int_0^r dr' \frac{k_B \alpha(r') \frac{dT(r')}{dr'} + m_\chi \frac{d\phi(r')}{dr'}}{k_BT(r')}\right] ,
		\end{equation}
		where $k_B$ is the Boltzmann constant, $T(r)$ is the temperature of the star, $\phi(r)$ is the gravitational potential within the star and $\alpha$ is the thermal diffusivity coefficient. The factors of $n_\chi(0)$ and $T(0)$ are the density and temperature in the core of the Sun, which normalise the total expression. The two terms in the exponential describe two physical processes holding the distribution in place. The $\frac{d\phi}{dr}$ term describes the gravitational pull of the Sun on the dark matter. The $\frac{dT}{dr}$ term has the opposite sign and describes the effect of conduction in the local thermal equilibrium limit on the distribution  of dark matter within the star. 
		
		The luminosity carried by dark matter scattering can be shown to be~\cite{gould90a}
		\begin{equation}
			\label{eq:Lchi}
			L_\chi(r) = 4\pi r^2 \kappa(r) n_\chi(r) l_\chi(r) \left[\frac{k_B T(r)}{m_\chi}\right]^{\frac{1}{2}} k_B \frac{dT}{dr},
		\end{equation}
		where $l_\chi = \left[\sum_i\sigma_in_i(\vec{r})\right]^{-1}$ is the mean free path of the dark matter particles. Here, the amount of energy carried is parametrised by $\kappa$, the dimensionless thermal conductivity. Both $\alpha$ and $\kappa$ are dependent on the model for dark matter-nucleon interactions. Therefore, computing $n_\chi$ and $L_\chi$ becomes an exercise in calculating $\alpha$ and $\kappa$.
		The thermal conductivity can be calculated from the first order expansion of a Boltzmann collision equation in the phase space distribution $F$ as~\cite{gould90a,vincent14}
		\begin{equation}
			\label{eq:alpha}
			\alpha = \frac{\langle y|C^{-1}|y^3 f_{0}^{0,0}\rangle}{\langle y|C^{-1}|y f^{0,0}_0\rangle},
		\end{equation}
		with the Dirac notation defined as
		\begin{equation}	
		\langle g|Q|f \rangle \equiv \int dydx~g(y)Q(y,x)f(x),
		\end{equation}
		normalised velocities as $\vec{x}=\vec{v}/\vec{v_T}$, $\vec{y}=\vec{u}/\vec{v_T}$ for incoming velocity $\vec{v}$, outgoing velocity $\vec{u}$ and thermal velocity $v_T = \sqrt{\frac{2k_B T(r)}{m_\chi}}$.
		$f^{j,m}_\nu$ is the $j,m$ spherical harmonics of the normalised phase space distribution expanded to the $\nu$th order about the parameter $l_\chi(r)\left|\nabla \ln T(r)\right| \ll 1$ (which imposes that the inter-scattering distance is smaller than the length scale of temperature change):
		\begin{equation}
			f_{\nu}^{j,m}(x,\vec{r})dx = \frac{1}{n_\chi(r)} F_{\nu}^{j,m}(\vec{v},\vec{r})dv.
		\end{equation}
		In particular, $f_{0}^{0,0}$ is the non-interacting isotropic Maxwell-Boltzmann velocity distribution and $f_{1}^{1,0}$ characterises the heat conduction through the presence of a dipole term. The collision operator $C$ between the DM particle and a given nuclear species is defined as 
		\begin{equation}
			\label{eq:captureoperator}
			CF = \int d^3 v C_{\mathrm{in}}(\vec{u},\vec{v},\vec{r},t)F(\vec{v},\vec{r},\vec{t}) - C_{\mathrm{out}}(\vec{u},\vec{r},t)F(\vec{u},\vec{r},t).
		\end{equation}
		In a similar manner, the thermal conductivity is defined by
		\begin{equation}
			\label{eq:kappa}
			\kappa = \frac{\sqrt{2}}{3}\langle y^3 |f^{1,0}_{1}\rangle,
		\end{equation}
		which is related to the conductivity by
		\begin{equation}
			\label{eq:alpha2kappa}
			(\alpha y - y^3) f_{0}^{0,0}(y,r) = \int dx C(y,x,r) f_{1}^{1,0}(x,r).
		\end{equation}
		$\alpha$ and $\kappa$ depend only on the ratio $\mu = \frac{m_\chi}{m_N}$. They can be generalized to multiple nuclear species via:
		\begin{equation}
			\label{eq:alphai}
			\alpha(r,t) = l_\chi(r,t) \sum_i \sigma_i n_i(r,t) \alpha_i(\mu)
		\end{equation}
		and
		\begin{equation}
			\label{eq:kappai}
			\kappa(r,t) = \left[l_\chi(r,t) \sum_i \frac{\sigma_i n_i(r,t)}{\kappa_i(\mu)}\right]^{-1}
		\end{equation}
		where $\sigma_i$ is the total, thermally averaged cross section for scattering from the $i$th nuclear species with number density $n_i$, and $\mu = m_\chi/m_N$ is the mass ratio for nuclear mass $m_N$. 
		
		
		Eqs.~(\ref{eq:nchi}) and (\ref{eq:Lchi}) provide a formalism for computing the transport due to dark matter in the Sun. The nature of the transport is governed by the parameters $\alpha_i$ and $\kappa_i$ in eq.~(\ref{eq:alphai}) and eq.~(\ref{eq:kappai}). Therefore, to compute the effect of dipole moment dark matter, it is necessary to compute $\alpha_i$ and $\kappa_i$ for each model.

		\subsubsection{Knusden vs. LTE Transport}
		
		The derivations of eq.~(\ref{eq:nchi}--\ref{eq:Lchi}) assume that the dark matter is in local thermal equilibrium with the surrounding material of the star. However, this approximation can break down when the mean free path of the interaction $l_\chi \gg r_\chi$, where $r_\chi$ is defined as the scale height
		\begin{equation}
			\label{eq:rchi}
			r_\chi = \left(\frac{3 k_B T_c}{2\pi G \rho_c m_\chi}\right)^{\frac{1}{2}}.
		\end{equation}
		When this condition is violated, the energy transfer shifts to the Knusden regime. Here, the particles essentially travel beyond the temperature scale height, and heat transport becomes non-local. The luminosity must therefore be modified to account for non-local interactions. An empirical correction gives the corrected luminosity as
		\begin{equation}
			L_{\chi,\mathrm{tot}}(r,t) = \mathfrak{f}(K)\mathfrak{h}(r,t)L_{\chi}(r,t)
		\end{equation}
		for $L_\chi$ defined in eq.~(\ref{eq:Lchi})~\cite{bottino02,scott09b}. Similarly, the dark matter distribution is modified as
		\begin{equation}
			n_{\chi}(r) = \mathfrak{f}(K)n_{\chi,\mathrm{LTE}} + [1 - \mathfrak{f}(K)] n_{\chi,\mathrm{iso}}
		\end{equation}
		for $n_{\chi,\mathrm{LTE}}$ from eq.~(\ref{eq:nchi}) and 
		\begin{equation}
			n_{\chi,\mathrm{iso}}(r) = N(t) \frac{e^{-\frac{r^2}{r_\chi^2}}}{\pi^{\frac{3}{2}}r_\chi^3},
		\end{equation}
		where $N(t)$ is the total number of dark matter particles.
		The parameters $\mathfrak{f}(K)$ and $\mathfrak{h}(r,t)$ are defined as
		\begin{align}
			\mathfrak{f}(K) =&~ \frac{1}{1+\left(\frac{K}{K_0}\right)^{\frac{1}{\tau}}};
			\\
			\mathfrak{h}(r,t) =&~ \left(\frac{r - r_\chi}{r_\chi}\right)^3 + 1,
		\end{align}
		for $K = \frac{l_\chi}{r_\chi}$ and empirically determined constants $K_0 = 0.4$ and $\tau = 0.5$~\cite{gould90c}.

		\subsubsection{Energy transport due to electric dipole dark matter}
		
			The simplest case is the electric dipole moment, for which the cross section eq.~(\ref{eq:edmcs}) is only dependent on $q^{-2}$:
			\begin{equation}
				\left(\frac{d\sigma}{d\cos\theta_{\mathrm{CM}}}\right)_{\mathrm{EDM}} = \frac{Z^2 e^2 \mathcal{D}^2 M_{\chi,N}^2}{4\pi q^2} = \frac{Z^2 e^2 \mathcal{D}^2}{4\pi v^2(1-\cos\theta_{\mathrm{CM}})}.
			\end{equation}
			The thermal parameters for the $q^{-2}$ dependent cross sections have already been determined by ref.~\cite{vincent14}. There, a generalised cross section is introduced:
			\begin{equation}
				\label{eq:q2n}
				\frac{d\sigma}{d\cos\theta_{\mathrm{CM}}} = \sigma_0 \left(\frac{q}{q_0}\right)^{-2}.
			\end{equation} 
			Hence, we set $\sigma_0 = \frac{Z^2 e^2 \mathcal{D}^2}{4\pi}$ and $q_0 = M_{\chi,N}$ and compute directly from the tables in ref.~\cite{vincent14}. However, an issue arises for the case where $\theta_{\mathrm{CM}}\rightarrow 0$, as the differential cross section diverges. To regulate the divergence, as in ref.~\cite{vincent14} we use the momentum transfer cross section, defined by~\cite{krstic99,tulin13}
			\begin{equation}
				\left(\frac{d\sigma}{d\cos\theta_{\mathrm{CM}}}\right)_T = (1-\cos\theta_{\mathrm{CM}})\frac{d\sigma}{d\cos\theta_{\mathrm{CM}}}.
			\end{equation}
			The use of the momentum transfer cross section is justified since the divergence only occurs for forward scattering, which for elastic scattering corresponds to no momentum transferred. Energy transport can only occur if momentum is transferred. The momentum transfer cross section removes the forward scattering, leaving behind the components which do transfer energy. 
			
			The thermally averaged cross section for electric dipole dark matter is
			\begin{equation}
				\langle\sigma_{ED}\rangle = Z^2 e^2 \mathcal{D}^2 \pi^{-1}(1+\mu)^{-1}v_T^{-2}.
			\end{equation}				
			
		\subsubsection{Energy transport due to magnetic dipole dark matter}
		
			The magnetic dipole cross section in eq.~(\ref{eq:mdmcs}) is necessarily more complicated than the electric dipole case; there is a $v^2 q^{-2} + constant$ dependence. Neither the mixed term $v^2 q^{-2}$ nor the combination of two terms is tabulated in refs.~\cite{gould90a,vincent14}. Combining two terms is non-trivial due to the inverse procedure to find $C^{-1}$ for eq.~(\ref{eq:alpha}). Therefore, it is necessary to compute new tables for $\alpha$ and $\kappa$ for the magnetic dipole moment.
			
			In order to calculate $\alpha$, an explicit expression for $C$ is required. By eq.~(\ref{eq:captureoperator}), there are two components, $C_{\mathrm{out}}$ and $C_{\mathrm{in}}$. $C_{\mathrm{out}}$ is relatively straightforward to calculate, and is given by
			\begin{equation}
				\label{eq:cout}
				C_{\mathrm{out}} = \int d^3 z \left|\vec{x}-\vec{z}\right| \hat{\sigma}_{\mathrm{tot}}(v_T|\vec{x}-\vec{z}|)F_{\mathrm{nuc}},
			\end{equation}
			where the thermal velocity distribution of the nuclei is
			\begin{equation}
				F_{\mathrm{nuc}}(\vec{z}) = (\pi\mu)^{-\frac{3}{2}}e^{-\frac{|\vec{z}|^2}{\mu}}
			\end{equation}
			with $\mu= \frac{m_\chi}{m_N}$ as the mass ratio, and  $\vec{x} = \frac{\vec{v}}{v_T}$ and $\vec{z} = \frac{\vec{v}_\mathrm{nuc}}{v_T}$ are dimensionless velocities defined about the thermal velocity $v_T$. Note that $\hat{\sigma}_{\mathrm{tot}}$ is the total dimensionless cross section, defined such that $\hat{\sigma}_{\mathrm{tot}}(v_T) = 1$. $C_{\mathrm{in}}$ is more complicated:
			\begin{equation}
			\begin{split}
				\label{eq:cin}
				C_{\mathrm{in}}^j(y,x,r) = (1+\mu)^4 \frac{y}{x}&~\int_0^\infty da \int_0^\infty db F_{\mathrm{nuc}}(\vec{z}) 2\pi b \langle P_j \hat{\sigma}\rangle \\&~\times\Theta(y-|a-b|)\Theta(a+b-y)\Theta(x-|a-b|)\Theta(a+b-x),
			\end{split}
			\end{equation}
			where $\Theta$ is a Heaviside step function and $\langle P_j\hat{\sigma}\rangle$ represents the angle-averaged product of the normalised differential cross section and the $j$th Legendre polynomial, expanded around the transverse scattering angles in the lab frame~\cite{gould90a}. To first order,
			\begin{equation}
				\label{eq:pjsigma}
				\langle P_j \hat{\sigma}\rangle = \langle P_1\hat{\sigma} \rangle = \frac{1}{2\pi} \int_0^{2\pi} d\phi\left( G+ B\frac{b^2}{xy}\cos\phi\right)\hat{\sigma}\left[(1+\mu)bv_T,A+B\cos\phi\right],
			\end{equation}
			where 
			\begin{align}
				A =&~ \frac{(x^2-a^2-b^2)(y^2-a^2-b^2)}{4a^2b^2},
				\\
				G =&~ \frac{(x^2+a^2-b^2)(y^2+a^2-b^2)}{4a^2xy},
				\\
				B^2 = &~ 1- \frac{A^2}{G^2} - G^2 + A^2\quad,
			\end{align}
			and the cross section is evaluated at $v_\mathrm{rel} = (1+\mu)bv_T$ and $\cos\theta = A+B\cos\phi$~\cite{vincent14}. Given the cross section from eq.~(\ref{eq:mdmcs})
			\begin{equation}
				\left(\frac{d\sigma}{d\cos\theta_{\mathrm{CM}}}\right)_{\mathrm{MDM}} = 	\frac{e^2\mu_\chi^2}{4\pi}
				\left[Z^2\left(\frac{2M_{\chi,N}^2v^2}{q^2} + \frac{1}{2(1+\mu)^2} - \frac{1}{2}\right) + 	\frac{I_N+1}{3I_N}\frac{M_{\chi,N}^2}{m_p^2}\frac{\mu_N^2}{\mu_p^2}\right],
			\end{equation}
			where the form factors are neglected as the momentum transfer is small,
			the angle-averaged Legendre Polynomial is 
			\begin{equation}
				\label{eq:p1sigmamagdip}
				\langle P_1\hat{\sigma}\rangle = \frac{G}{2} - \frac{\left(2GA + B^2\frac{b^2}{xy}\right)\left(1+(2S-1)(1+\mu)^2\right)}{4(2S+1)(1+\mu)^2+4},
			\end{equation}
			where $S$ is defined as
			\begin{equation}
				\label{eq:S}
				S = \frac{1}{Z^2}\frac{I_N+1}{3I_N} \frac{M_{\chi,N}^2}{m_p^2} \frac{\mu_N^2}{\mu_p^2},
			\end{equation}			
			which arises only if the scattering target has a non-zero magnetic dipole moment $\mu_N$. The non-trivial values of $S$ used in our simulations are:
			\begin{subequations}
			\label{eq:Svalues}
			\begin{flalign}
			&\text{Hydrogen-1:} &
				S_{\mathrm{H}^1} =&~ 2.79285 \frac{\mu^2}{(1+\mu)^2}. &
				\\
			&\text{Nitrogen-14:} &
				S_{\mathrm{N}^{14}} =&~ 1.076696 \frac{\mu^2}{(1+\mu)^2}. &
				\\
			&\text{Sodium-23:} &
				S_{\mathrm{Na}^{23}} =&~ 5.385992 \frac{\mu^2}{(1+\mu)^2}. &
				\\
			&\text{Aluminium-27:}  &
				S_{\mathrm{Al}^{27}} =&~ 7.330418 \frac{\mu^2}{(1+\mu)^2}. &
			\end{flalign}
			\end{subequations}
			The magnetic dipoles for each element are taken from the tabulated and collated results of ref.~\cite{stone05}. Note that the magnetic dipole also requires the use of the momentum transfer cross section to regulate the forward scattering divergence, in a similar manner to the electric dipole moment. Finally, the thermally averaged cross section for magnetic dipole dark matter is
			\begin{equation}
				\langle\sigma_{MD}\rangle = \frac{e^2 \mu_\chi^2}{2\pi}\left[Z^2\left(1+\frac{1}{(1+\mu)^2}\right) + \frac{I_N + 1}{3I_N} \frac{M_{\chi,N}^2}{m_p^2}\frac{\mu_N^2}{\mu_p^2}\right].
			\end{equation}

		\subsubsection{Energy transport due to anapole dark matter}
		
			The anapole moment cross section in eq.~(\ref{eq:admcs}) presents a similar problem to the magnetic dipole moment. Here, the dependence is $q^2 + v^2$, which means that there is a non-trivial combination of two terms for $C^{-1}$. Therefore, the values of $\alpha$ and $\kappa$ need to be retabulated for this case as well. Given the cross section from eq.~(\ref{eq:admcs}),
			\begin{equation}
				\label{eq:anapolethetaCM}
				\left(\frac{d\sigma}{d\cos\theta_{\mathrm{CM}}}\right)_{\mathrm{ADM}} = \frac{M_{\chi,N}^2}{2\pi}\frac{e^2 g^2}{\Lambda^4}\left[Z^2\left( v^2 - \frac{q^2}{4M_{\chi,N}^2}\right) + \frac{I_N+1}{3I_N} \frac{q^2}{2m_N^2}\frac{\mu_N^2}{\mu_p^2}\frac{m_N^2}{m_p^2}\right],
			\end{equation}
			the angle averaged Legendre polynomial is
			\begin{equation}
				\label{eq:p1sigmaanapole}
				\langle P_1 \hat{\sigma}\rangle = \frac{\langle P_1 \hat{\sigma}_{v^2}\rangle}{1 + \frac{1}{2}\left(2S-1\right)} + \frac{\langle P_1 \hat{\sigma}_{q^2}\rangle}{{2\left(2S-1\right)^{-1}} + 1},
			\end{equation}
			with
			\begin{align}
				\label{eq:p1v2}
				\langle P_1 \hat{\sigma}_{v^2}\rangle = &~ \frac{1}{2}b^2 G(1+\mu)^2;
				\\
				\label{eq:p1q2}
				\langle P_1 \hat{\sigma}_{q^2}\rangle = &~ \frac{1}{2}b^2(1+\mu)^2 \left[G(1-A), -\frac{b^2B^2}{2xy}\right]
			\end{align}			
			as in ref.~\cite{vincent14}. Lastly, the thermally averaged cross section is 			
			\begin{equation}
				\langle\sigma_{AN}\rangle = \frac{M_{\chi,N}^2 e^2 g^2}{\Lambda^4} \frac{3}{4\pi}(1+\mu)v_T^2\left[Z^2 + \frac{I_N+1}{3I_N}\frac{2M_{\chi,N}^2}{m_N^2}\frac{\mu_N^2}{\mu_p^2}\frac{m_N^2}{m_p^2}\right].
			\end{equation}
			
		\subsubsection{Tabulation of diffusivity and conductivity coefficients}
		\label{sec:tabulation}
		
			\begin{figure}[h]
				\centering
					\IfFileExists{../ak_files/alpha.eps}{
						\subfloat{\includegraphics[width=0.47\textwidth]{../ak_files/alpha.eps}}
						\subfloat{\includegraphics[width=0.47\textwidth]{../ak_files/kappa.eps}}
					}{	
						\subfloat{\includegraphics[width=0.47\textwidth]{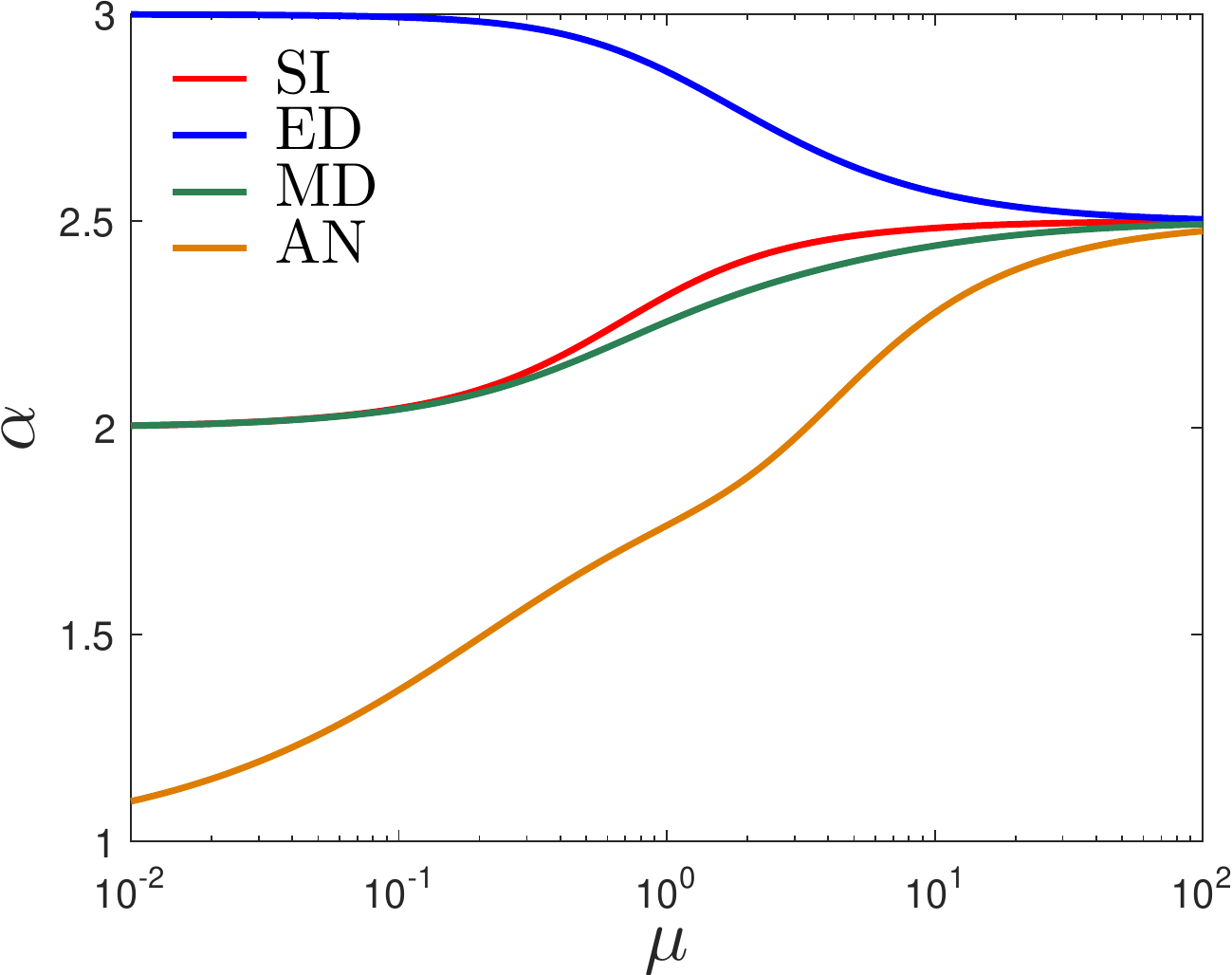}}
						\subfloat{\includegraphics[width=0.47\textwidth]{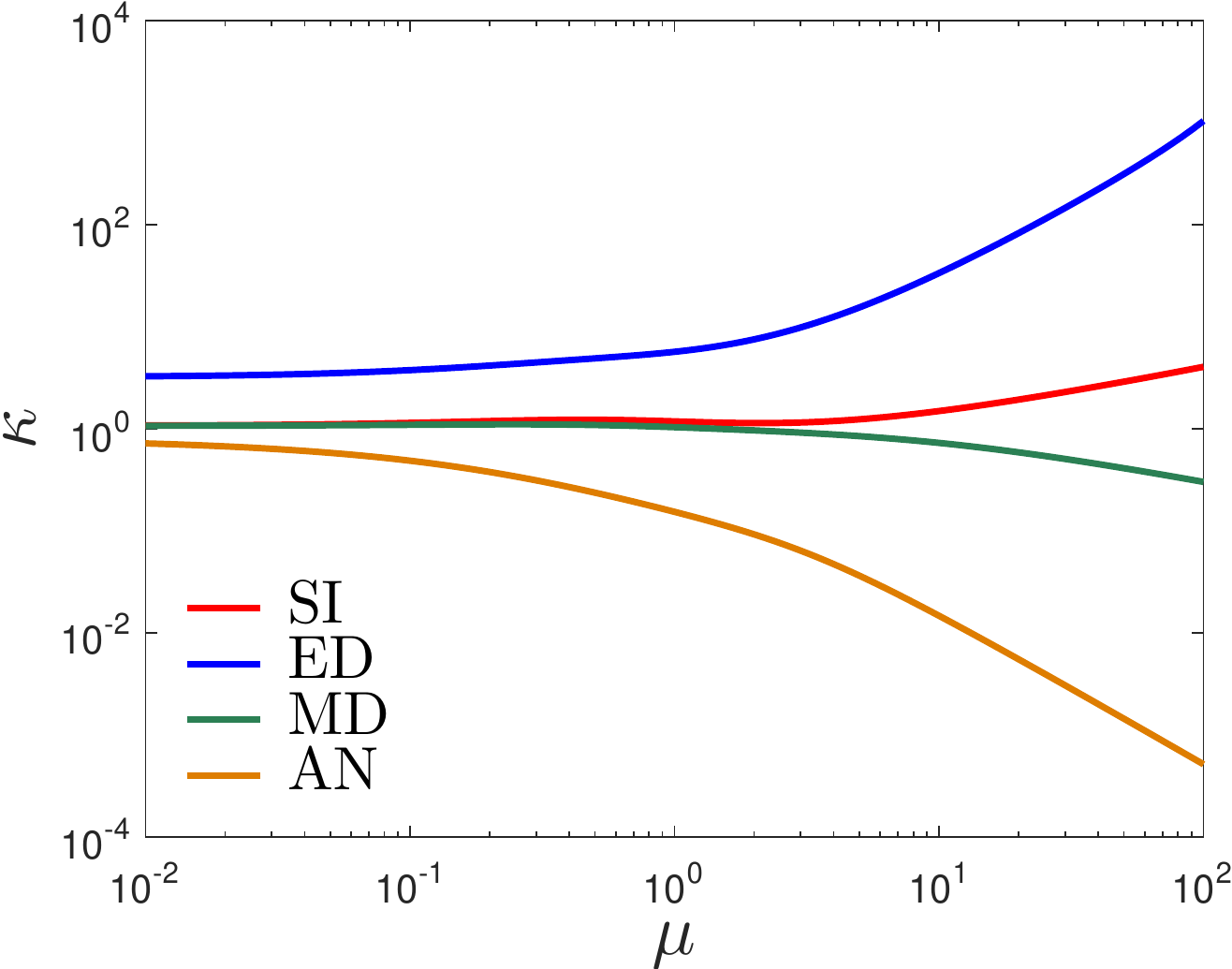}}
					}
					\IfFileExists{../ak_files/kappa.eps}{
						\subfloat{\includegraphics[width=0.47\textwidth]{../ak_files/alphamdm.eps}}
						\subfloat{\includegraphics[width=0.47\textwidth]{../ak_files/kappamdm.eps}}
					}{
						\subfloat{\includegraphics[width=0.47\textwidth]{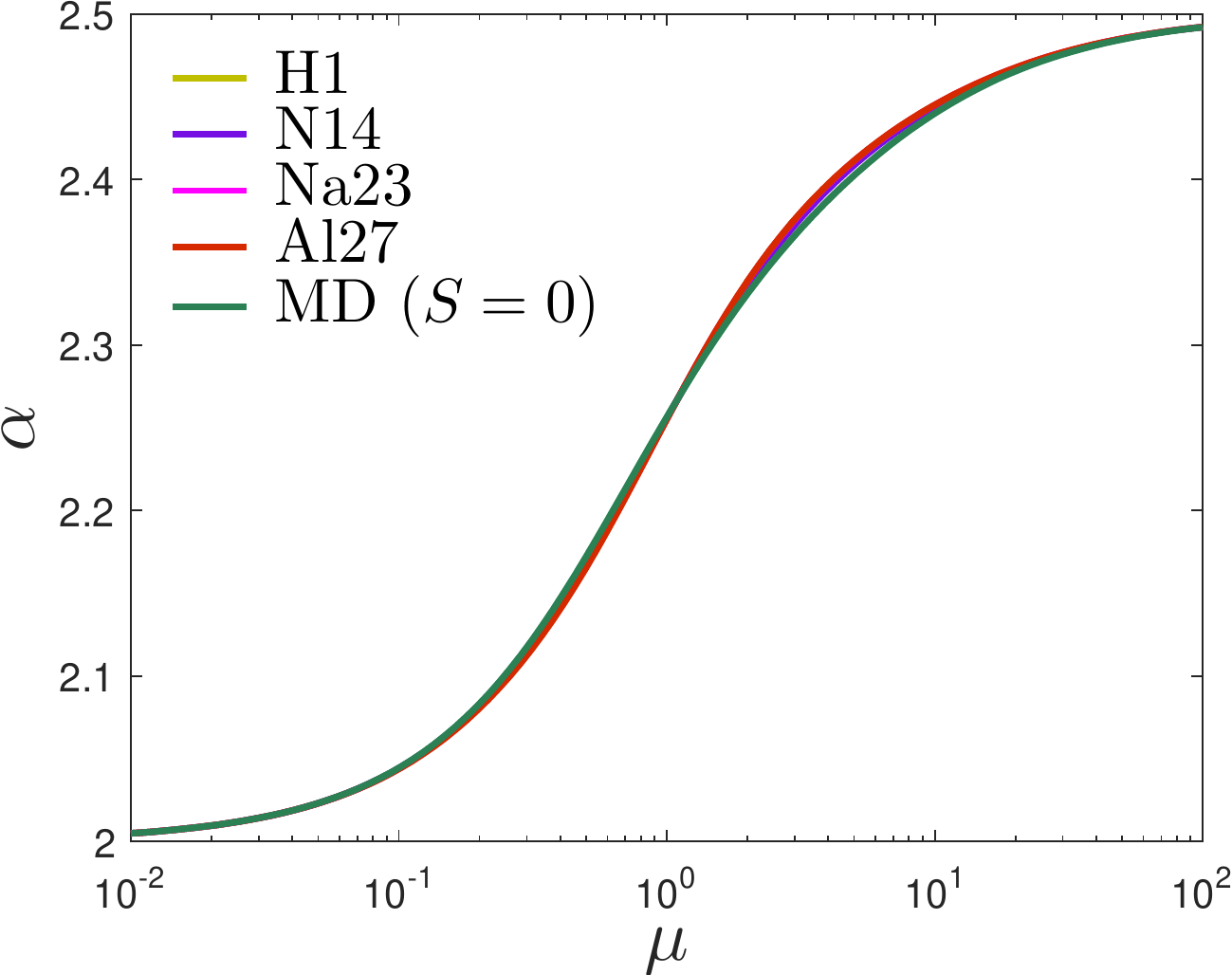}}
						\subfloat{\includegraphics[width=0.47\textwidth]{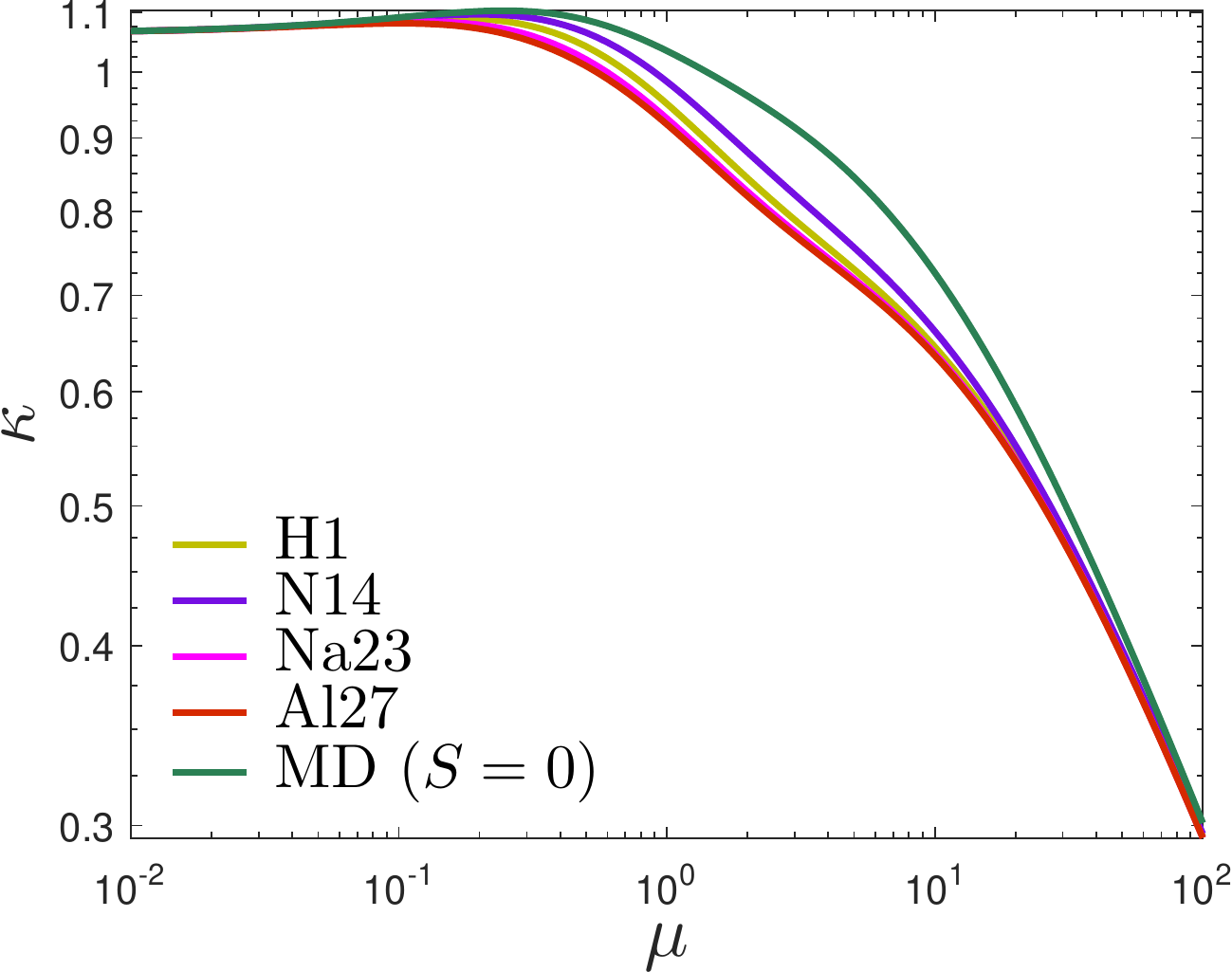}}
					}
					\IfFileExists{../ak_files/kappa.eps}{
						\subfloat{\includegraphics[width=0.47\textwidth]{../ak_files/alphaan.eps}}
						\subfloat{\includegraphics[width=0.47\textwidth]{../ak_files/kappaan.eps}}
					}{
						\subfloat{\includegraphics[width=0.47\textwidth]{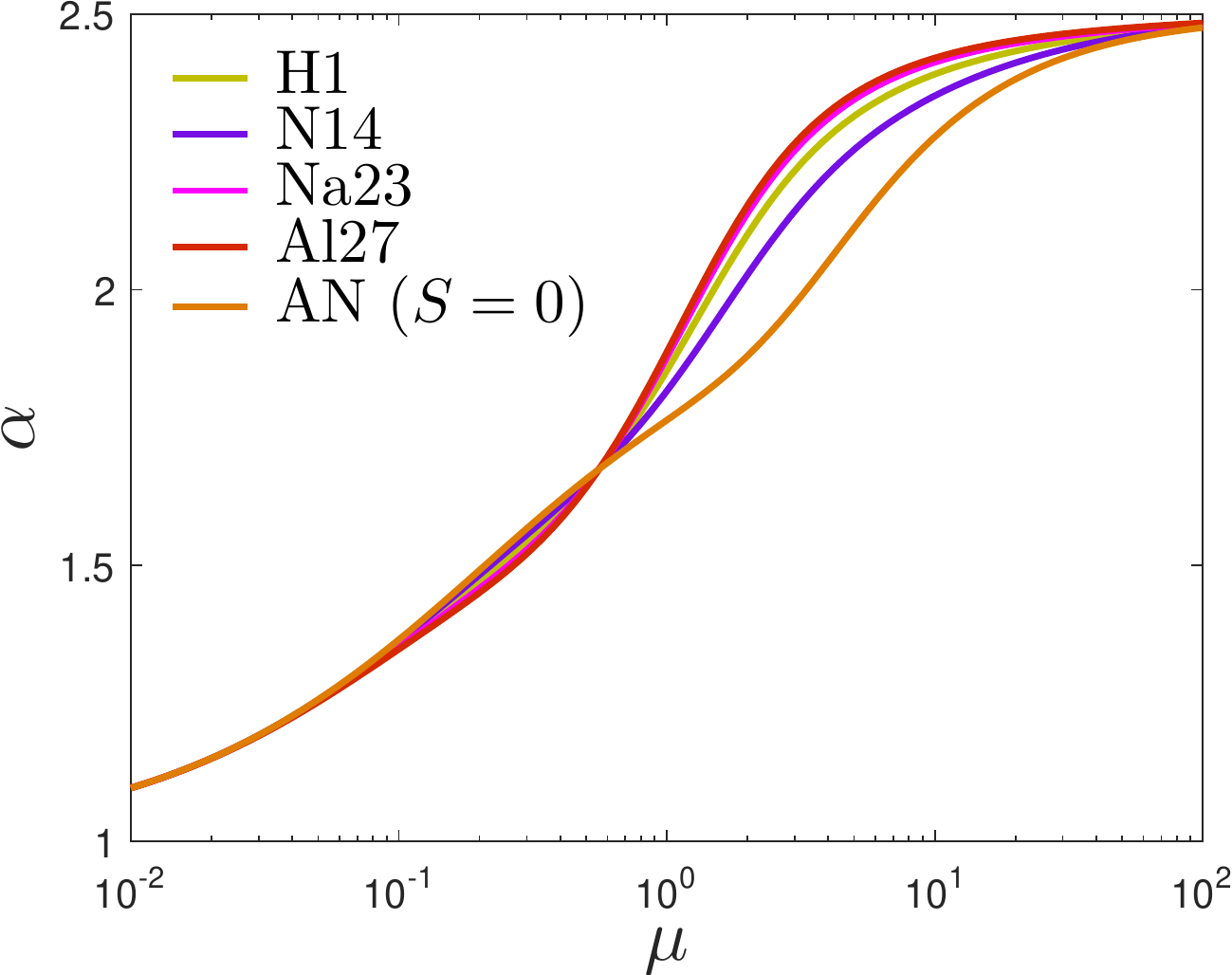}}
						\subfloat{\includegraphics[width=0.47\textwidth]{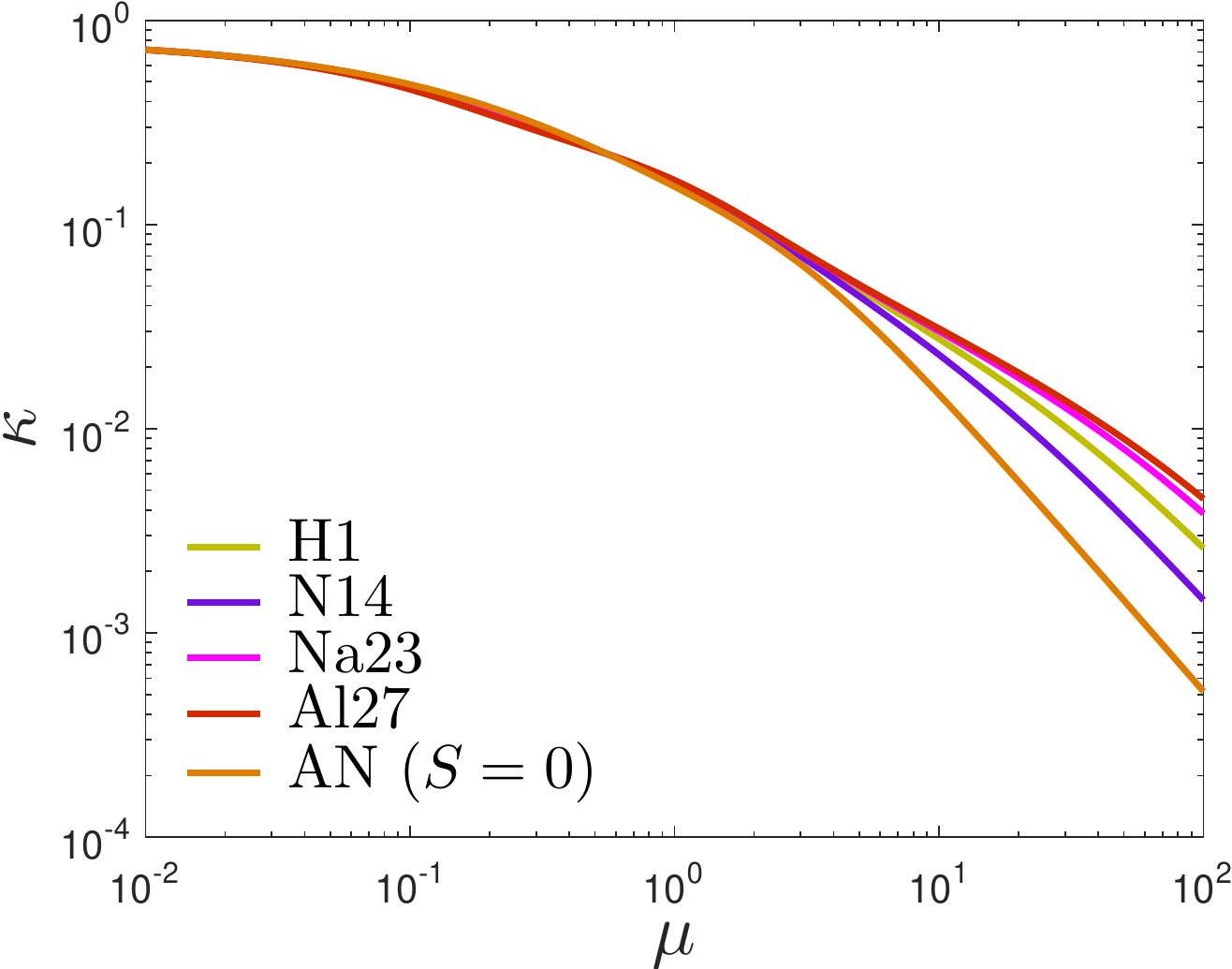}}
					}
				\caption{Dimensionless thermal diffusivity $\alpha$ (left) and conductivity $\kappa$ (right) as a function of mass ratio $\mu$. The plots at the top compare values for spin-independent (SI), electric dipole moment (ED), magnetic dipole moment (MD), and anapole moment (AN), with $S=0$ for the latter two. The middle and bottom graphs compare the $\alpha$ (left) and $\kappa$ (right) for non-zero values of $S$ for the magnetic dipole (middle) and anapole (bottom) respectively (see Eqs.~(\ref{eq:p1sigmamagdip}) and (\ref{eq:p1sigmaanapole})).}
				\label{fig:alphakappa}
			\end{figure}
			
			Using the definitions of $C_{\mathrm{in}}$ and $C_{\mathrm{out}}$ in Eqs.~(\ref{eq:cout}) and (\ref{eq:cin}), it is possible to calculate $\alpha$ and $\kappa$ using Eqs.~(\ref{eq:alpha}) and (\ref{eq:kappa}) respectively. In order to compute these values, we discretise the lower order spherical harmonics of the phase space distribution $f_{0}^{0,0}$ as a 500 element vector of the normalised velocity distribution, with velocities ranging from $y = 0$ to $y = 5$, i.e. five times the average thermal velocity of the Sun. The collision operator $C$ is then formed as a $500\times500$ element matrix, which is explicitly inverted to find $\alpha$. The collision operator, combined with $\alpha$, then may be used to calculate the first order spherical harmonics of the phase space distribution $f_{1}^{1,0}$ using the eq.~\ref{eq:alpha2kappa}.
			From here, a value for $\kappa$ may be obtained from eq.~(\ref{eq:kappa}). Dimensions can then be re-inserted using Eqs.~(\ref{eq:alphai}) and (\ref{eq:kappai}). We calculate for a range of the mass ratio $\mu$ of $10^{-2}$ to $10^2$.
			
			The result of the calculations is a tabulation of $\alpha$ and $\kappa$ as a function of the mass ratio $\mu$. For the electric dipole moment, this table is identical to ref.~\cite{vincent14}. For the magnetic dipole moment and anapole moment, the tabulation is new. The thermal cross sections for the magnetic dipole and anapole moments contain a term $S$ which only appears if the nucleus has non-zero spin. Thus, for the four elements with non-zero spin tracked in our simulations, a separate table must be produced as each element produces a different value of $S$. The values of $\alpha(\mu)$ and $\kappa(\mu)$ are plotted in figure~\ref{fig:alphakappa}.
		
			The values of $\alpha$ and $\kappa$ may be compared between the electromagnetic dipole models and the standard constant cross section, spin-independent case. 
			For the electric dipole moment, $\alpha$ is enhanced relative to the spin independent case for small $\mu$, whereas for large $\mu$, $\kappa$ is greatly enhanced. Therefore, for small masses, the distribution is spread, with less momentum transfer; but for large masses the distribution is more compact, with more momentum transfer. The behaviour arises due to the $q^{-2}$ dependence in the electric dipole cross section. The $q^{-2}$ component derives from the photon propagator, and is a signature of electromagnetic theories. It does not appear in the standard 4-point spin-independent contact interaction, where the new heavy mediator that has been integrated out is implicitly assumed to have a mass much larger than the transferred momentum.

			For the magnetic dipole moment, the diffusivity $\alpha$ shows only minimal deviation from the constant cross section, spin independent model. Because the magnetic dipole cross section contains no velocity-dependence, the result is hardly surprising. Similarly for the conductivity $\kappa$, there is minimal variation from the constant case for small $\mu$. For larger $\mu$, $\kappa$ decreases, indicating a suppression in energy transport. In this region, the second term in eq.~(\ref{eq:p1sigmamagdip}) becomes non-negligible. The term arises from the angular dependence of the magnetic dipole cross section. When there is angular dependence, not all collisions transfer the full potential amount of momentum.  Thus, on average less energy can be transferred for each collision. 
			The angular dependence in the magnetic dipole cross section arises from the requirement that the dark matter particle should have a component of its spin parallel to the incoming momentum of a nucleus for maximal interaction with the magnetic field. The velocity dependence in the Lorentz force is partially cancelled by the momentum dependence of the photon propagator. For lighter masses, the momentum transfer cross section is dominated by the electric field induced by the changing magnetic field that is the motion of the  dipole. The electric field component has no angular dependence and so the total energy transferred is greater for light masses.
			
			For the anapole interaction, there is a significant suppression of $\alpha$ for small values of $\mu$, while there is a suppression of $\kappa$ for large values of $\mu$. As expected, the behaviour very closely follows both the generalised $q^2$ and $v^2$ cross sections analysed by ref.~\cite{vincent14}. Velocity and momentum-dependent terms with positive powers result in an increase in the collision operator, as the difference between the centre of momentum frame, where the evaluation occurs, and the frame of the Sun is greater. So, for low masses, the distribution of dark matter will be very compact, as there will be little outward diffusivity along the temperature gradient of the star. The greater cross section, and hence collision rate, for larger masses means that the particles cannot travel as far before their energy is lost, resulting in a lowering of the thermal conductivity $\kappa$. The reasoning applies as the calculations are made in the local thermal equilibrium limit.
			
			The dependence of the conduction coefficients on the value of $S$ is shown in the middle and lower plots for the magnetic and anapole moments respectively. 
			The effect of the spin-related term has a minimal effect on the distribution of dark matter particles within the Sun, described by the diffusivity $\alpha$. When the spin term is included for the magnetic dipole, the values for $\kappa$ decrease, related to the size of $S$ in eq.~(\ref{eq:Svalues}), in the range around $\mu = 1$. The implication is a marginal reduction in the efficiency of heat transport. The contrary is true for the anapole: the introduction of the  spin term causes a marginal increase in energy transport, since more of the normalised cross section in eq.~(\ref{eq:p1sigmaanapole}) is weighted towards the velocity-dependent scattering. The result is a smaller effect of the angular dependence. Note that the $q^2$-dependent terms in eq.~(\ref{eq:anapolethetaCM}) have opposite signs.			
		
		\subsubsection{Effectiveness of energy transport}
		
			\begin{figure}[t]
				\centering
					\IfFileExists{../plots/SIenergy.eps}{
						\subfloat{\includegraphics[width=0.48\textwidth]{../plots/SIenergy.eps}}
						\subfloat{\includegraphics[width=0.48\textwidth]{../plots/EDenergy.eps}}
					}{
						\subfloat{\includegraphics[width=0.48\textwidth]{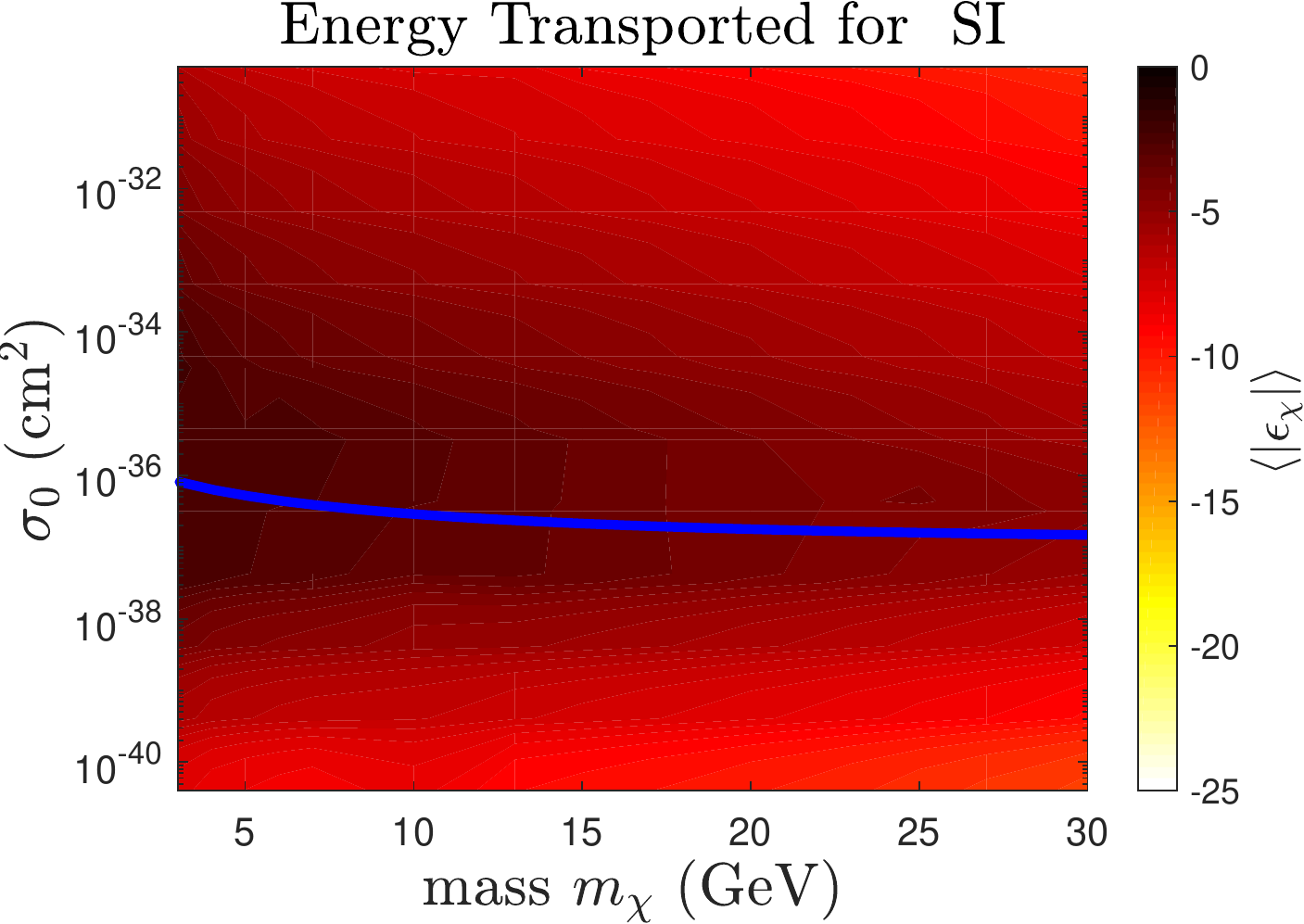}}
						\subfloat{\includegraphics[width=0.48\textwidth]{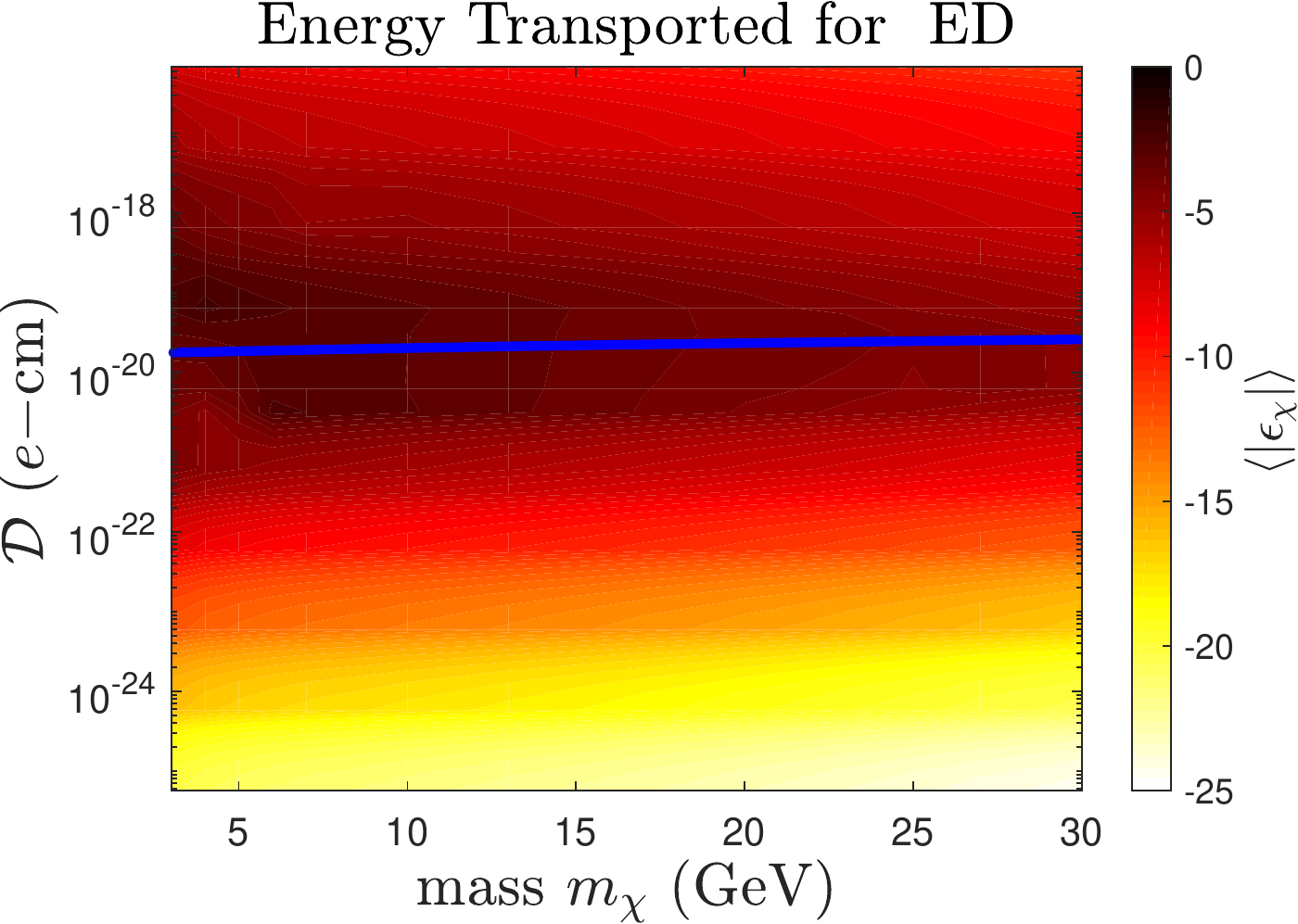}}
					}
					\IfFileExists{../plots/EDenergy.eps}{
						\subfloat{\includegraphics[width=0.48\textwidth]{../plots/MDenergy.eps}}
						\subfloat{\includegraphics[width=0.48\textwidth]{../plots/ANenergy.eps}}
					}{

						\subfloat{\includegraphics[width=0.48\textwidth]{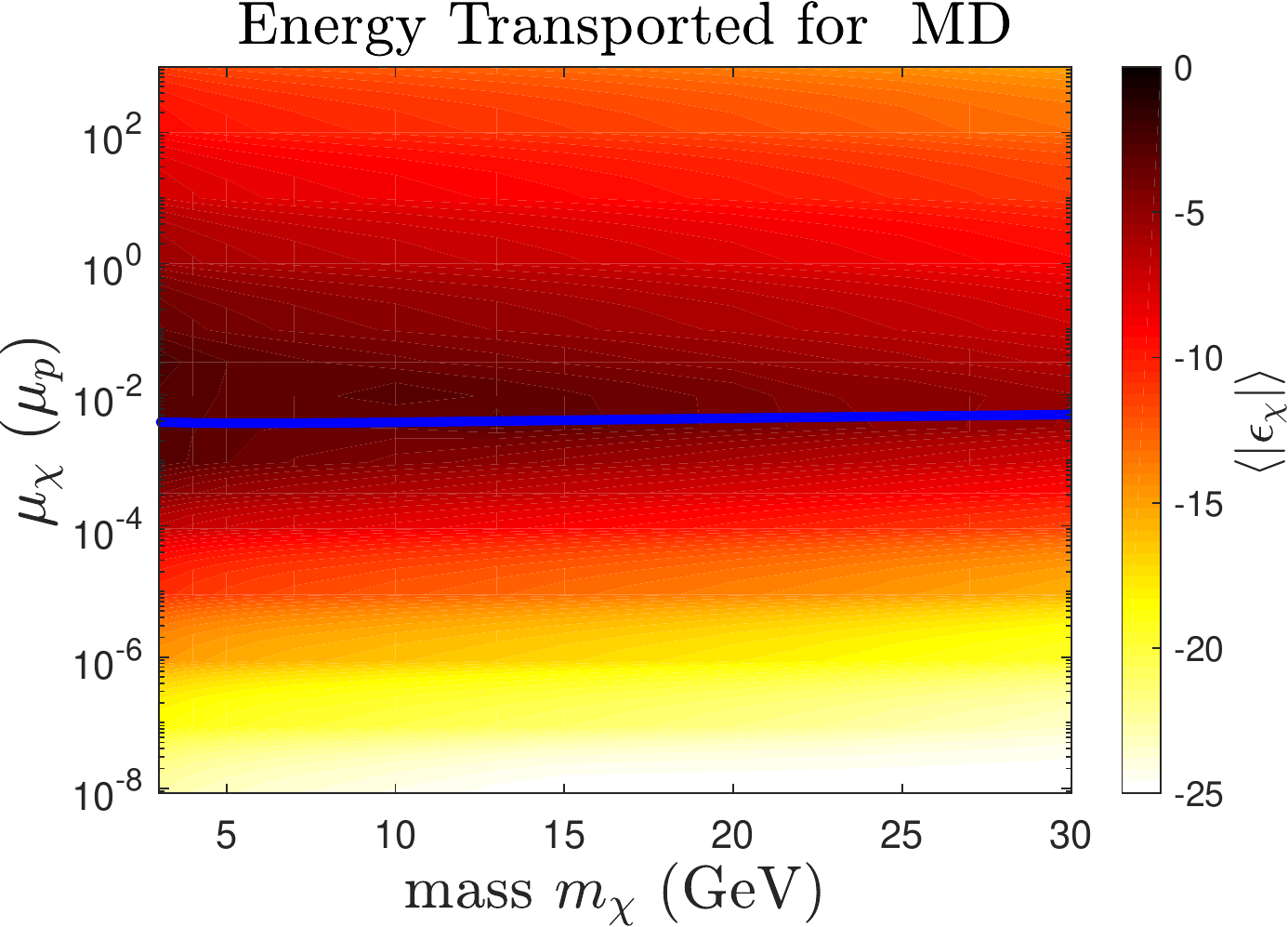}}
						\subfloat{\includegraphics[width=0.48\textwidth]{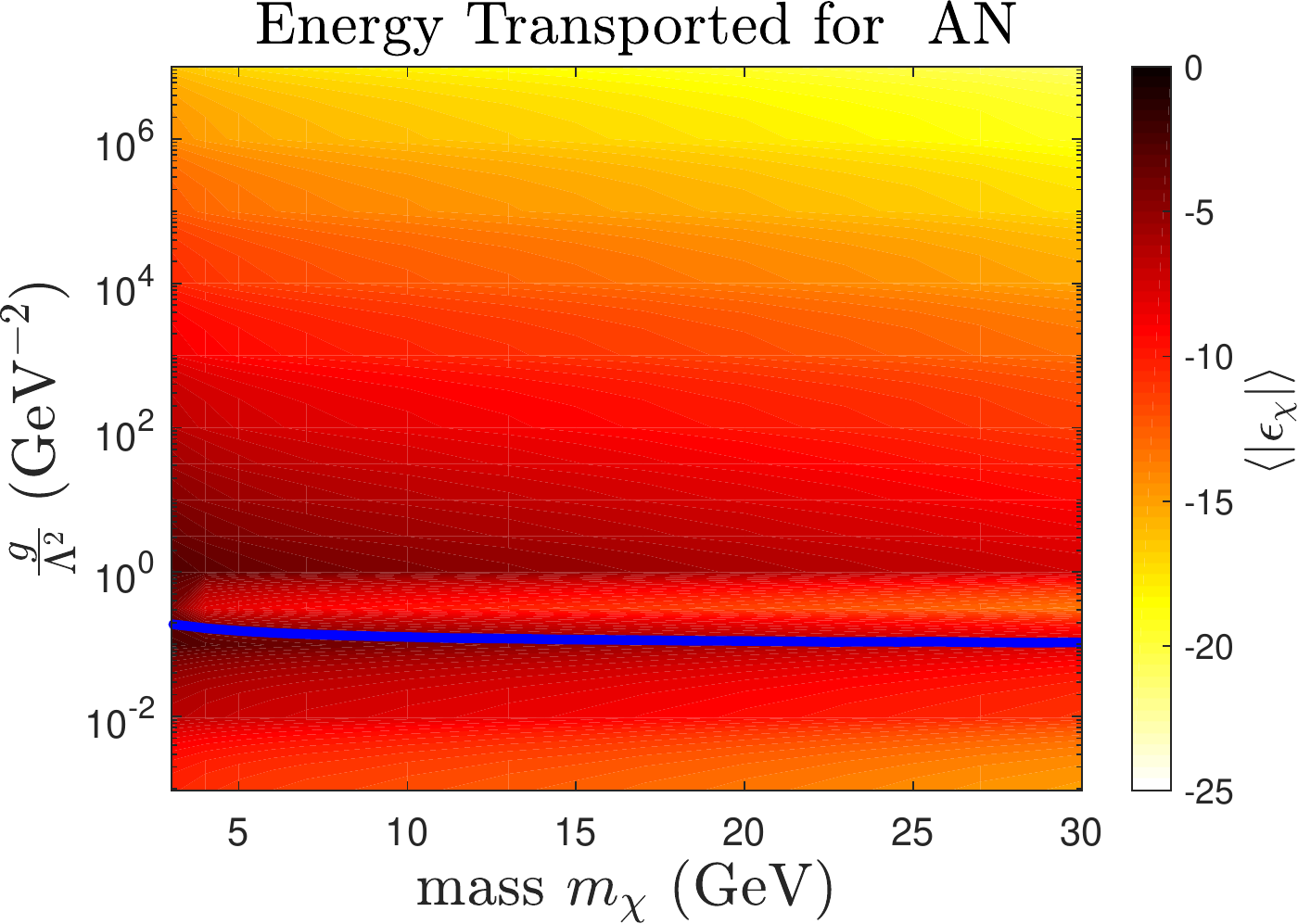}}
					}
				\caption{Average absolute value of the rate of energy transfer $\left|\epsilon\right|$ for spin independent dark matter (top left), electric dipole dark matter (top right), magnetic dipole dark matter (bottom left) and anapole dark matter (bottom right). The blue line illustrates the saturation capture rate, above which the population of dark matter within the star is roughly constant. %
				The colour scale is measured in units of $\log_{10}\left(\epsilon ~[\erg\g^{-1}\s^{-1}]\right)$. %
				}
				\label{fig:energy}
			\end{figure}		
		
			To facilitate discussion on the effectiveness or otherwise of each model at transporting energy via dark matter, the rate of energy transported per unit mass can be defined as 
			\begin{equation}
				\epsilon_{\chi} = \frac{1}{4\pi r^2 \rho(r)} \frac{dL_\chi(r)}{dr},
			\end{equation}
			where $\rho(r)$ is the radial density profile of the Sun, and $L_\chi$ is defined in eq.~(\ref{eq:Lchi}). The average rate of energy transport is then given by
			\begin{equation}
				\langle|\epsilon_\chi|\rangle = \frac{\int_0^{R_\odot} r^2 |\epsilon_\chi| dr}{\int_0^{R_\odot} r^2 dr}
			\end{equation}
			for a spherically symmetric star. The average rate of energy transfer for each model considered in the present investigation is plotted in figure~\ref{fig:energy}, for a Sun-like star at an age of $4.57 \Gy$ based on the AGSS09ph model~\cite{asplund09}. 
			
			There are multiple factors influencing the regions of the parameter space that have a high energy transport rate. For each model, there exists a narrow range of coupling strengths where the transport is maximised. The energy transported drops off by several orders of magnitude above and below each band. The region above the band is that of local thermal equilibrium; the dark matter particles are at the same temperature as the surrounding material. In this regime, a decreasing interaction strength actually increases the energy transported. Models with a high interaction strength collide more frequently and so the change in energy over long distances is diminished. Contrastingly, below the band is the Knusden regime~\cite{gould90c}, where the particles travel long distances without interaction. Here, the mean free path is much larger than the scale factor defined in eq.~(\ref{eq:rchi}), and particles may pass through the Sun multiple times before scattering. The overall result is a region of transition between Knusden transport and local thermal equilibrium, which corresponds to the largest possible energy transport. 
			
			Another observation is the relationship between the energy transport and capture rate. The luminosity in eq.~(\ref{eq:Lchi}) is directly proportional to the dark matternumber of particles in the Sun. So when the capture rate is below the saturation value, the total energy transport also decreases. To aid comparison, the saturation cut-off for each model is plotted in blue in figure~\ref{fig:energy}. In the spin-independent  and magnetic dipole cases, the cut-off occurs at or near the maximum of the energy transport. The implication is that the unsaturated models suppress the transition to Knusden transport, in part or entirely. Contrastingly, for the electric dipole and anapole, the cut-off occurs well clear from the transition, meaning almost all electric dipole and anapole models of interest will occur at saturated values of the capture rate. 
		
			Having now developed the required theory for capture and energy transport, the effects on the Sun may be evaluated computationally. The total capture rate in eq.~(\ref{eq:capturerate}) and luminosity transfer in eq.~(\ref{eq:Lchi}) are sufficient to describe the alterations to the solar models. Both have a dependence on the elemental composition of the Sun, and so both are a function of the age of the Sun. Therefore, a complete calculation requires the evolution through the entire history of the Sun, starting from a protostar and ending at the present day.

		\section{Simulations}
\label{sec:simulations}

	\subsection{Numerical code}
	
		We simulate dark matter within the Sun using a modified version of the \texttt{DarkStec} code~\cite{vincent15a,vincent15b,vincent16}. The \texttt{DarkStec} code is an amalgamation of \texttt{GARSTEC} (Garching Stellar Evolution code) and \texttt{DarkStars}.
		\texttt{GARSTEC} is directly descended from the Kippenhahn~\cite{kippenhahn67} code and has been continuously developed by various authors \cite{thomas67,weiss89,wagenhuber94,weiss00,weiss08,serenelli11}. \texttt{DarkStars} has itself evolved from the \texttt{STARS} evolution code~\cite{eggleton71,eggleton72,pols95,paxton04}, modified to incorporate dark matter physics from \texttt{DarkSUSY}~\cite{gondolo04,fairbairn08,scott08b,scott08a,scott09a,scott09b}. The inputs, evolution and convergence of the \texttt{DarkStec} code are described in ref.~\cite{vincent15b}.

		We calculate five different models: one for each of the three electromagnetic dipole moments outlined in chapter~\ref{sec:dipolemoments}, one model with dark matter with a constant, spin-independent cross section, and one without any dark matter at all. For the first four cases, a solar model is simulated on a grid of points in the parameter space, which comprises of the dark matter mass $m_\chi$ and the coupling strength parameter $\mathcal{D}$, $\mu_\chi$, $\frac{g}{\Lambda^2}$ or $\sigma_0$, depending on the model. The masses were selected between $3 \GeV$ and $30 \GeV$. For masses below $\sim4\GeV$, evaporation of dark matter may become significant, so the results should be treated with caution until a full evaporation calculation can be implemented~\cite{busoni16}. The range of parameter space for the coupling strength was motivated by current observational bounds. Initial runs were taken at coarsely selected parameter points, allowing for a relatively quick scan over the parameter space of interest. For the regions in the parameter space which showed promising or unusual results, a higher resolution of simulations was selected. If no region of the parameter space showed any interesting result, the bounds of the grid were widened. This was the case for the electric dipole and anapole models. 
		
	\subsection{Calibration of solar models}
	
		Solar models computed in this work are evolutionary models calibrated to the minimal set of present-day observables: the solar luminosity $L_\odot = 3.8418\times 10^{33} \erg\s^{-1}$, the solar radius $R_{\odot}=6.9598\times 10^{10}\cm$ and the surface photospheric metal to hydrogen ratio $(Z/X)_{\odot} = 0.0181$~\cite{asplund09}. The solar model starts as a pre-main sequence, fully mixed, $1 M_{\odot}$ star and its evolution is followed up to the solar system age $\tau_\odot = 4.57 \Gy$. At this stage, a successful solar model has to satisfy the constraints above to better than a part in $10^4$. In order to achieve this, a set of three free parameters in the model is adjusted iteratively by means of a Newton-Raphson scheme. The adjustable, or calibration, parameters are: the initial helium abundance ($Y_{\mathrm{ini}}$) and metallicity ($Z_{\mathrm{ini}}$) of the models and the mixing length parameter $\alpha_{\mathrm{MLT}}$, associated with the efficiency of convection.
		
	\subsection{Solar observables}
	
		Once a simulation has been completed, it must be compared with observations to determine its viability. From here, the goodness of fit can be determined. There are several observables available that show moderate to severe discrepancies from the Standard Solar Model~\cite{serenelli11,vincent15b,vincent16}. Not all observables are necessarily independent, nor provide a complete description of the fit to the Sun. 
		
		\subsubsection{Solar neutrino fluxes}
		
			One class of tight constraints on the solar model is provided by the observed solar neutrino fluxes. Most neutrinos are emitted through protons fusing into deuterium via the process $p + p \rightarrow d + e^+ + \nu_e$. However, a small fraction are emitted through higher order processes, notably $\ce{^7Be} + e^- \rightarrow \ce{^7Li + \nu_e}$ and $\ce{^8B} \rightarrow \ce{^8Be + e^+ + \nu_e}$. The neutrinos produced in these reactions characteristically have higher energies, and may be distinguished from neutrinos from proton fusion in detection experiments. They also have a sufficiently high flux at Earth and appropriate spectrum to be measured relatively precisely. An aggregation of results gives the flux of neutrinos from $\ce{^8B}$ decays as $\phi_{B,\mathrm{obs}} =5.16 \times 10^6 \cm^{-2}\s^{-1}$ and the flux of neutrinos from electron capture of $\ce{^7Be}$ as $\phi_{Be,\mathrm{obs}} = 4.80 \times 10^9 \cm^{-2}\s^{-1}$ , with observational errors of $5\%$ and $13\%$ respectively~\cite{antonelli13,bergstrom16}. There is an additional error due to the modelling effects, estimated to be $\sim14\%$ and $\sim7\%$ respectively~\cite{vincent15b,vincent16}. 
			
			The rates of these reactions, especially $\ce{^8B}$, have a strong dependence on the temperature of the core of the Sun. Analytical approximations indicate that the ratio between the fluxes goes as $\frac{\phi_B}{\phi_{Be}} \sim T^{13.5}_c$ where $T_c$ is the core temperature~\cite{castellani93,castellani94}. Any fit to the neutrino fluxes is therefore very sensitive to changes in the core temperature. As the objective of incorporating dark matter into the solar model is to provide additional energy transport, the model must ensure that the core temperature is not substantially affected.
			
			\begin{figure}[t]
				\centering
				\IfFileExists{../plots/SIBe7neutrinos.eps}{
					\subfloat{\includegraphics[width=0.49\textwidth]{../plots/SIBe7neutrinos.eps}}
					\subfloat{\includegraphics[width=0.49\textwidth]{../plots/EDBe7neutrinos.eps}}
				}{
					\subfloat{\includegraphics[width=0.49\textwidth]{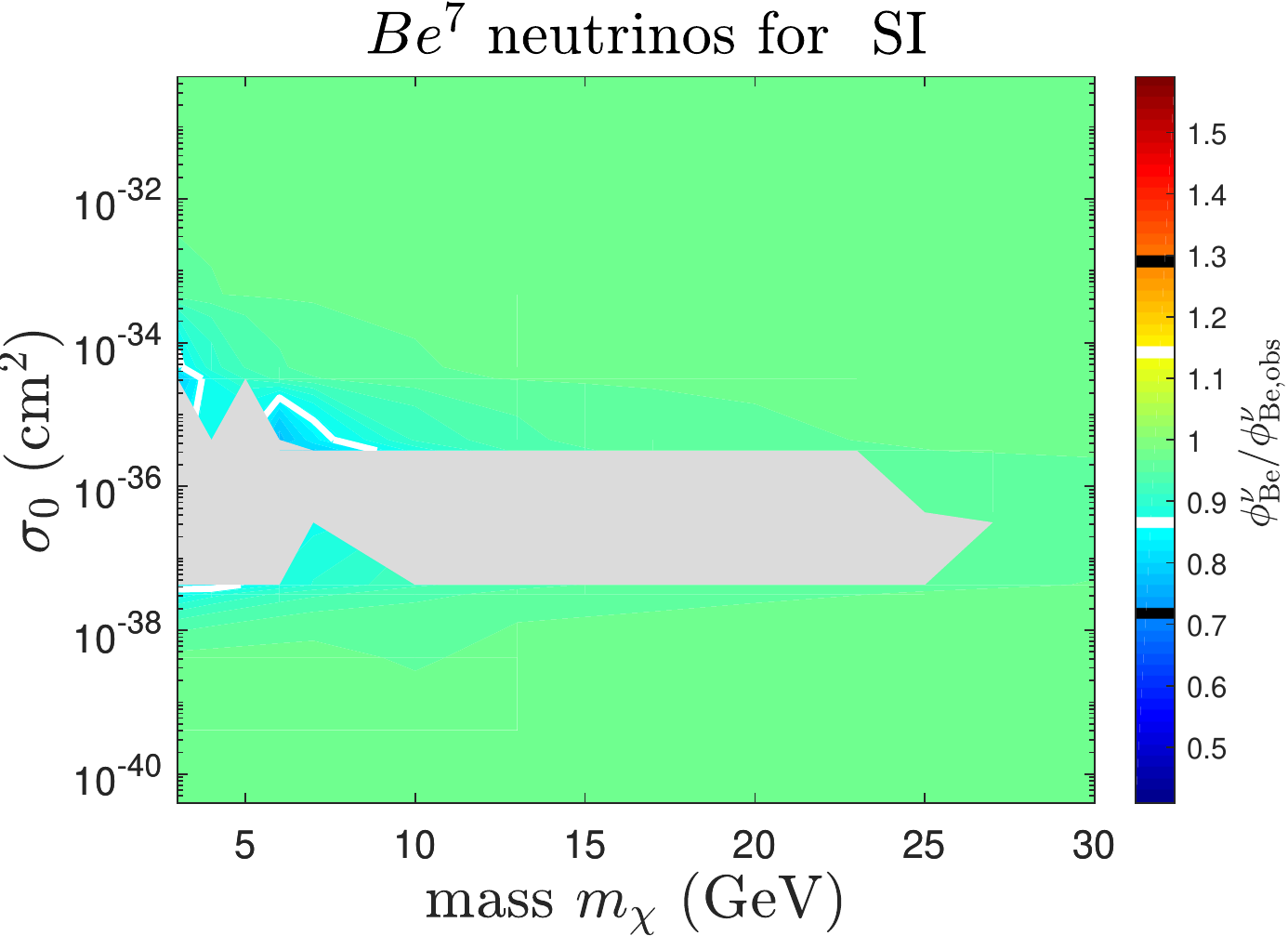}}
					\subfloat{\includegraphics[width=0.49\textwidth]{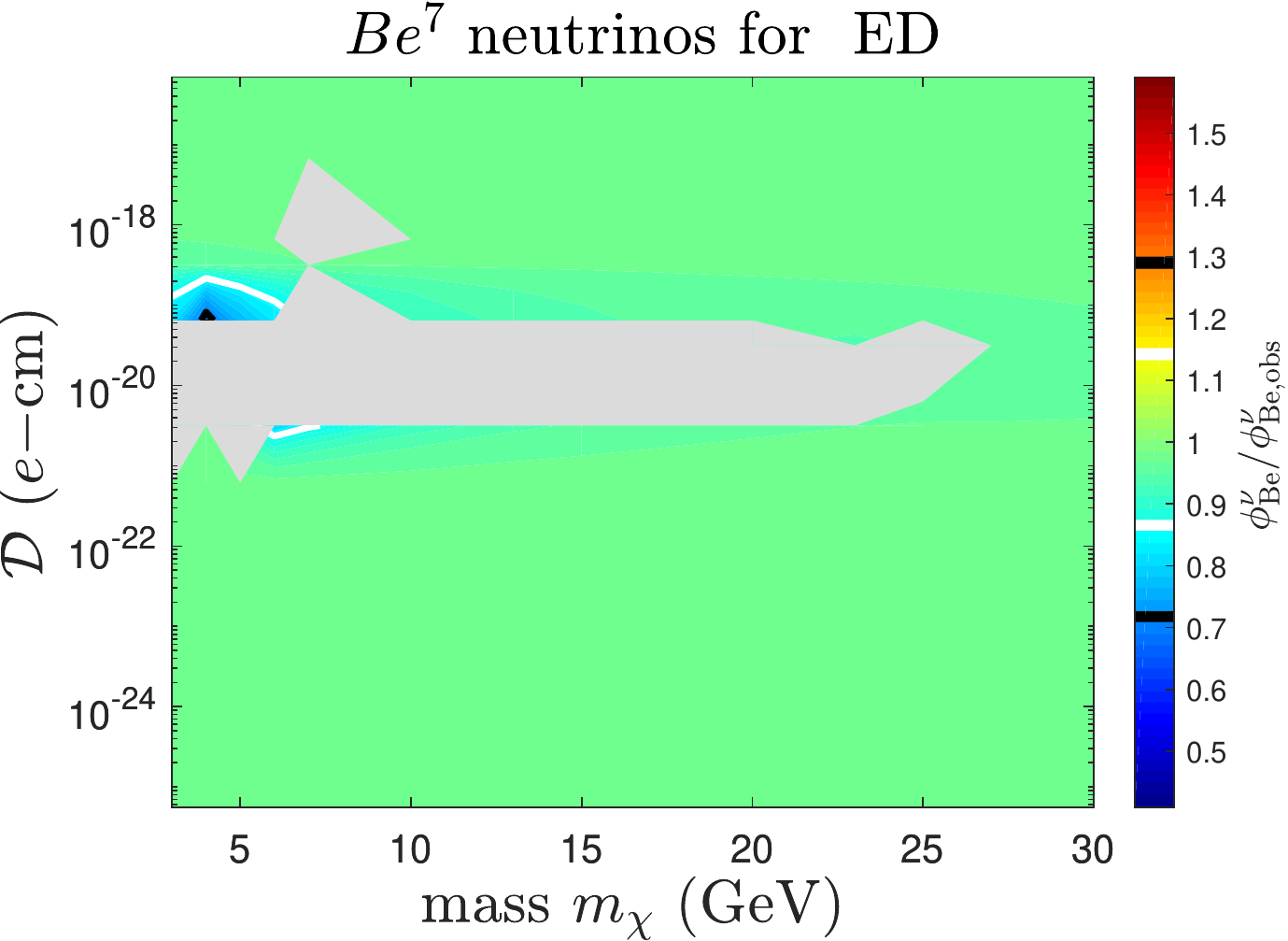}}
				}
				\IfFileExists{../plots/EDBe7neutrinos.eps}{
					\subfloat{\includegraphics[width=0.49\textwidth]{../plots/MDBe7neutrinos.eps}}
					\subfloat{\includegraphics[width=0.49\textwidth]{../plots/ANBe7neutrinos.eps}}
				}{
					\subfloat{\includegraphics[width=0.49\textwidth]{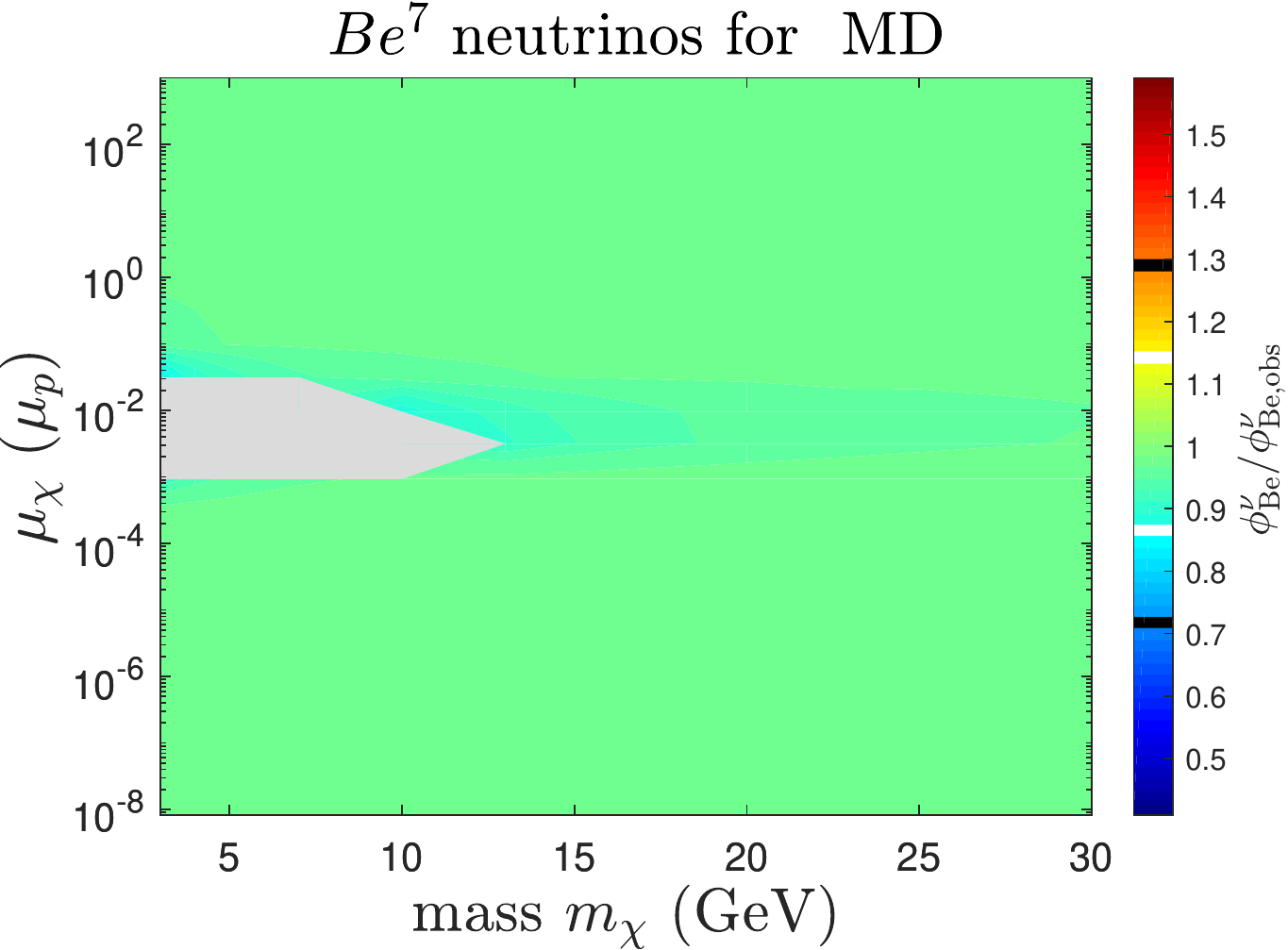}}
					\subfloat{\includegraphics[width=0.49\textwidth]{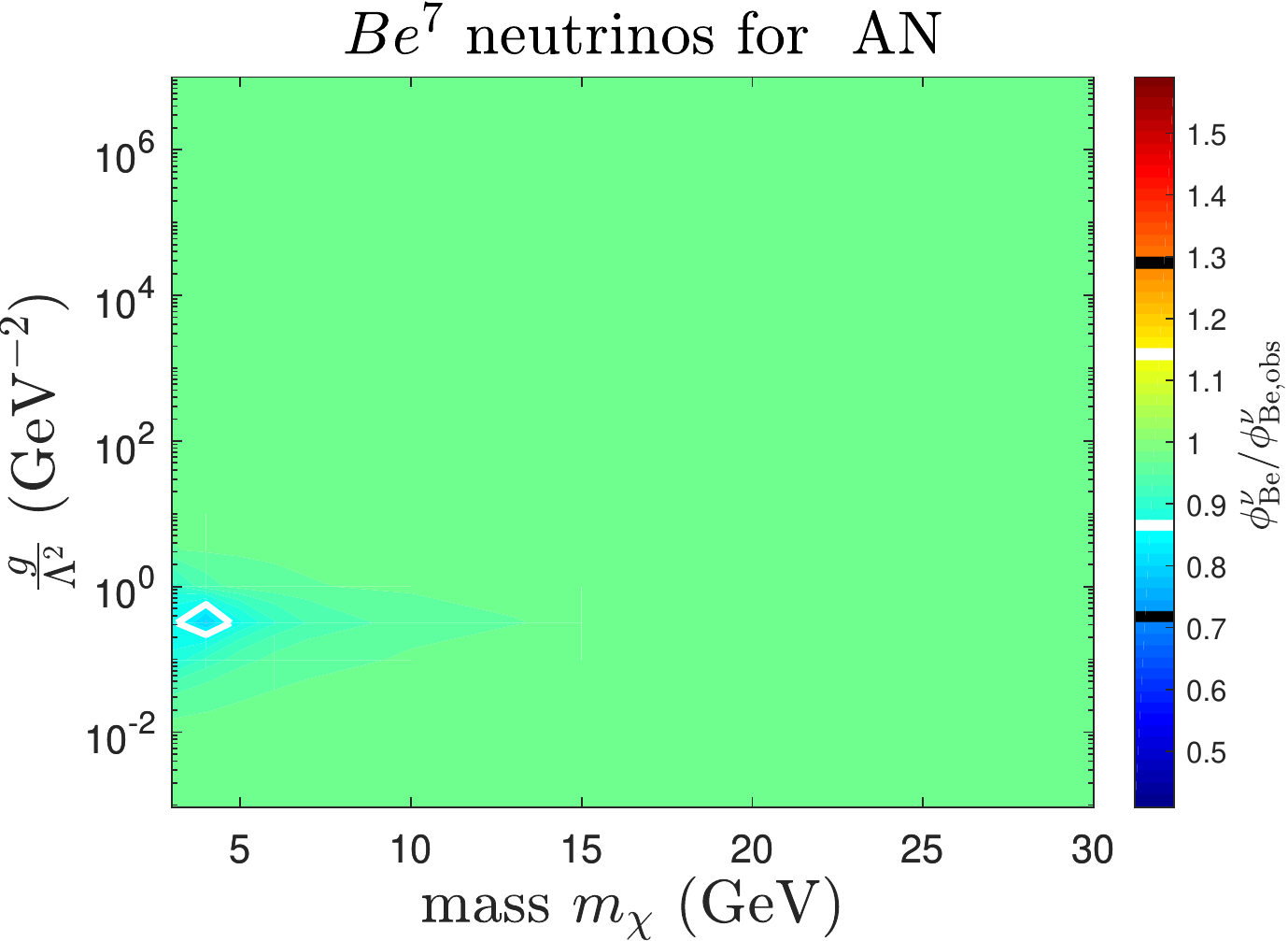}}
				}	
				\caption{Ratio of predicted $\ce{^7Be}$ neutrino flux to the measured value $\phi_{Be,\mathrm{obs}} = 4.82 \times 10^9 \cm^{-2}\s^{-1}$ for spin independent dark matter (top left), electric dipole dark matter (top right), magnetic dipole dark matter (bottom left) and anapole dark matter (bottom right). The expected value is 1. The white
				contours show the regions where the flux is $1\sigma$
				above/below the expected value
				. Simulations in the grey regions did not converge.}
				\label{fig:Be7}
			\end{figure}			
			\begin{figure}[t]
				\centering
				\IfFileExists{../plots/SIB8neutrinos.eps}{
					\subfloat{\includegraphics[width=0.49\textwidth]{../plots/SIB8neutrinos.eps}}
					\subfloat{\includegraphics[width=0.49\textwidth]{../plots/EDB8neutrinos.eps}}
				}{
					\subfloat{\includegraphics[width=0.49\textwidth]{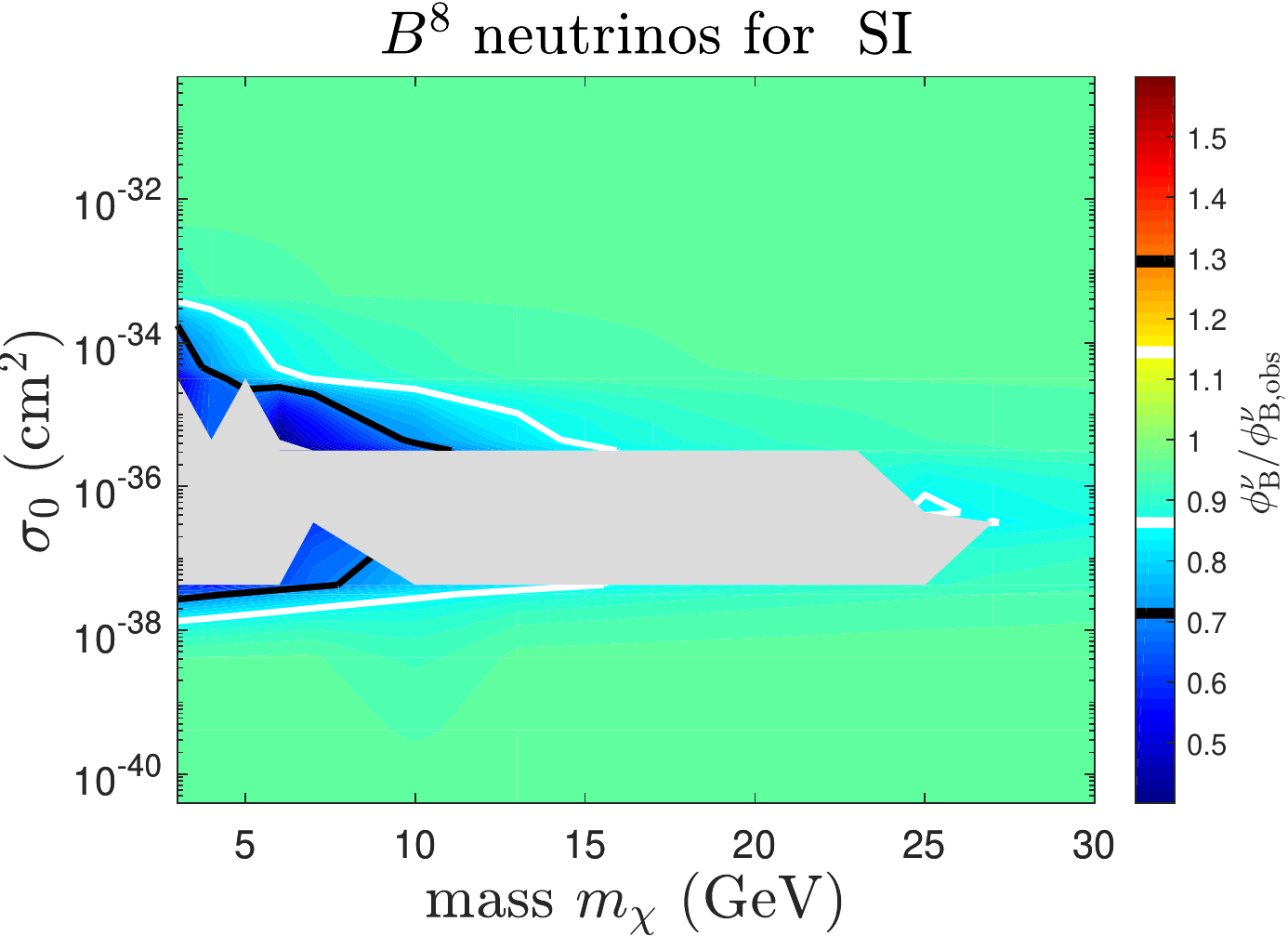}}
					\subfloat{\includegraphics[width=0.49\textwidth]{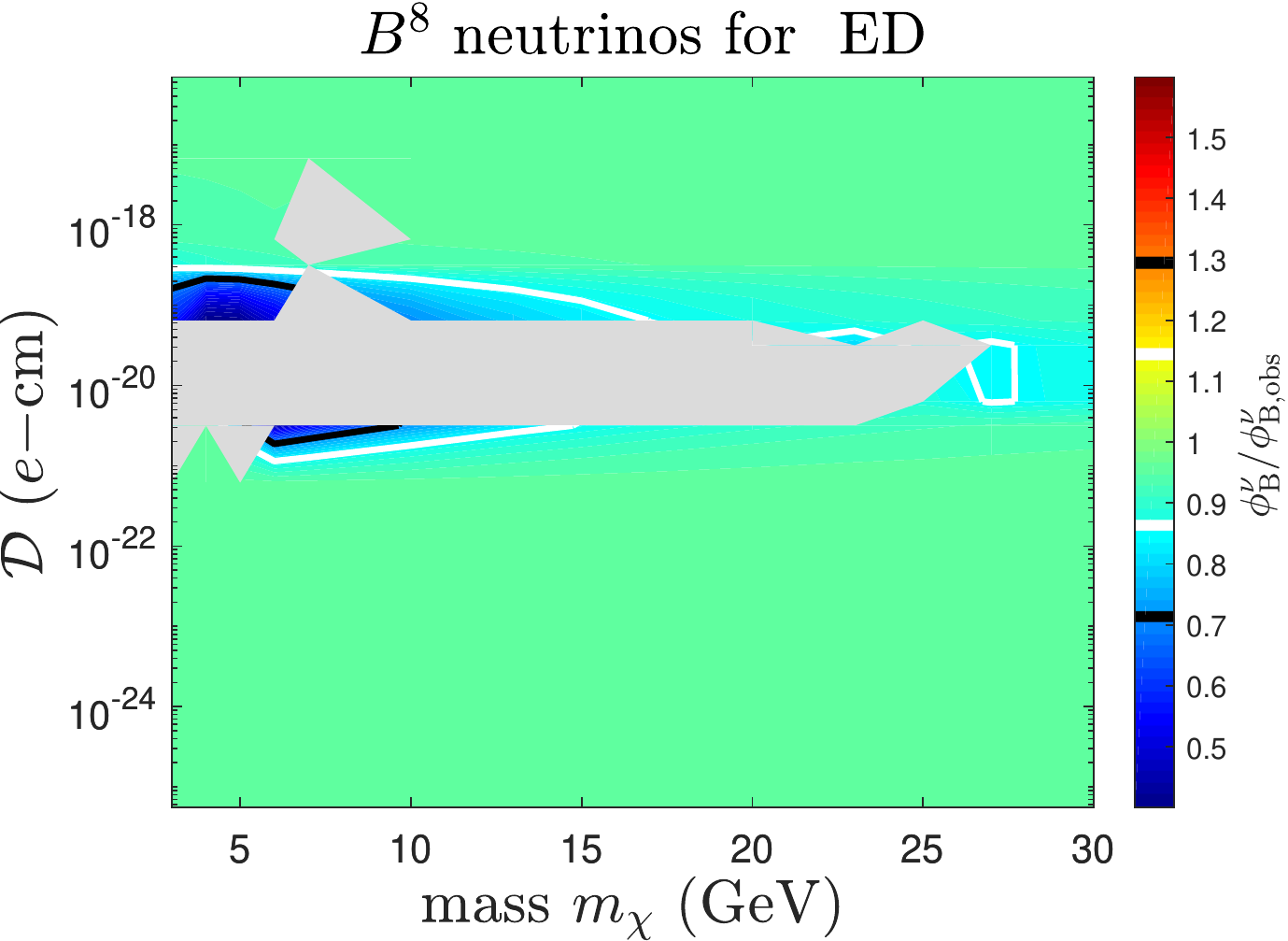}}
				}
				\IfFileExists{../plots/EDB8neutrinos.eps}{
					\subfloat{\includegraphics[width=0.49\textwidth]{../plots/MDB8neutrinos.eps}}
					\subfloat{\includegraphics[width=0.49\textwidth]{../plots/ANB8neutrinos.eps}}
				}{
					\subfloat{\includegraphics[width=0.49\textwidth]{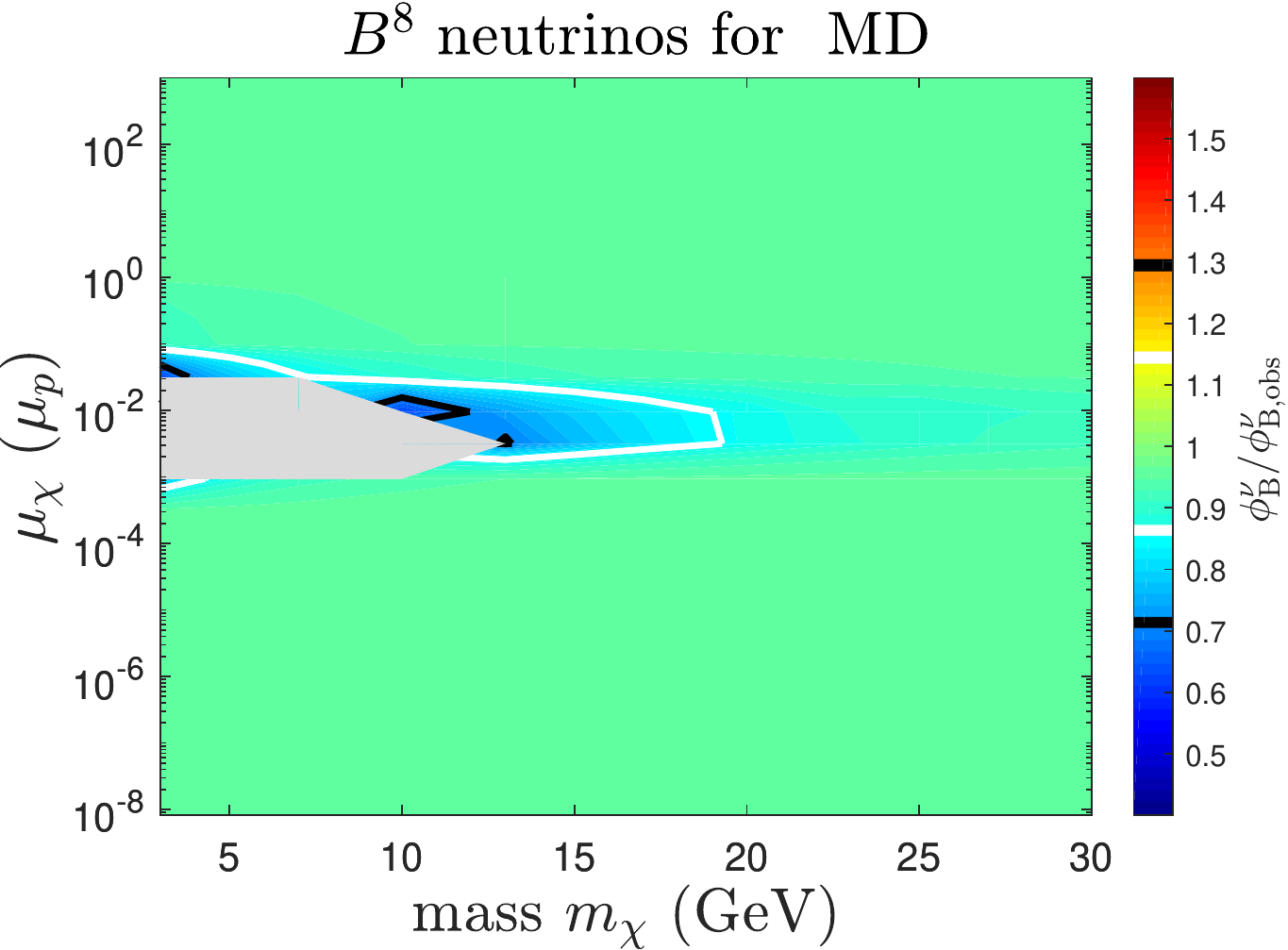}}
					\subfloat{\includegraphics[width=0.49\textwidth]{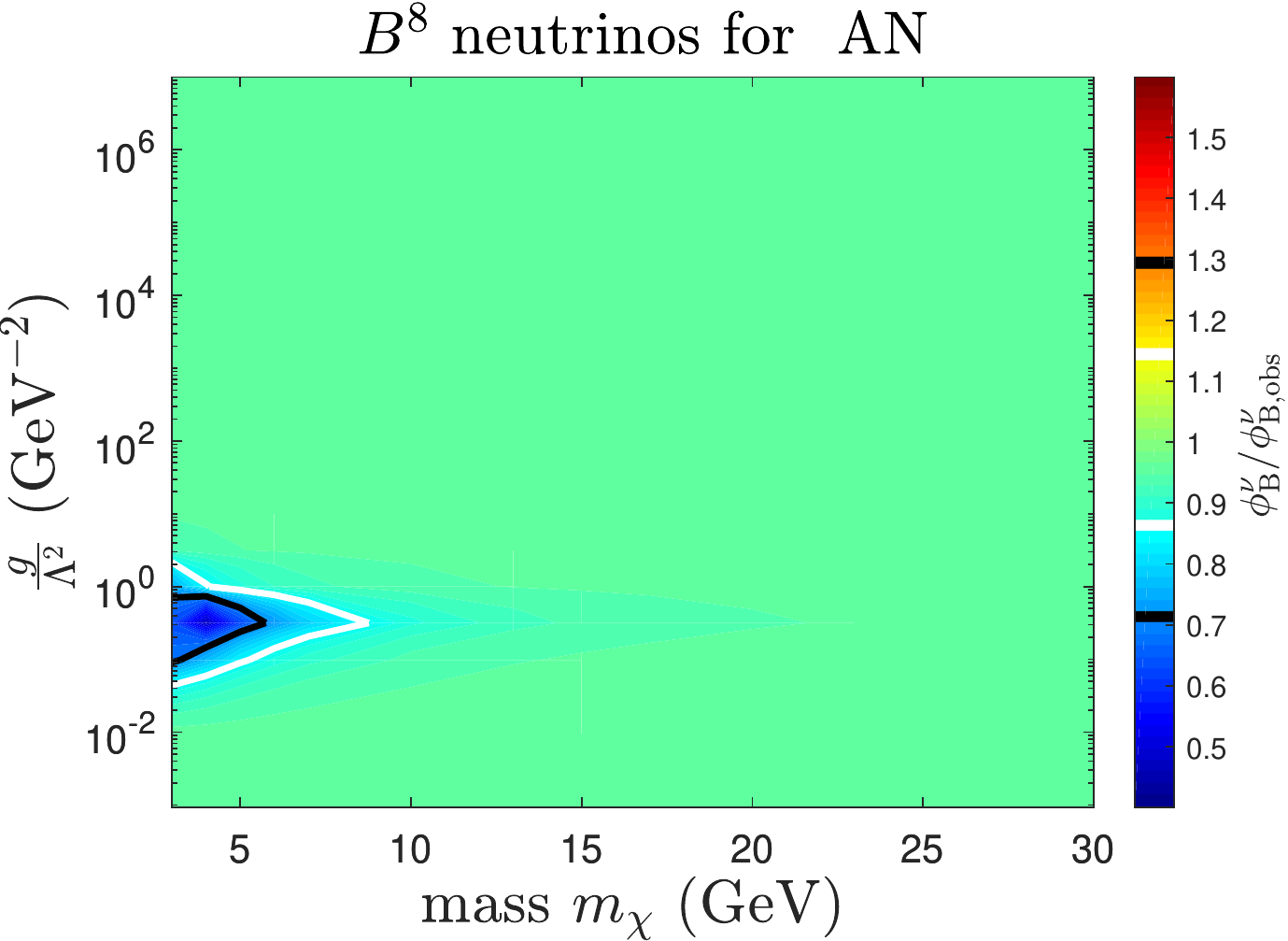}}
				}
				\caption{Ratio of predicted $\ce{^8B}$ neutrino flux to the measured value $\phi_{B,\mathrm{obs}} =5.00 \times 10^6 \cm^{-2}\s^{-1}$ for spin independent dark matter (top left), electric dipole dark matter (top right), magnetic dipole dark matter (bottom left) and anapole dark matter (bottom right). The expected value is 1. The white and black contours show the regions where the flux is $1\sigma$ and $2\sigma$ above/below the expected value respectively. Simulations in the grey regions did not converge.}
				\label{fig:B8}
			\end{figure}			
		
			Figures~\ref{fig:Be7} and \ref{fig:B8} show the ratios of the calculated to the measured neutrino flux for $\ce{^7Be}$ and $\ce{^8B}$ respectively.
			
			Models without dark matter are already in excellent agreement with the observed neutrino fluxes. Incorporating dark matter tends to have the effect of worsening this fit. The green regions in figures~\ref{fig:Be7} and \ref{fig:B8} fit exceptionally well with the observations. For each of the models, there is a band of approximately constant coupling strengths where the fit either falls more than $2\sigma$ below the observed value or fails to converge. For the magnetic dipole and anapole models, the band is truncated for $m_\chi > 15\GeV$ and $m_\chi > 5\GeV$ respectively. The band occurs for the same coupling strengths for both the $\ce{^7Be}$ and $\ce{^8B}$ neutrino fluxes. These regions, shown in blue, are where the flux is much lower than the observed value, indicating that the temperature of the core regions is being substantially reduced.
			The result is much stronger for the $\ce{^8B}$ flux, because of its stronger temperature dependence. 
			
			The link between the reduction in neutrino fluxes and the presence of dark matter is clear when one compares the neutrino flux plots to the magnitude of energy transfer in figure~\ref{fig:energy}. There is a clear correlation between the regions of significant energy transfer (dark regions in figure~\ref{fig:energy}) and the regions of reduced neutrino flux (blue regions in figures~\ref{fig:Be7} and \ref{fig:B8}). Because the peak energy transport in the spin-independent model is broadly smeared over the chosen parameter space, there is a correspondingly sizeable region where neutrino fluxes are reduced. 
			Interestingly, the flux falls off rapidly for the electric dipole over a relatively small range to a relatively wide-band of non-convergence.  

		\subsubsection{Helioseismology}
		\label{sec:helioseismology}
		
			Helioseismology, in particular the sound-speed profile, has provided an unprecedented probe of the inner structure of the Sun. Surface oscillations of the Sun as measured by Doppler movements give the amplitude and frequency of a variety of modes of pressure waves. The identification of large numbers of modes provides an accurate description of the medium within which the waves propagate, in this case the solar interior. The commonly presented measure, obtained from inversions of helioseismological data, is the sound-speed profile: the speed at which waves propagate through the Sun as a function of the internal radius. It is possible to compare the speed inferred from inversions of helioseismic measurements to the predicted values from solar models. The sound speed profile is dependent on both the temperature and the mean molecular weight in the Sun.
			
			Early solar models~\cite[e.g.][]{grevesse98,bahcall04} suggested moderate agreement with the observed sound speed profile~\cite{bahcall05a}. However, more recent heavy-element abundances in the Sun are lower by nearly a factor of two~\cite{allendeprieto01,allendeprieto02,asplund04,asplund05b,scott06,Socas07,Koesterke08, melendez08,scott09c,asplund09,scott15a,scott15b,grevesse15}. This is in part due to the advent of new three-dimensional hydrodynamic models of the photosphere, but mostly due to big improvements in the treatment of departures from local thermodynamic equilibrium in atomic level populations giving rise to absorption lines \cite[e.g.][]{Pereira09,Lind11,Bergemann12,Mashonkina12}, and in the fundamentals of spectroscopy: atomic data, line selection and equivalent widths.  Alternative, higher, abundances have been advocated on the basis of other 3D models \cite{caffau11} and measurements of the solar wind \cite{vSZ16}.  These have however been debunked on the basis of their spectroscopic fundamentals \cite{scott15a,scott15b,grevesse15} and neglected fractionation effects \cite{serenelli16}, respectively, and their results have not been adopted by the astrophysical community.
			
			However, solar models computed with the updated surface composition \cite[e.g.][]{asplund09,serenelli09,serenelli11} generally do not agree with helioseismological observations. Near the core of the Sun ($R \sim 0.2 R_\odot$), models predict a sound speed larger than observed, while in the range $0.2 $ $\lesssim$ $ R $ $\lesssim 0.7$ this is inverted. To a very good approximation $c^2$ scales as $\sqrt{T/\mu}$, where $\mu$ is the mean molecular weight. A suitable modification to the temperature profile due to an additional energy transport mechanism, accompanied by the associated changes to $\mu$ in order to satisfy the constraints imposed on the solar models by $L_\odot$, might push the models closer to the observations. If the AGSS09ph abundances~\cite{asplund09,serenelli09,serenelli11} are appropriate even in the core, there is a strong implication for some mechanism for transporting energy from the core to the radiative zone, in order to push the models closer to the observations.

			In the present analysis, the simulated sound-speed profile generated by \texttt{DarkStec} is compared to the helioseismological inversions presented in ref.~\cite{basu09}. The errors arise from two sources; the errors from modelling are taken from ref.~\cite{vincent15b}, and the errors from the inversions are taken from ref.~\cite{deglinnoccenti97}. Both errors are added in quadrature. An effective $\chi^2$ value is defined as~\cite{vincent15b}
			\begin{equation}
			\label{eq:chi2cs}
				\chi_{c_s}^2 = \sum_{r_i} \frac{(c_{s,\mathrm{model}}(r_i) - c_{s,\mathrm{hel}}(r_i))^2}{\sigma^2_{c_{s},\mathrm{hel}}(r_i)}.
			\end{equation}
			The data are sampled at 5 equally-spaced points $r_i$ between $R = 0.1 R_\odot$ and $R = 0.67 R_\odot$. They are not sampled at more points in order to maintain the approximation of statistical independence. Helioseismological values from inversions at radii less than this range are rather uncertain. 
			Note that the sound-speed profile is not necessarily statistically independent from the frequency separation ratios defined in the next section, and hence the likelihoods of the two sets of observables should note be directly combined.
			

			\begin{figure}[t]
				\centering
				\IfFileExists{../plots/SIchi2CsB.eps}{
					\subfloat{\includegraphics[width=0.49\textwidth]{../plots/SIchi2CsB.eps}}
					\subfloat{\includegraphics[width=0.49\textwidth]{../plots/EDchi2CsB.eps}}
				}{
					\subfloat{\includegraphics[width=0.49\textwidth]{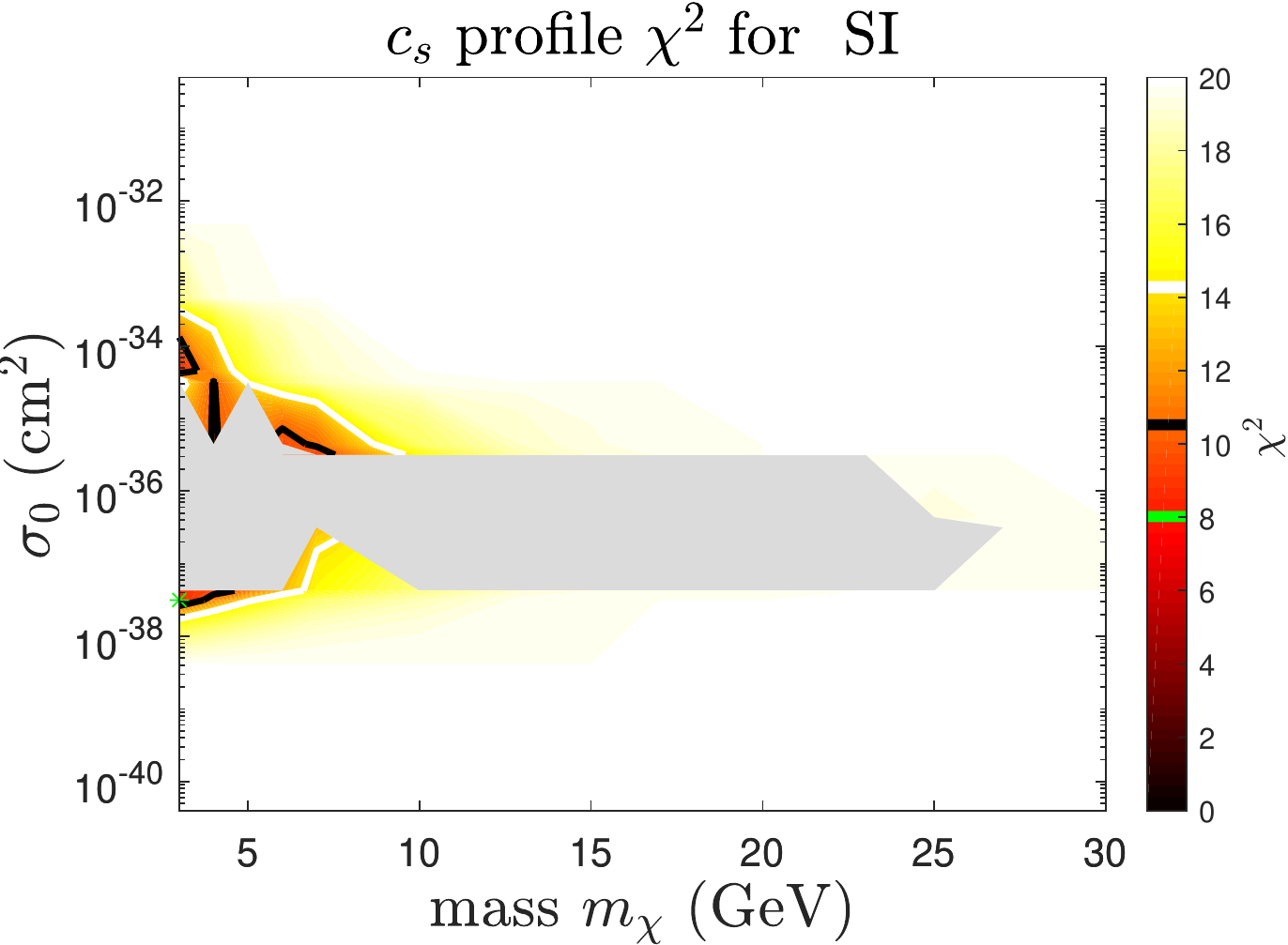}}
					\subfloat{\includegraphics[width=0.49\textwidth]{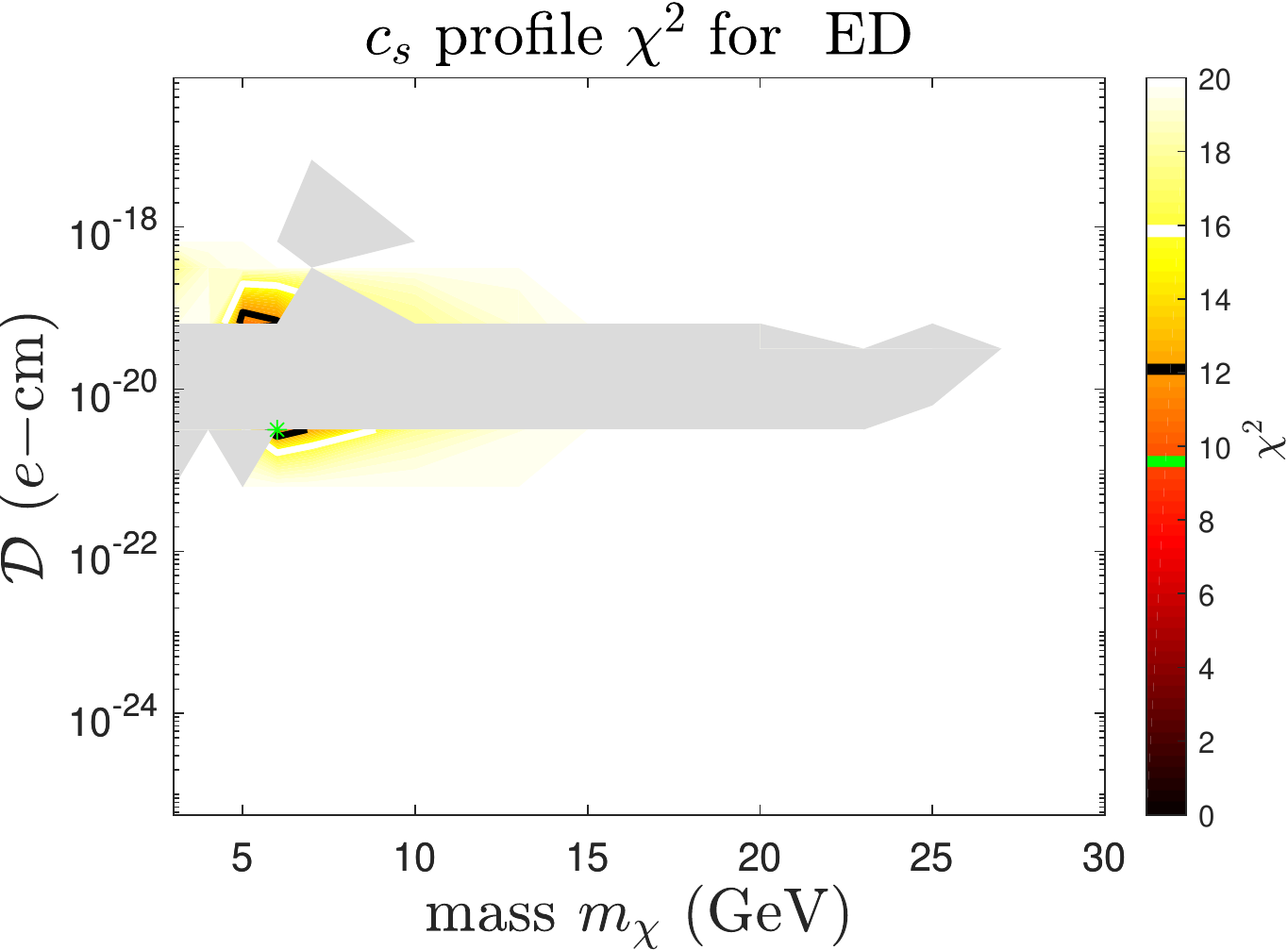}}
				}
				\IfFileExists{../plots/EDchi2CsB.eps}{
					\subfloat{\includegraphics[width=0.49\textwidth]{../plots/MDchi2CsB.eps}}
					\subfloat{\includegraphics[width=0.49\textwidth]{../plots/ANchi2CsB.eps}}				
				}{
					\subfloat{\includegraphics[width=0.49\textwidth]{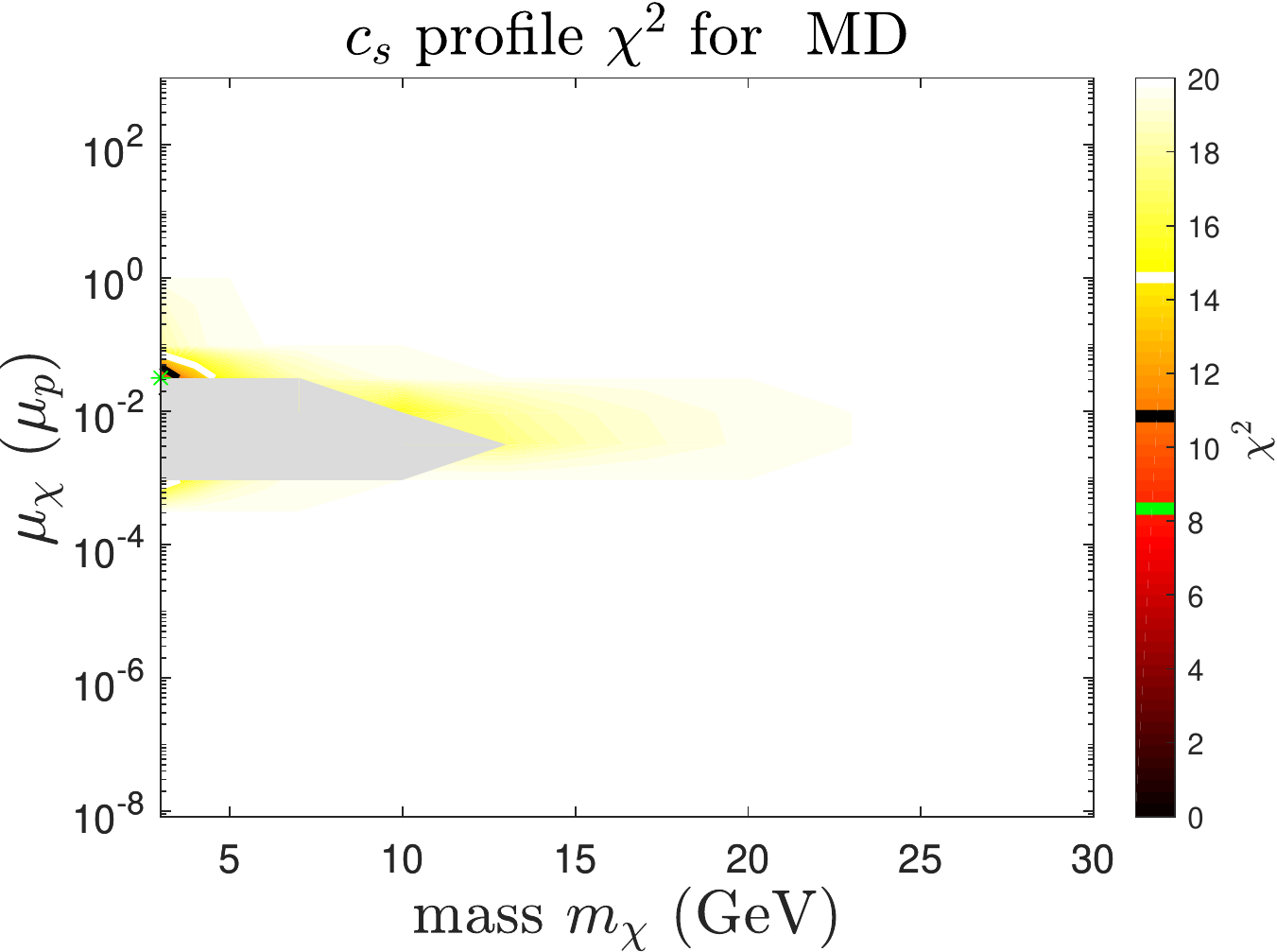}}
					\subfloat{\includegraphics[width=0.49\textwidth]{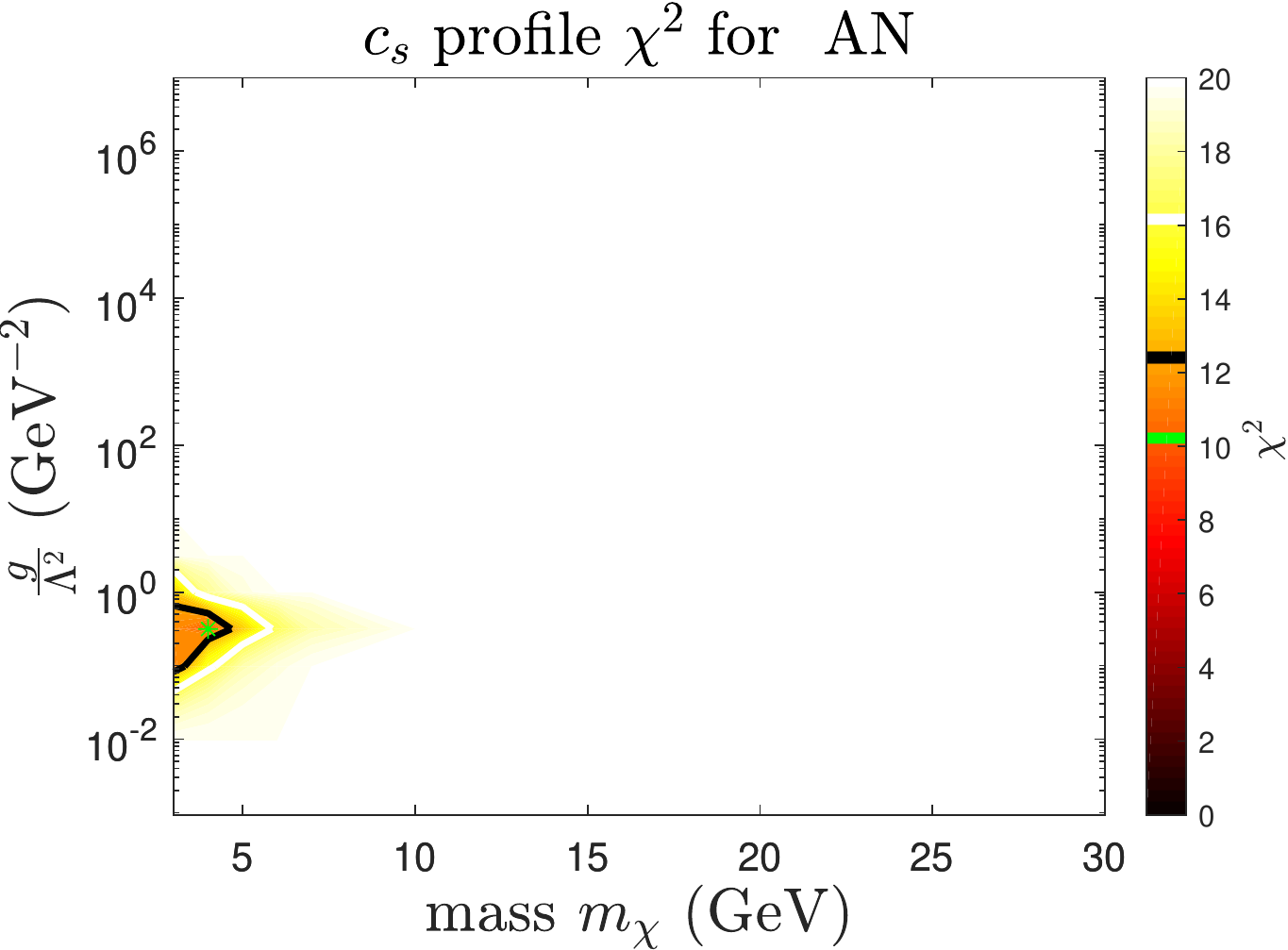}}
				}
				\caption{Combined likelihood $\chi^2$ of the sound-speed profile defined in eq.~(\ref{eq:chi2cs}) for spin independent dark matter (top left), electric dipole dark matter (top right), magnetic dipole dark matter (bottom left) and anapole dark matter (bottom right). The green star shows the best-fit $\chi^2$ and the black and white contours show the preferred regions at $1$ and $2\sigma$ respectively, corresponding to $\Delta\chi^2 = 2.3$ and $6.18$ respectively. Simulations in the grey regions did not converge.}
				\label{fig:cslikelihood}
			\end{figure}
			
			\begin{figure}[t]
				\centering
				\IfFileExists{../plots/csbestfit.eps}{
					\includegraphics[width=0.8\textwidth]{../plots/csbestfit.eps}
				}{
					\includegraphics[width=0.8\textwidth]{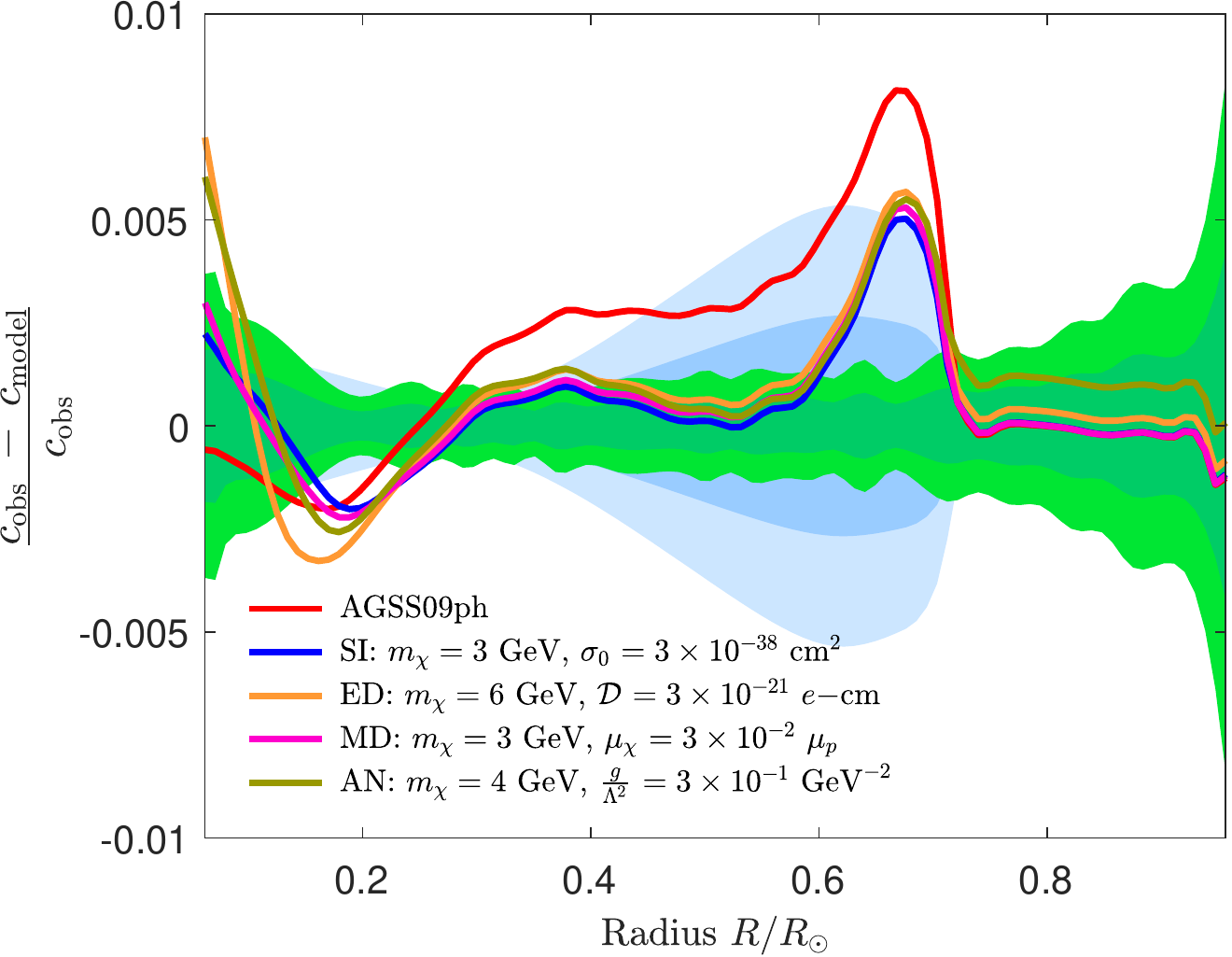}
				}
				\caption{Best fit profile of the sound speed for spin independent dark matter (SI), electric dipole dark matter (ED), magnetic dipole dark matter (MD) and anapole dark matter (AN). The light blue regions are the $1\sigma$ and $2\sigma$ errors from modelling. The green regions are the $1\sigma$ and $2\sigma$ errors from inversions. The red profile is the AGSS09ph~\cite{asplund09} without dark matter.}
				\label{fig:csprofile}
			\end{figure}
			
			Figure~\ref{fig:cslikelihood} shows the calculated $\chi_{c_s}^2$ from eq.~(\ref{eq:chi2cs}) for each of the simulated models. The darker red regions correspond to better fits to the sound-speed profile. Most of the regions that show an improvement are the same regions that show significant decreases in neutrino fluxes in figures~\ref{fig:Be7} and \ref{fig:B8}, since both phenomena are due to the temperature of the core being reduced.
			
			Having calculated the likelihood, it is possible to compare the sound-speed profile for the model with dark matter with the best fit to that without. The sound-speed profiles for the models of best fit are shown in figure~\ref{fig:csprofile}. 
			It is possible to discriminate the features of good or poor fits. All best fits occur at light masses, the regime where evaporation effects may become significant. The spin-independent, magnetic dipole and anapole fits do show some remediation of the tension at the base of the convective zone, $R\sim0.6R_\odot$. But all are characterised by an increasing speed of sound near the core of the Sun. Even though the profile in the inner regions is still within the helioseismological errors, the result is heavily constrained by the neutrino flux. Introducing dark matter can reduce the predicted neutrino flux by up to $35\%$ warranting caution in the results. The similarity between the magnetic dipole, anapole and spin-independent models can be explained by the fact that, for low masses, the thermal conductivities $\kappa$ are almost identical (see figure~\ref{fig:alphakappa}). However, the thermal diffusivity differs significantly between the anapole and other models, indicating that the distribution of dark matter for the anapole is more compact. The results suggest that the conductivity $\kappa$ can have a more pronounced effect on the sound-speed profile than the diffusivity $\alpha$. It is interesting to note that the finer resolution of points in the parameter space investigated here has found better fits to both the spin-independent and magnetic dipole models than the best fits presented for the same models elsewhere in the literature~\cite{cumberbatch10,lopes14}, due to the improved model of energy transport~\cite{vincent15b,vincent16}.
			
								
		\subsubsection{Frequency separation ratios}
		
			\begin{figure}[t]
				\centering
					\IfFileExists{../plots/r02bestfit.eps}{
						\subfloat{\includegraphics[width=0.49\textwidth]{../plots/r02bestfit.eps}}
					}{
						\subfloat{\includegraphics[width=0.49\textwidth]{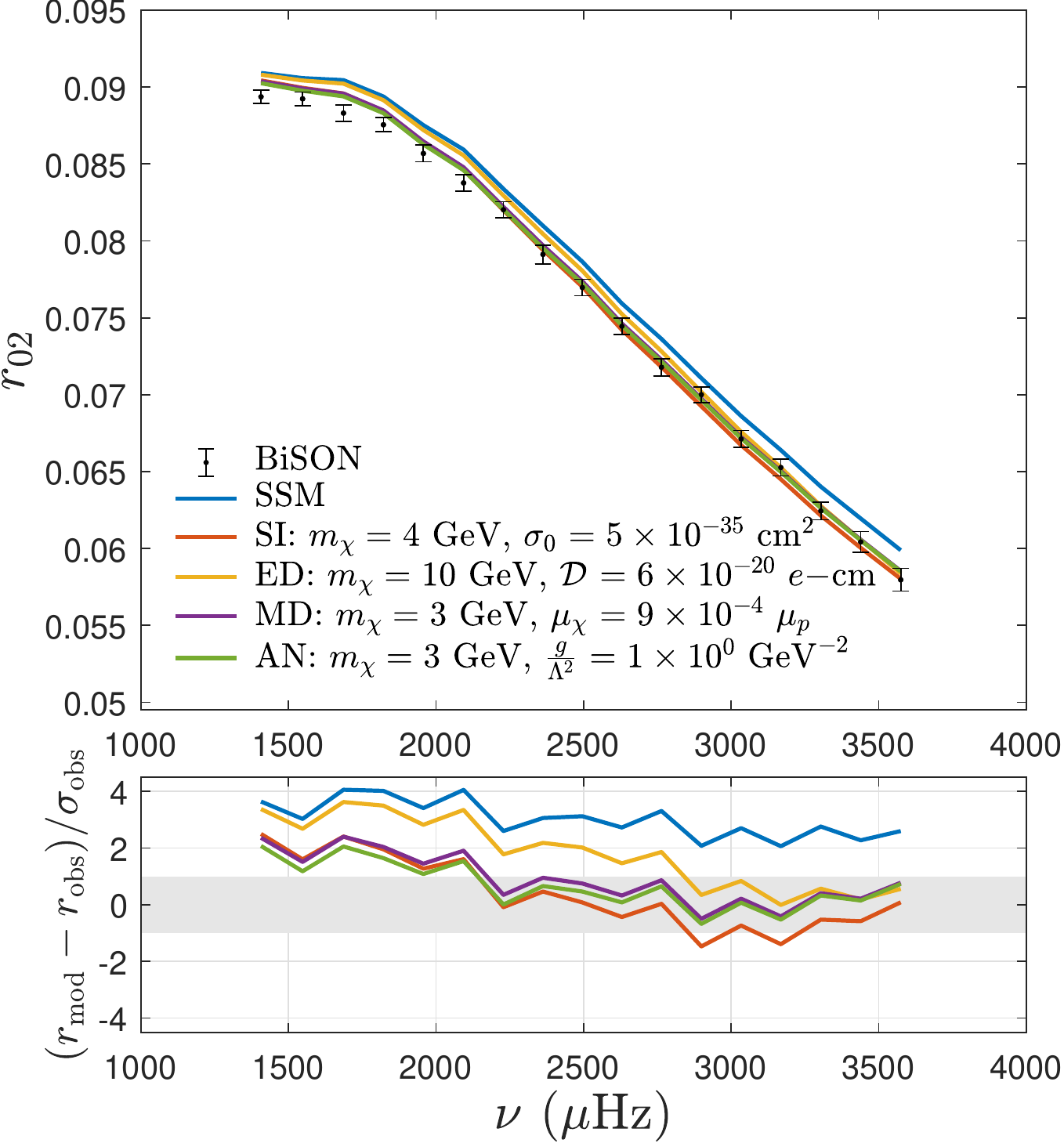}}
					}
					\IfFileExists{../plots/r13bestfit.eps}{
						\subfloat{\includegraphics[width=0.49\textwidth]{../plots/r13bestfit.eps}}
					}{
						\subfloat{\includegraphics[width=0.49\textwidth]{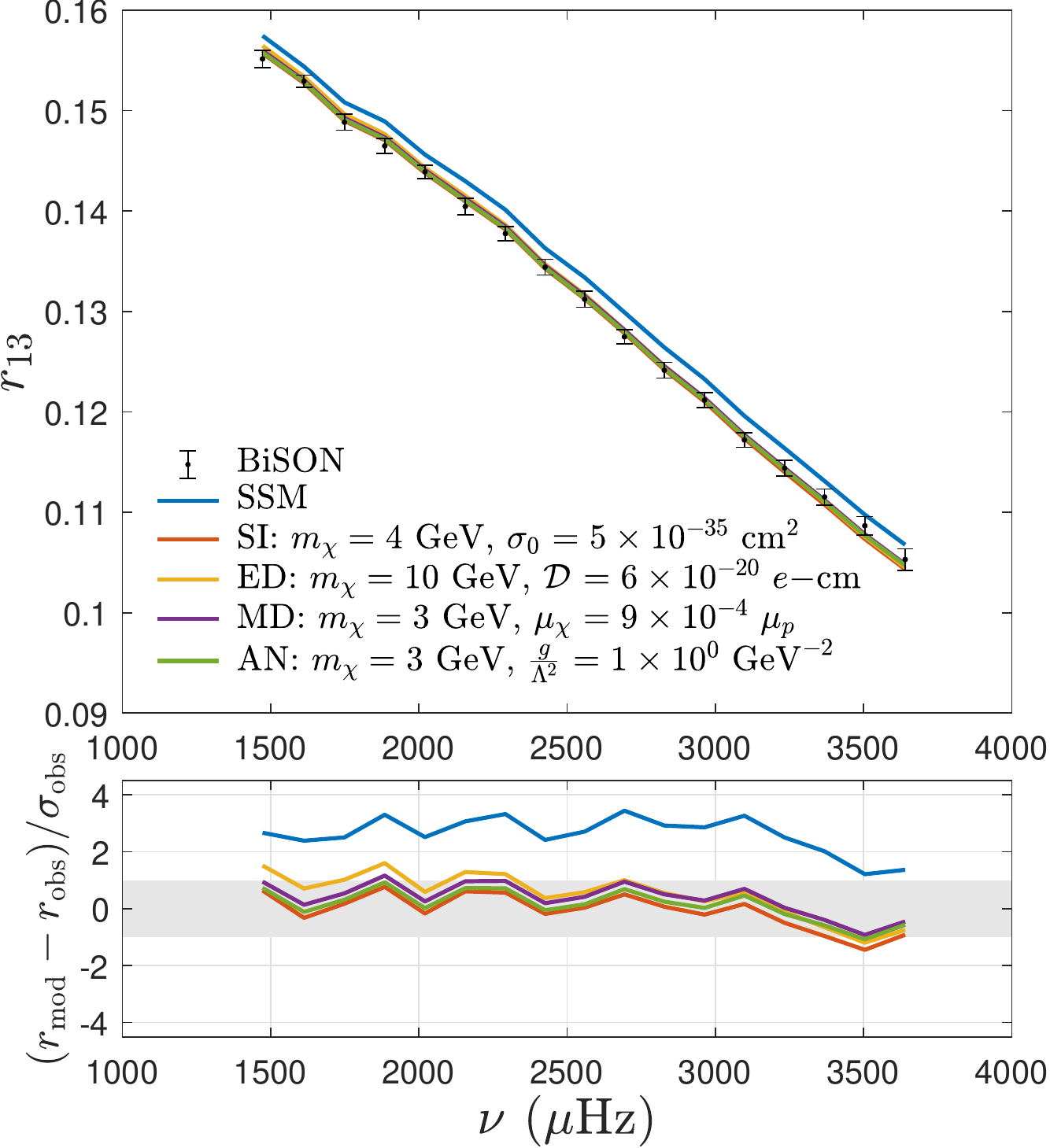}}
					}
				\caption{Small frequency separations $r_{02}$ (left) and $r_{13}$ (right) for the best-fit models to helioseismological observations. Data is compared to the Standard Solar Model (SSM) and BiSON experiment~\cite{basu07}. The error bars correspond to observational and modelling error~\cite{vincent15b}. Below each figure are the residuals with respect to BiSON data, in units of the total error.
				}
				\label{fig:freqsepbestfit}
			\end{figure}		
			\begin{figure}[t]
				\centering
				\IfFileExists{../plots/SIratiosB.eps}{
					\subfloat{\includegraphics[width=0.49\textwidth]{../plots/SIratiosB.eps}}
					\subfloat{\includegraphics[width=0.49\textwidth]{../plots/EDratiosB.eps}}
				}{
					\subfloat{\includegraphics[width=0.49\textwidth]{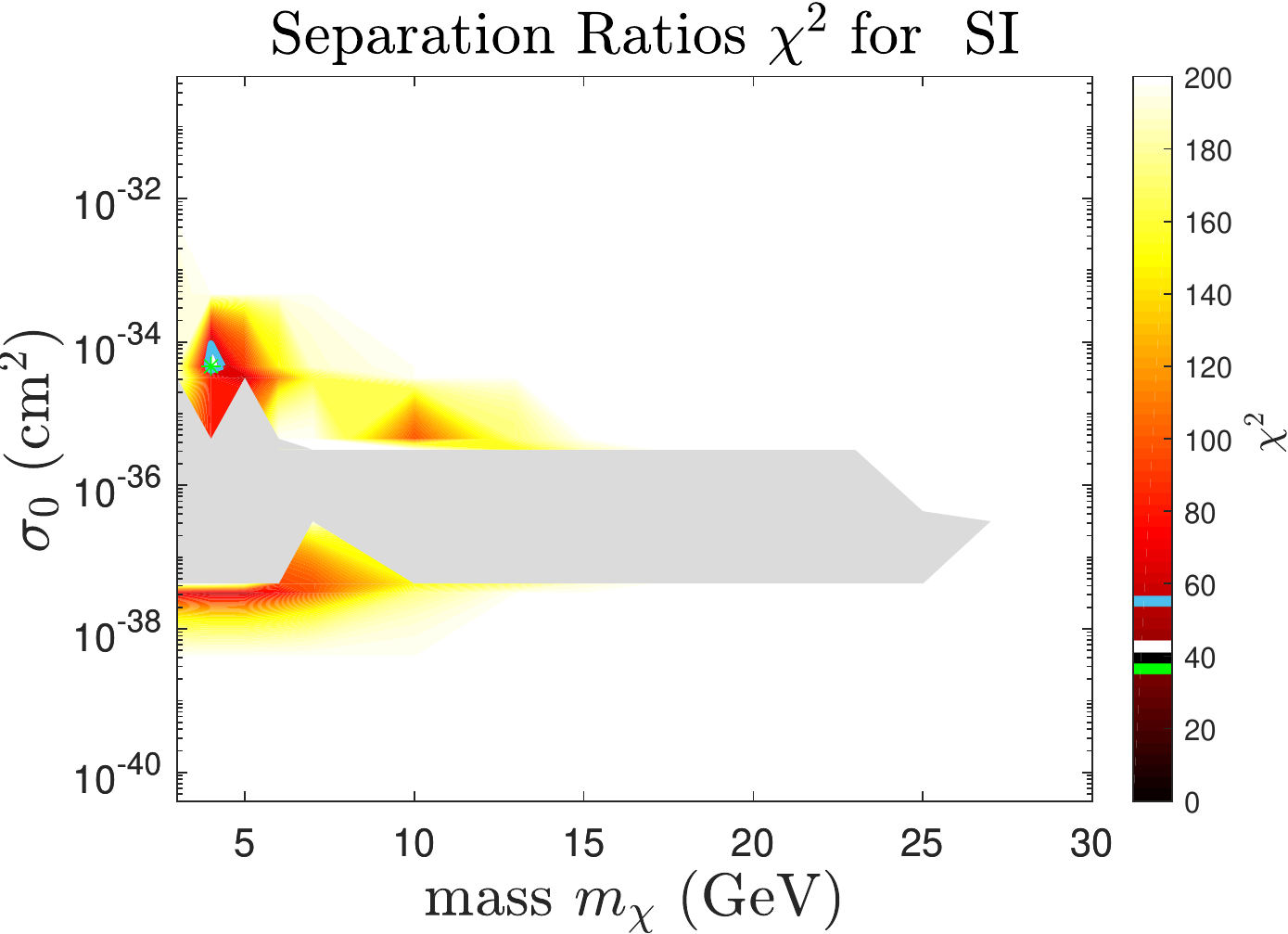}}
					\subfloat{\includegraphics[width=0.49\textwidth]{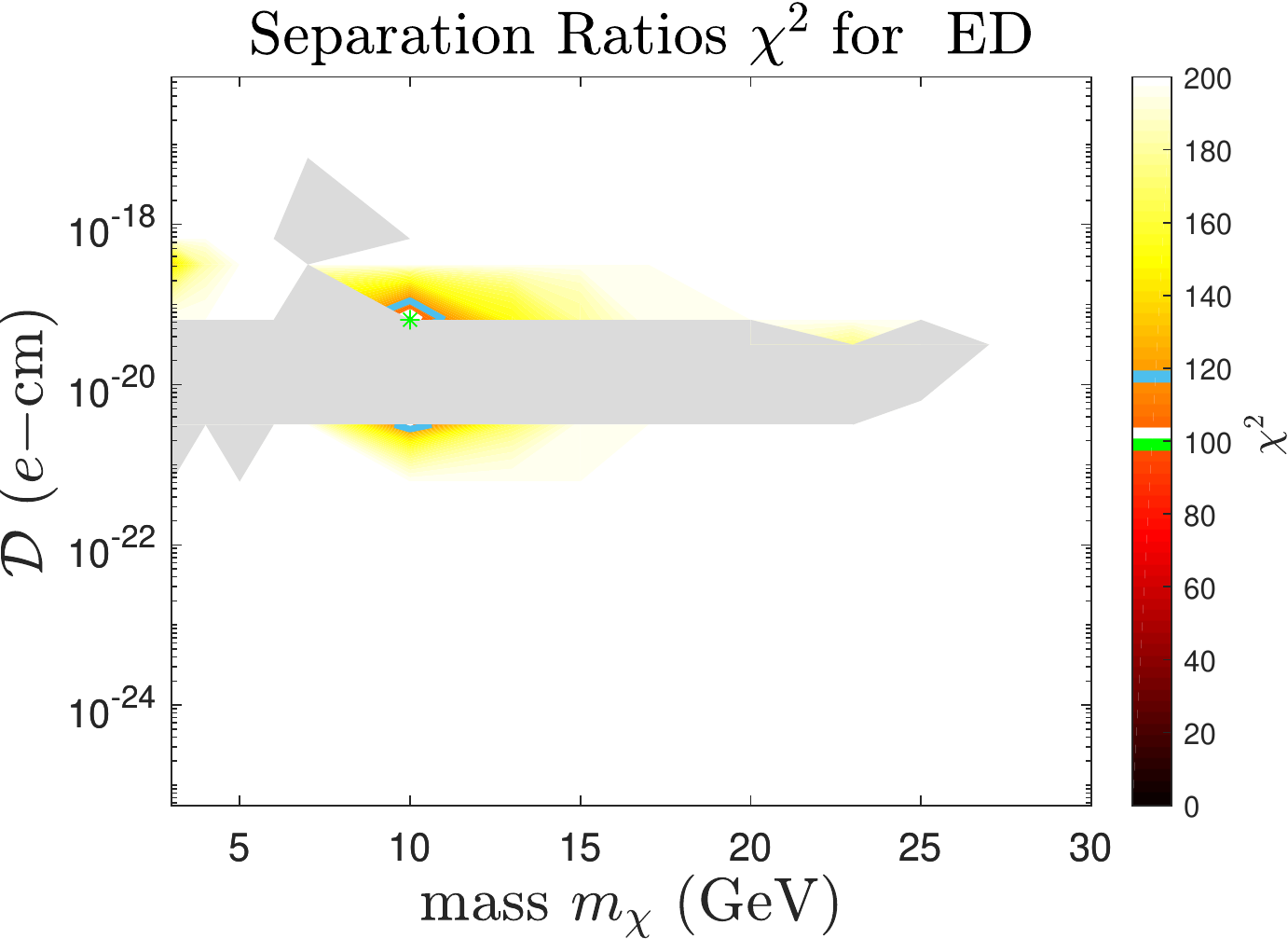}}
				}
				\IfFileExists{../plots/EDratiosB.eps}{
					\subfloat{\includegraphics[width=0.49\textwidth]{../plots/MDratiosB.eps}}
					\subfloat{\includegraphics[width=0.49\textwidth]{../plots/ANratiosB.eps}}
				}{
					\subfloat{\includegraphics[width=0.49\textwidth]{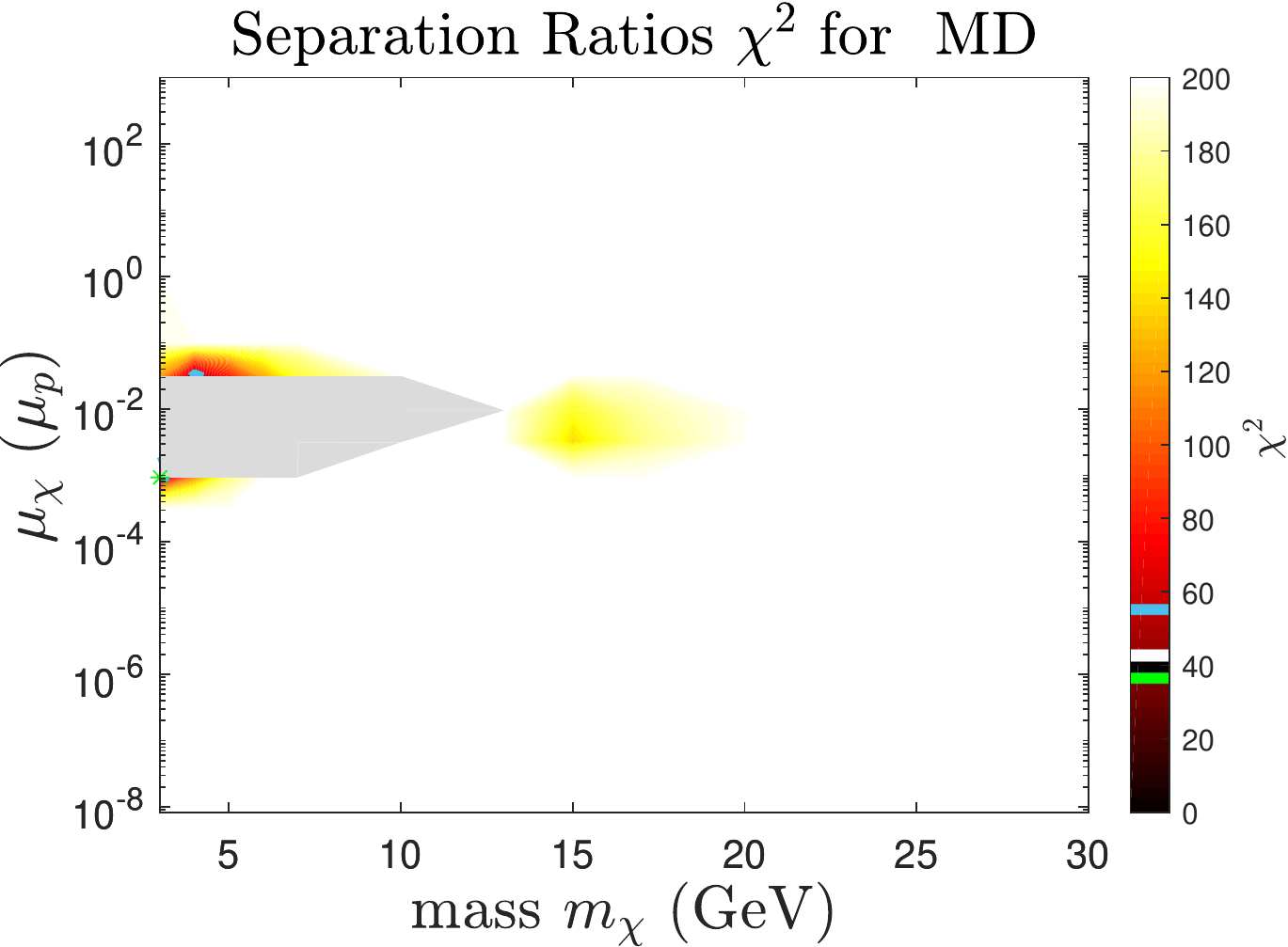}}
					\subfloat{\includegraphics[width=0.49\textwidth]{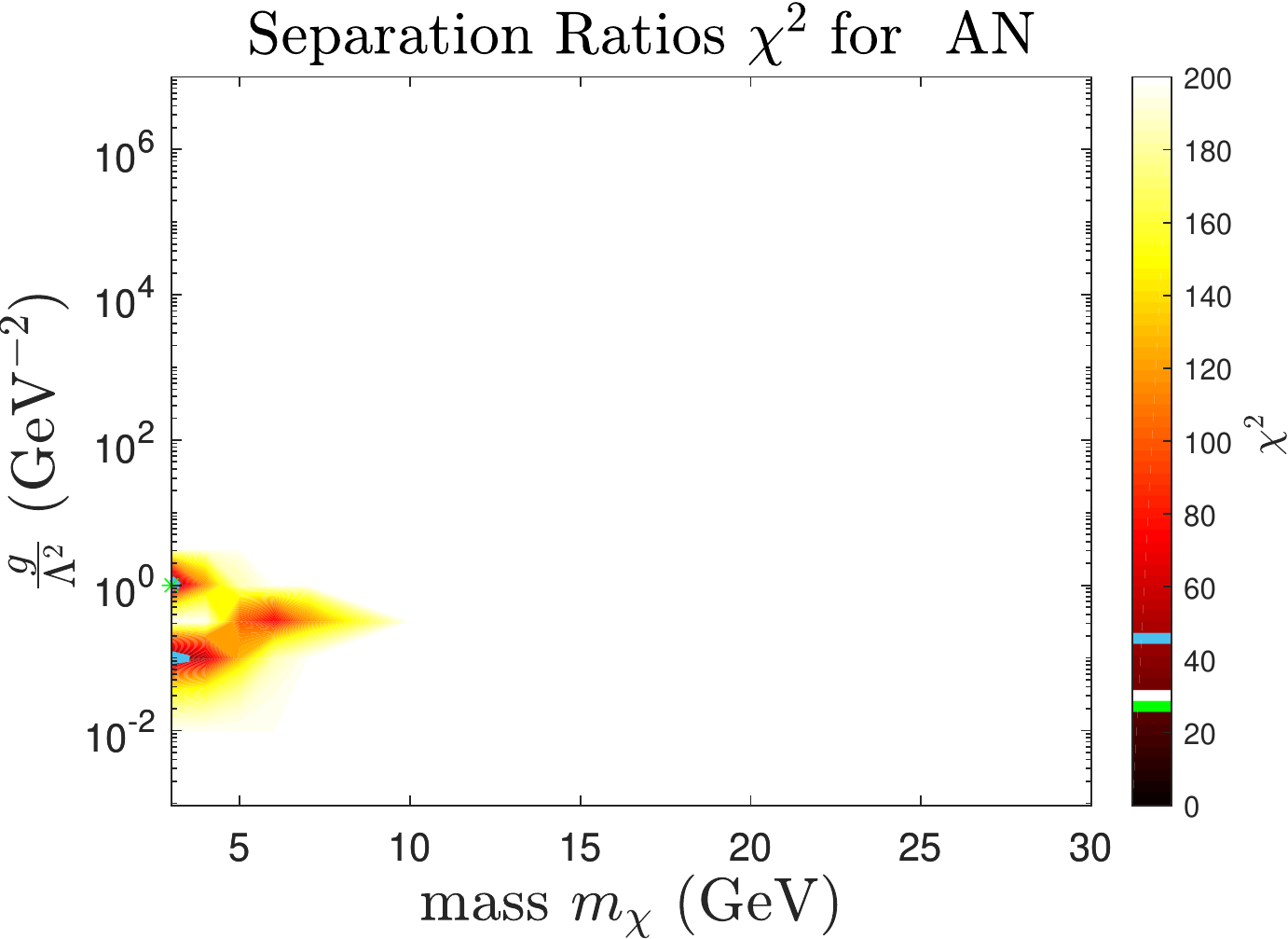}}
				}
				\caption{
				Combined likelihood $\chi^2$ of the small frequency separation ratios defined in eq.~(\ref{eq:chi2ratio}) for spin independent dark matter (top left), electric dipole dark matter (top right), magnetic dipole dark matter (bottom left) and anapole dark matter (bottom right). The green star shows the best-fit $\chi^2$ and the black, white and cyan contours show the preferred regions at $1$, $2$ and $4\sigma$ respectively, corresponding to $\Delta\chi^2 = 2.3$, $6.18$ and $19.33$ respectively. Simulations in the grey regions did not converge.}
				\label{fig:freqsep}
			\end{figure}

			Uncertainties of the sound speed profile are large near the core of the sun, and the region can be better probed by the so-called frequency separation ratios, which are not dependent on the solar surface as described by solar models~\cite{roxburgh03}. These ratios are sensitive to changes in the solar core, while remaining largely independent of systematic errors found in sound speed inversions~\cite{basu07,chaplin07}. The two ratios are calculated using
			\begin{equation}
				r_{02}(n) = \frac{d_{02}(n)}{\Delta_1(n)},\quad r_{13}(n) = \frac{d_{13}(n)}{\Delta_0(n+1)}
			\end{equation}
			where
			\begin{equation}
				d_{l,l+2}(n) \equiv \nu_{n,l}-\nu_{n-1,l+2}\simeq -(4l + 6)\frac{\Delta_l(n)}{4\pi^2\nu_{n,l}}\int_0^{R_\odot} \frac{dc_s}{dr} \frac{dr}{r}
			\end{equation}
			and $\Delta_l(n) = \nu_{n,l}-\nu_{n-1,l}$.
		
			Figure~\ref{fig:freqsepbestfit} shows the simulated models which best fit the small-frequency ratios $r_{02}$ and $r_{13}$ observed by the BiSON experiment~\cite{basu07,chaplin07}. Dipole moment dark matter can significantly improve the fit to observation, with respect to the SSM. The overall fit is quantified by the combined chi-squared $\chi_{r_{02}}^2 + \chi_{r_{13}}^2$, where
			\begin{equation}
			\label{eq:chi2ratio}
				\chi_{r_{l,l+1}}^2 = \sum_n \frac{\left[r_{l,l+2,\mathrm{th.}}(n) - r_{l,l+2,\mathrm{obs.}}(n)\right]^2}{\sigma_{\mathrm{obs.}}^2(n) + \sigma_{\mathrm{th.}}^2(n)} .
			\end{equation}
			Figure~\ref{fig:freqsep} shows the combined $\chi_{r_{02}}^2 + \chi_{r_{13}}^2$ for each of the simulated models. The parameters which best fit the frequency separations are not necessarily the same parameters which form the best fit to the sound speed profile, but in general the regions of improvement overlap.

		\subsubsection{Depth of the convection zone}
		
			The depth of the convection zone $R_{\mathrm{cz}}$ is an observable that describes the boundary between two different mechanisms of heat transfer, known as the tachocline. It therefore has important implications for the structure of the Sun~\cite{spiegel92}. 
			The depth of the convection zone is also significant as it delineates the region of the Sun where convective mixing homogenises the chemical composition~\cite{christensendalsgaard91}. 
			
			$R_{\mathrm{cz}}$ can be determined precisely because it is the depth at which the temperature gradient reaches the adiabatic value and convection sets in, leading to a discontinuity in the second derivative of the temperature with radius.  The localised variation of the thermal properties of the Sun introduces a signal in the oscillation pattern that allows for accurate extraction of $R_{\mathrm{cz}}$.
			
			$R_{\mathrm{cz}}$ is determined by calculating the discontinuity in the temperature gradient inferred from the sound-speed profile (see section~\ref{sec:helioseismology}). The measured value is calculated to be $R_{\mathrm{cz}} = (0.713 \pm 0.001) R_\odot$~\cite{christensendalsgaard91,basu97,basu04,basu98}, but models currently suggest a $\sim 3\sigma$ overestimate.
			
			\begin{figure}[t]
				\centering
				\IfFileExists{../plots/SIRcz.eps}{
					\subfloat{\includegraphics[width=0.49\textwidth]{../plots/SIRcz.eps}}
					\subfloat{\includegraphics[width=0.49\textwidth]{../plots/EDRcz.eps}}
				}{
					\subfloat{\includegraphics[width=0.49\textwidth]{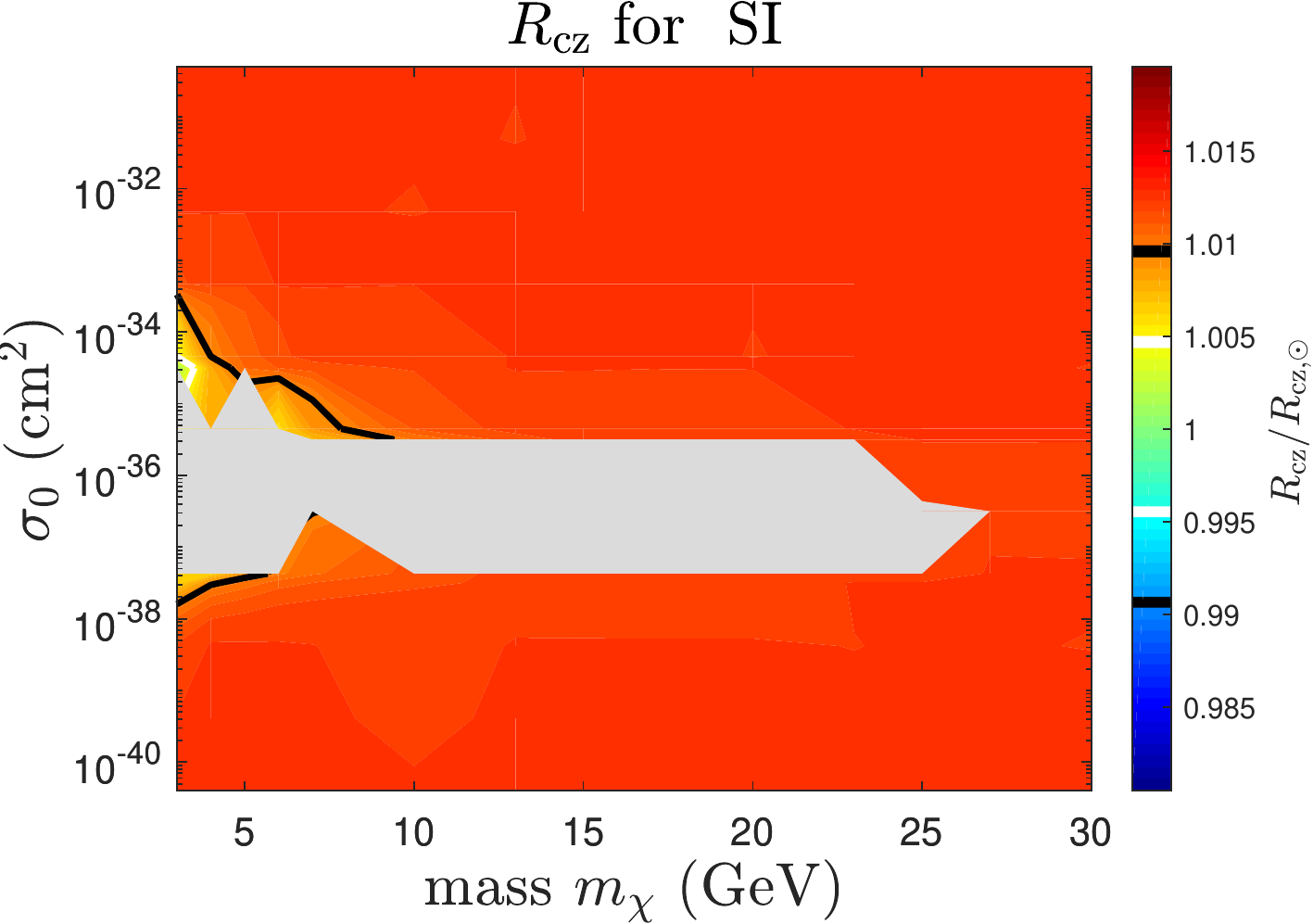}}
					\subfloat{\includegraphics[width=0.49\textwidth]{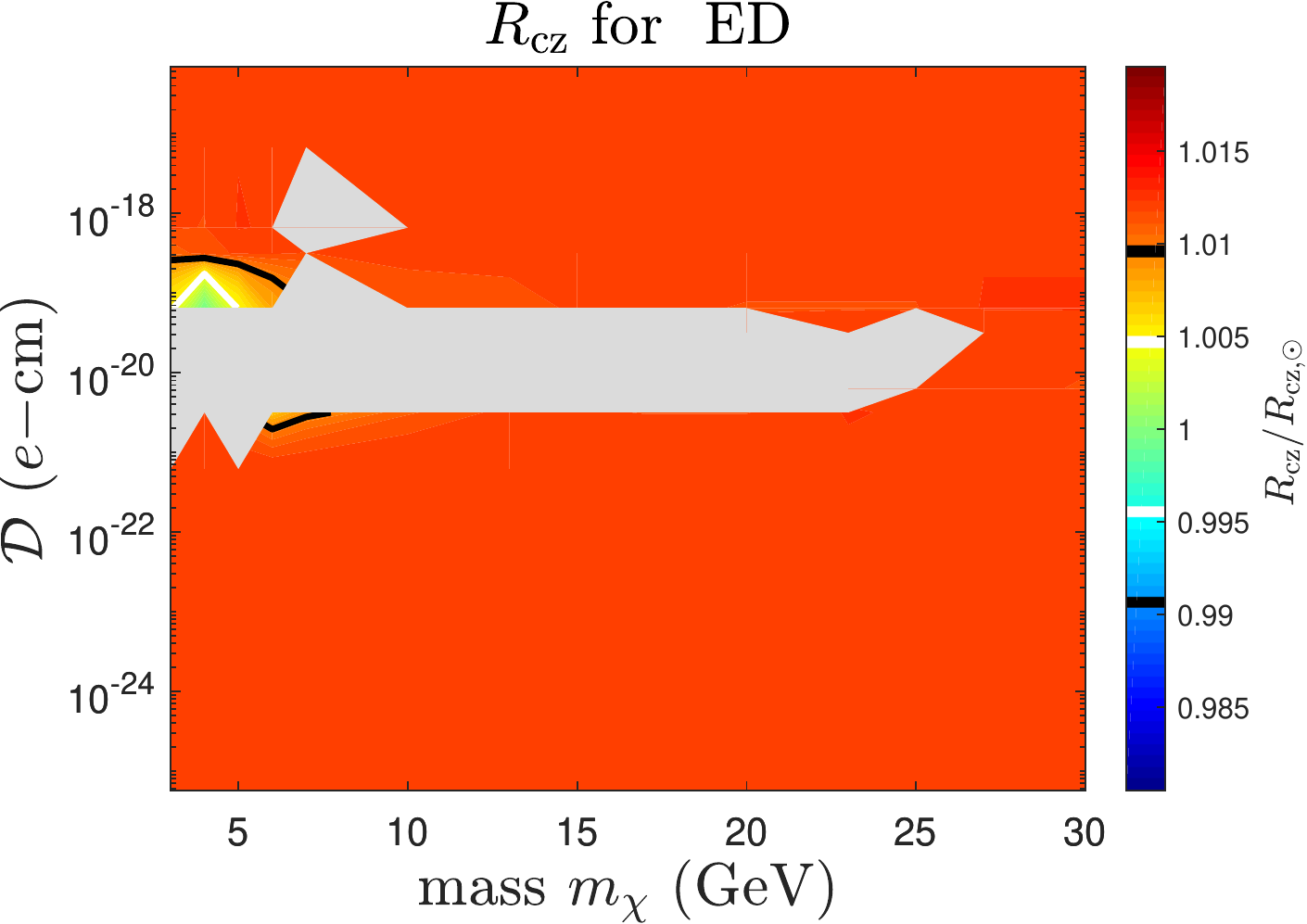}}
				}
				\IfFileExists{../plots/EDRcz.eps}{
					\subfloat{\includegraphics[width=0.49\textwidth]{../plots/MDRcz.eps}}
					\subfloat{\includegraphics[width=0.49\textwidth]{../plots/ANRcz.eps}}
				}{
					\subfloat{\includegraphics[width=0.49\textwidth]{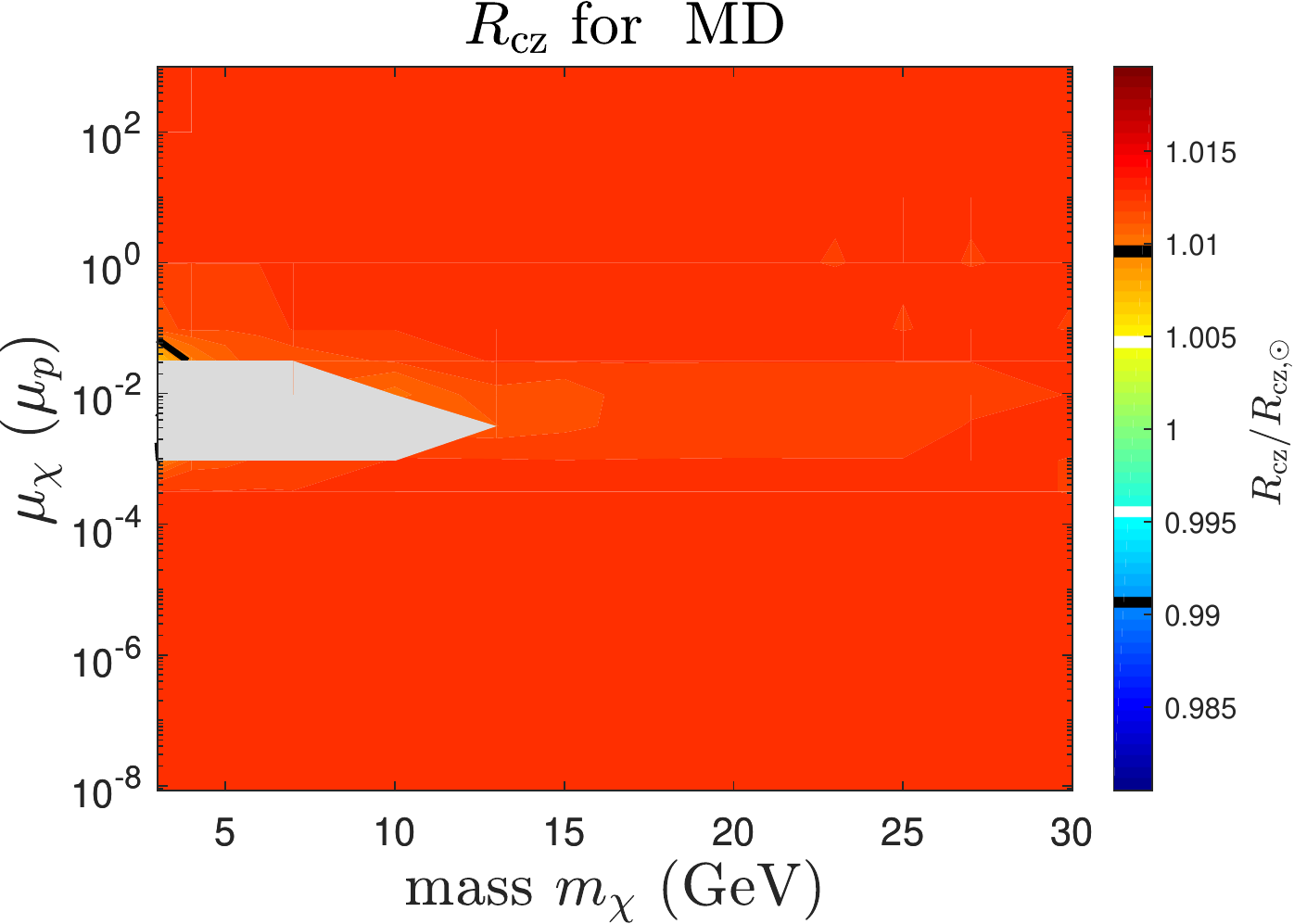}}
					\subfloat{\includegraphics[width=0.49\textwidth]{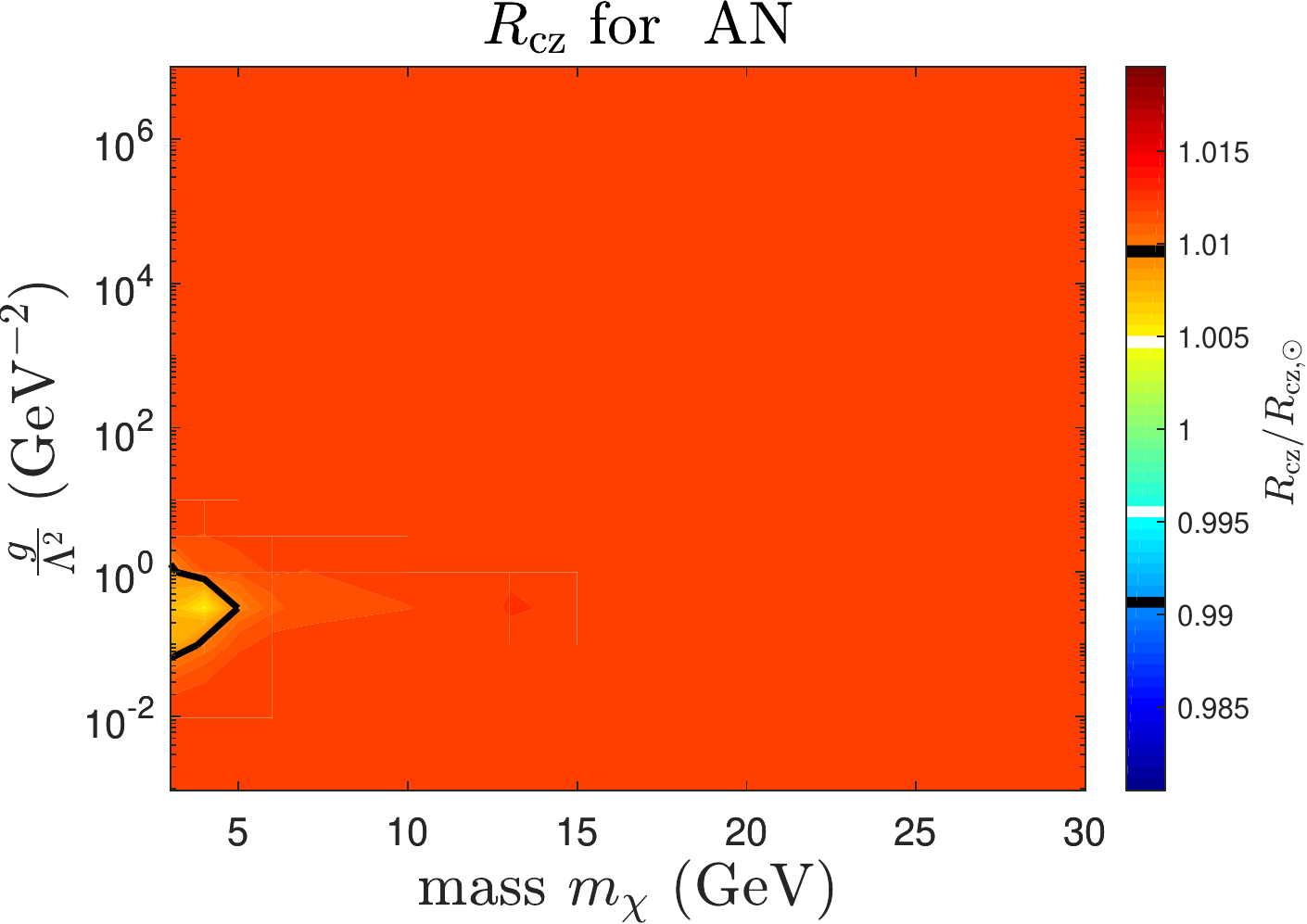}}
				}
				\caption{Ratio of predicted convective zone depth to the measured value $R_{\mathrm{cz}} = 0.713 R_\odot$ for spin independent dark matter (top left), electric dipole dark matter (top right), magnetic dipole dark matter (bottom left) and anapole dark matter (bottom right). The expected value is therefore 1. The white and black contours show the regions where the $R_{\mathrm{cz}}$ is $1\sigma$ and $2\sigma$ above/below the expected value respectively. Simulations in the grey regions did not converge.}
				\label{fig:Rcz}
			\end{figure}		
			
			The ratio of the predicted to the measured convective zone depth is shown in figure~\ref{fig:Rcz}. 
			 The unmodified background appears as red. The models which provide a better fit appear more yellow/green and correspond to the better fits to the sound-speed profile in figure~\ref{fig:cslikelihood}.
			Models that predict a convective zone depth within $2\sigma$ of the observed value typically produce sound-speed profiles that alleviate the larger discrepancy at around $0.65R_\odot$ in the sound-speed profile, including all of the best fit models to the sound-speed profile in figure~\ref{fig:csprofile}. The physical insight as to why these models are effective is therefore the same as that for the sound-speed profiles: more energy deposited in the radiative zone marginally increases the temperature gradient, causing it to exceed the adiabatic gradient at a slightly lower depth. 			
			
			Again, the regions of slight improvement correspond to the regions for which the energy transport in figure~\ref{fig:energy} is maximised. However, $R_{\mathrm{cz}}$ is an independent observable to the neutrino fluxes. In most cases, a better $R_{\mathrm{cz}}$ fit corresponds to a worse neutrino fit.		
			Unlike the neutrino fluxes, which drop off rapidly on the borders of the non-converged regions, the best fit regions of $R_{\mathrm{cz}}$ do not necessarily relate to the convergence. Therefore, the lack of convergence is more likely to be due to effects within the core, rather than in the outer regions. 

		\subsubsection{Surface helium abundance}
		
			An additional property that can be extracted from helioseismic analysis is the abundance of helium in the solar convective envelope. Second ionization of helium occurs in a localised region of the Sun, around $R = 0.98 R_\odot$, and causes a sharp depression in the so-called adiabatic index $\Gamma_1 = \left.\frac{\partial \log{P}}{\partial \log{\rho}}\right|_{\mathrm{ad}}$. Oscillation frequencies are sensitive to both the location and amplitude of this depression, and from the latter the abundance of helium can be extracted. Convective mixing in the solar convective envelope occurs on time scales of the order of a month, much shorter than the time scales over which the solar structure is altered. The mixing ensures that the convective envelope is chemically well mixed and the surface, or photospheric, helium abundance matches the one extracted from helioseismology.
			
			As already discussed, one of the effects of including energy transport due to the presence of dark matter in solar models is to decrease the temperature in the solar core, where nuclear energy is released. If this were the only change in solar models, then the integrated nuclear energy release would be lower as well, as a consequence of the cooler core. However, solar models are forced by construction to match $L_\odot$. Therefore, a decrease in the core temperature is accompanied by an increase in the hydrogen content of the core such that the integrated nuclear energy release reproduces $L_\odot$. The lower initial helium in the models is then propagated to the present day surface helium abundance. Values in table~\ref{tab:results} show that the different dark matter models lead to a reduction of the surface helium abundance in the range 1-2\%, comparable to about 1$\sigma$ of the total error budget. The implication is a slightly worse agreement between the helium abundance of dark matter models and helioseismology, but it is a change that is overcome by the much larger improvement seen in other helioseismic diagnostics.

		\subsubsection{Total likelihood}
		
			We construct a total combined likelihood from the two neutrino fluxes, convective zone radius, surface helium abundance and frequency separation ratios. All four sets of observables are independent, so the combined $\chi^2$ value is defined as
			\begin{equation}
			\label{eq:chi2total}
				\chi^2 = \frac{(\phi_{B} - \phi_{B,\mathrm{obs}})^2}{\sigma_B^2} + \frac{(\phi_{Be} - \phi_{Be,\mathrm{obs}})^2}{\sigma_{Be}^2} + \frac{(R_{\mathrm{cz}} - R_{\mathrm{cz,obs}})^2}{\sigma_{R_{\mathrm{cz}}}^2} + \frac{(Y_{\mathrm{surf}} - Y_{\mathrm{surf,obs}})^2}{\sigma_{Y_{\mathrm{surf}}}^2} + \chi^2_{r_{02}} + \chi^2_{r_{13}}
			\end{equation}
			with the uncertainties taken from each of the measurements~\cite{vincent15b}. The sound speed profile is not included as it is correlated with the small frequency separations.
			The total $\chi^2$ is plotted in figure~\ref{fig:totallikelihood}. The likelihoods for each of the models which best fit $\chi^2_{\mathrm{total}}$ are presented in table~\ref{tab:results}. For the electric dipole model, the best fit is the same as the fit to the sound-speed profile and small frequency separations. For the magnetic dipole and anapole models, the best fit parameters differ slightly from the best fit to the sound-speed profile.

			\IfFileExists{appendices/resultstable.tex}{
				\begin{table}[t]
\centering
\begin{tabular}{|l|l|l|l|l|l|l|l|l|l|}
	\hline
	Model                                    & $m_\chi$ & Coupling                            & $\nu_{\mathrm{Be}^7}$ & $\nu_{\mathrm{B}^8}$ & $ R_{\mathrm{cz}}$ & $Y_s$ & $\chi^2_{r}$ & $\chi^2_{\mathrm{total}}$ & $p_{\mathrm{tot}}$ \\
	\hhline{|=|=|=|=|=|=|=|=|=|=|} no DM     & -        & -                                   & 4.71                  & 4.95                 & 0.722              & 0.236 & 276          & 287                       & $< 10^{-10}$         \\
	SI ($\sigma_0$)                          & 4        & $5\times10^{-35}~ \mathrm{cm}^2 $   & 4.37                  & 3.80                 & 0.720              & 0.234 & 34.7         & 51.8                      & 0.04                 \\
	ED ($\mathcal{D}$)                       & 10       & $6\times10^{-20}~ e$-$\mathrm{cm} $   & 4.41                  & 3.69                 & 0.721              & 0.235 & 96.8         & 115                       & $3\times10^{-10}$    \\
	MD ($\mu_\chi$)                          & 3        & $9\times10^{-4}~ \mu_p $            & 4.42                  & 3.96                 & 0.720              & 0.234 & 35.0         & 50.6                      & 0.05                 \\
	AN ($\frac{g}{\Lambda^2}$)               & 3        & $1\times10^{0}~ \mathrm{GeV}^{-2} $ & 4.38                  & 3.88                 & 0.720              & 0.238 & 23.9         & 40.4                      & 0.28                 \\
	\hhline{|=|=|=|=|=|=|=|=|=|=|} Obs.      & -        & -                                   & 4.80                  & 5.16                 & 0.713              & 0.249 & -            & -                         & -                    \\
	\hhline{|=|=|=|=|=|=|=|=|=|=|} Obs. Err. & -        & -                                   & 5\%                   & 13\%                 & 0.001              & 0.003 & -            & -                         & -                    \\
	Mod. Err.                                & -        & -                                   & 14\%                  & 7\%                  & 0.004              & 0.003 & -            & -                         & -                    \\ \hline
\end{tabular}
\caption{Table of parameters for best fit models. The run types are: Spin Independent (SI), Electric Dipole (ED), Magnetic Dipole (MD), Anapole (AN). The mass is in units of $\GeV$. The beryllium-7 neutrino flux is in units $ 10^{-6} \mathrm{cm}^{-2} \mathrm{s}^{-1} $, the boron-8 neutrino flux is in units $ 10^{-9} \mathrm{cm}^{-2} \mathrm{s}^{-1} $ and the convective zone radius is in units $ R_{\odot} $. $\chi^2_r$ represents the likelihood of the small frequency separations defined in eq.~(\ref{eq:chi2ratio}). The total $\chi^2$ value is defined in eq.~(\ref{eq:chi2total}).}
\label{tab:results}
\end{table}
	
			}{
				
			}
			
			\begin{figure}[t]
				\centering
				\IfFileExists{../plots/SIchi2totalB.eps}{
					\subfloat{\includegraphics[width=0.49\textwidth]{../plots/SIchi2totalB.eps}}
					\subfloat{\includegraphics[width=0.49\textwidth]{../plots/EDchi2totalB.eps}}
				}{
					\subfloat{\includegraphics[width=0.49\textwidth]{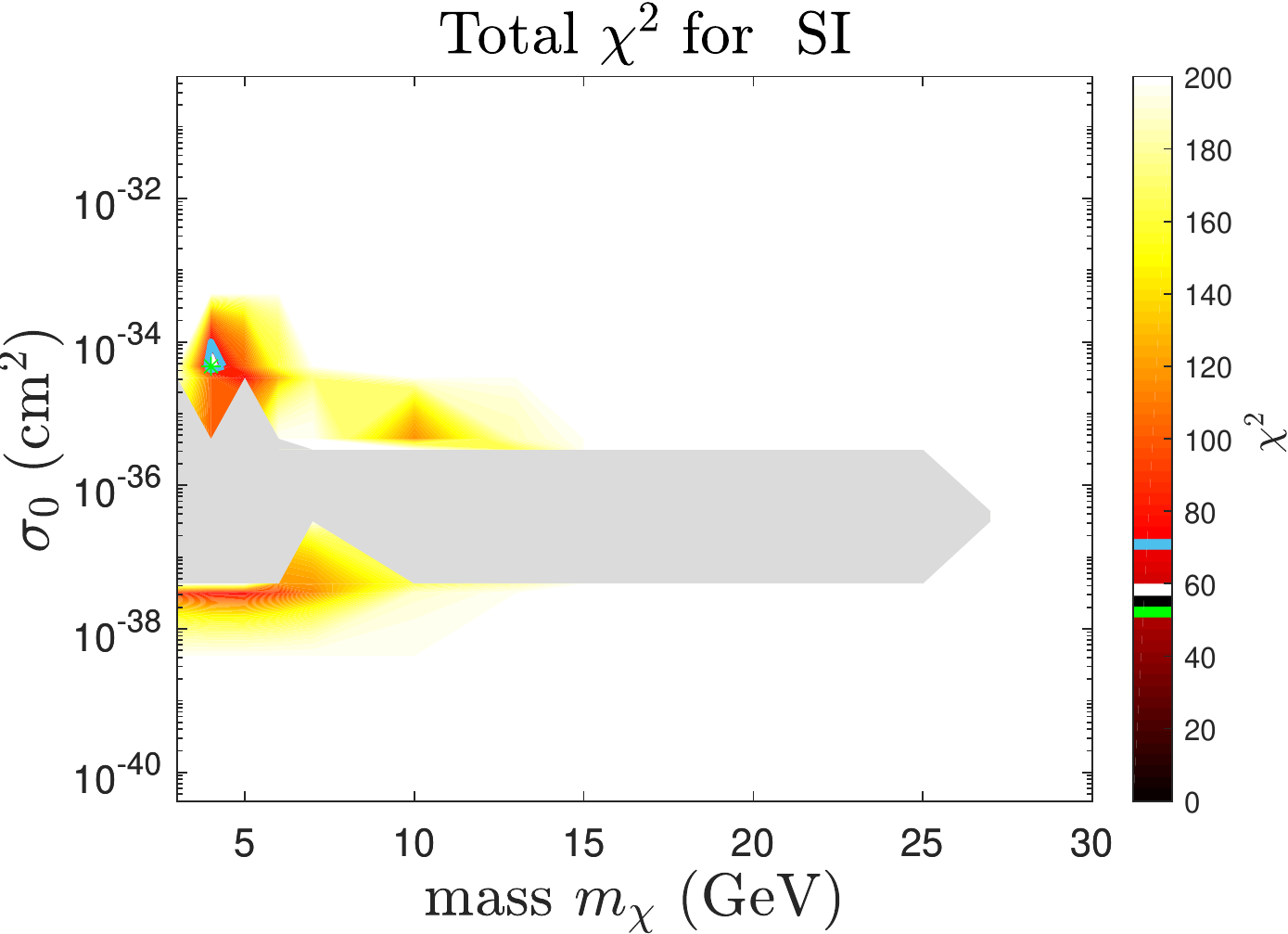}}
					\subfloat{\includegraphics[width=0.49\textwidth]{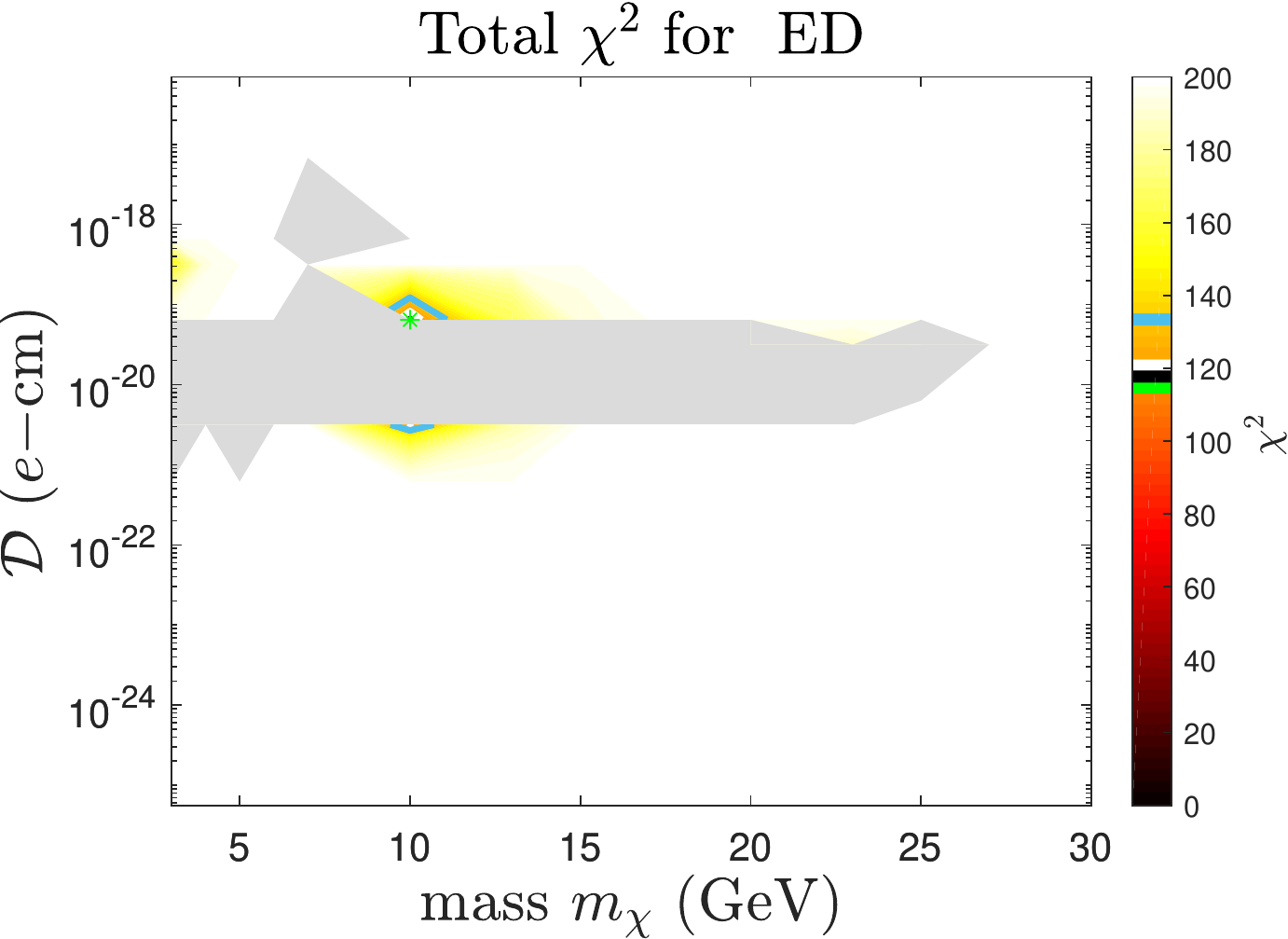}}
				}
				\IfFileExists{../plots/EDchi2totalB.eps}{
					\subfloat{\includegraphics[width=0.49\textwidth]{../plots/MDchi2totalB.eps}}
					\subfloat{\includegraphics[width=0.49\textwidth]{../plots/ANchi2totalB.eps}}
				}{
					\subfloat{\includegraphics[width=0.49\textwidth]{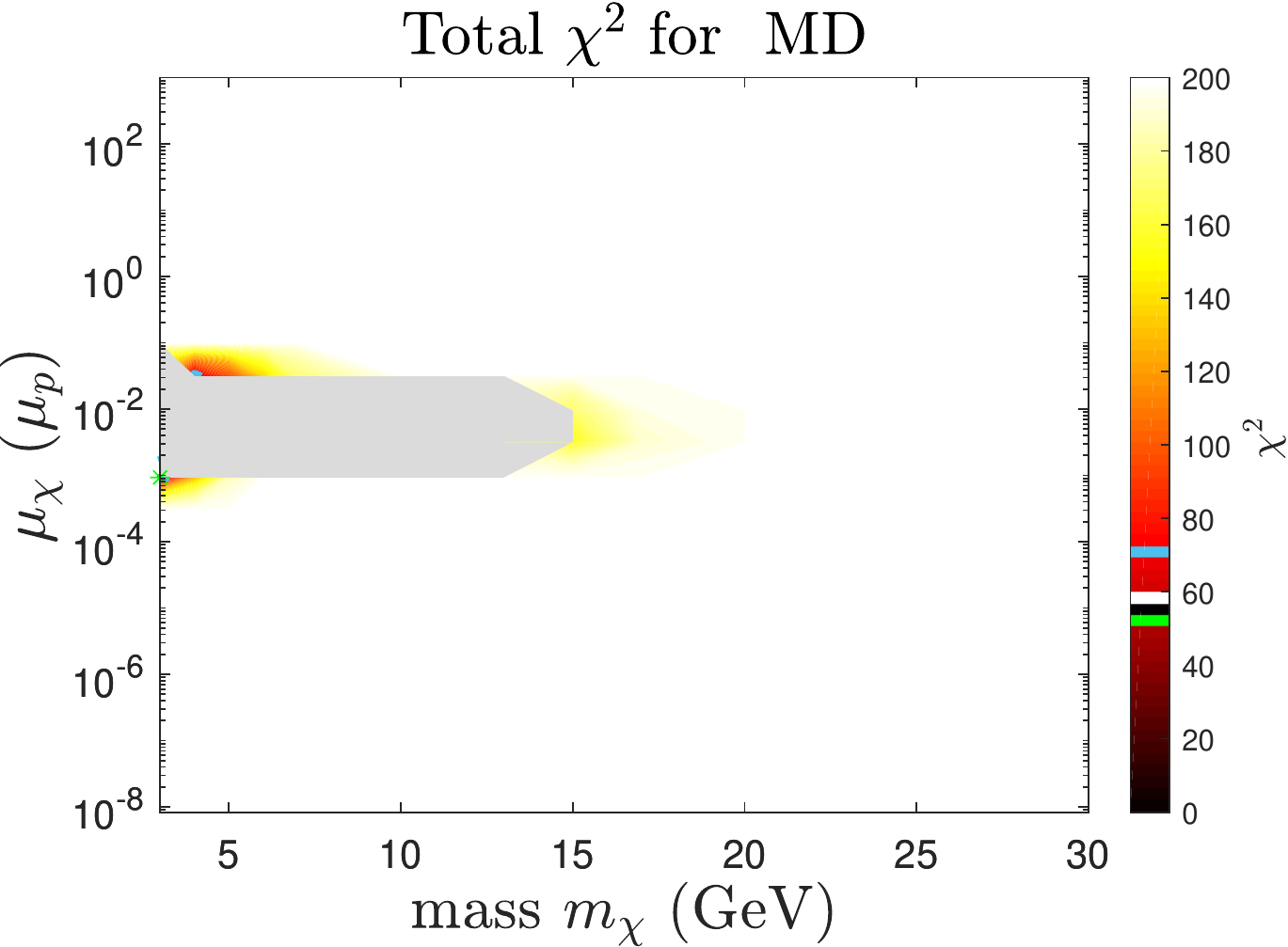}}
					\subfloat{\includegraphics[width=0.49\textwidth]{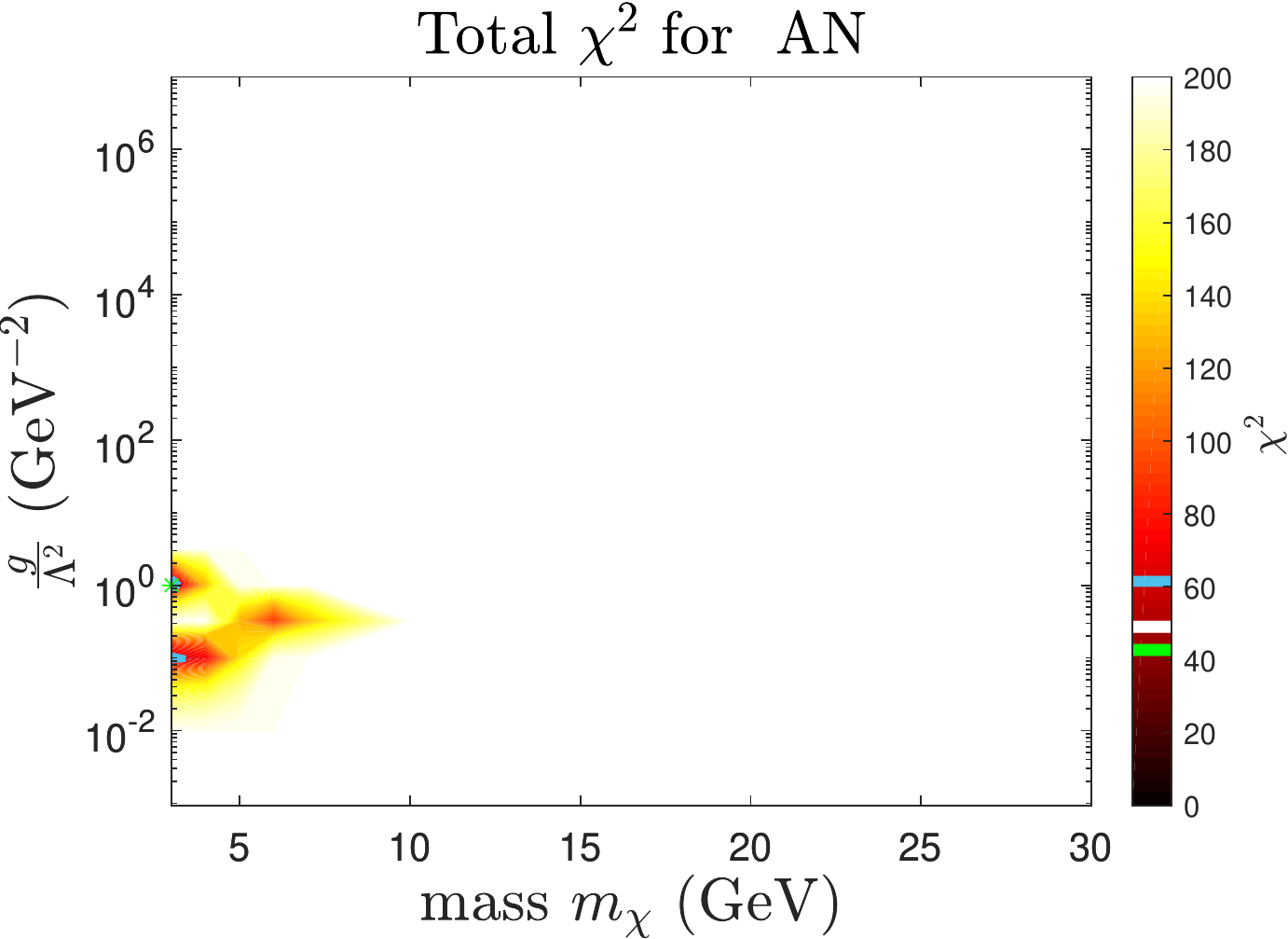}}
				}
				\caption{Combined $\chi^2$ of neutrino fluxes, convective zone radius, surface helium abundance and frequency separation ratios matching observed values as given by eq.~(\ref{eq:chi2total}) for spin independent dark matter (top left), electric dipole dark matter (top right), magnetic dipole dark matter (bottom left) and anapole dark matter (bottom right). The green star shows the best-fit $\chi^2$ and the black, white and cyan contours show the preferred regions at $1$, $2$ and $4\sigma$ respectively, corresponding to $\Delta\chi^2 = 2.3$, $6.18$ and $19.33$ respectively. Simulations in the grey regions did not converge.}
				\label{fig:totallikelihood}
			\end{figure}

			Introducing energy transport through electromagnetic dipole dark matter improves the combined likelihood compared to the SSM with no dark matter. The reduced neutrino fluxes are offset by the improved frequency separation ratios, in particular for the electric dipole and anapole models. The best fit parameters for each model are displayed in table~\ref{tab:results}. Every one of the electromagnetic dipole models outperforms both the SSM and constant-cross section spin-independent dark matter. In particular, electric dipole and anapole models are favoured. Note, however, that the best-fit models typically occur for small masses, where the impact of evaporation is likely to be most pronounced.
			

		\section{Discussion}
\label{sec:discussion}

	The prospect of solving the discrepancies in the SSM by introducing specific models of dark matter is promising, yet major issues remain. In particular, a model of dark matter with an anapole moment of $1\GeV^{-2}$ and mass $3\GeV$ can modify the sound-speed profile to agree with the helioseismological observations. The magnetic dipole model performs comparably well to the constant cross section model, and both are a significant improvement over models without dark matter.
	
	However, significant problems remain. The favoured models are well within the exclusion regions of all direct detection, beam dump and collider experiments that have reported constraints on electromagnetic dipole dark matter~\citep{masso09,delnobile14,mohanty15}. For example, the upper bound on the anapole moment and magnetic dipole momnet are $\sim 10^{-4}\GeV^{-2}$ and $10^{-4}\mu_p$, considerably lower than that required by our simulations. It is therefore difficult to reconcile our results with the direct detection limits.
	

	The implication is that, even though the model is excluded, it is correctly replicating a potential set of physics that could explain the discrepancies in the Standard Solar Model.
	Other problems also arise; the strong agreement of standard models with observed neutrino fluxes is weakened, and the surface helium abundance is not corrected. 
	Also, generous assumptions have also been made on the number of dark matter particles within the Sun. Both the annihilation and evaporation of particles in and from the Sun have been neglected. Both processes reduce the number of particles within the Sun, and thence any effect on the solar physics. In particular, evaporation is expected to significantly deplete the solar dark matter population for dark matter masses less than $\sim 4\GeV$, precisely where the best fit results lie. Nevertheless, the results here show that it is possible to find a coherent solution; whether through a lighter particle, alternate interaction or equivalent astrophysical process.
		
	
	One past investigation has analysed the effect of magnetic dipole dark matter on the Sun. Ref.~\citep{lopes14} used helioseismology to exclude models with a magnetic dipole moment greater than $1.5\times10^{-3}\mu_p$. However, the energy transport formalism used, based on the work of ref.~\citep{spergel85}, has been shown to be incorrect~\citep{vincent15b}. Ref.~\citep{lopes14} integrates the differential cross section in eq.~(\ref{eq:mdmcs}) and proceeds assuming a constant cross section proportional to the dipole moment. We have implemented the momentum and velocity-dependent components of the cross section using the thermal diffusivity $\alpha$ and the thermal conductivity $\kappa$; the calculation of each depends on the functional form of the differential cross section, leading to significantly different energy transport than assumed in ref.~\citep{lopes14}. 
	In addition, anapole dark matter has not previously been confronted with constraints from solar physics, although the electric dipole moment is equivalent to the $q^{-2}$-dependent generalised example considered by refs.~\citep{vincent15b,vincent16}.
	
	Because the introduction of dark matter in general improved the overall fit to solar observables compared to the SSM, we cannot place new exclusion bounds on the parameter space. Also, a large area of the parameter space, especially where the rate of energy transfer was not maximised, provided no overall improvement to the observables. In these cases, any change was too minor to be observed. In some rare cases, the overall fit was weakened, but such regions are too small to make general statements about new exclusion bounds.

	There are a number of ways that the result can be extended to improve precision, accuracy and viability. Removing the assumptions relating to evaporation is an obvious first step. A theoretical framework does exist for evaporation~\citep{gould87a,gould90b}, but has not yet been modified to account for momentum and velocity-dependent dark matter. The annihilation is only a function of the dark matter density profile and can also be included. It would serve to deplete the dark matter population and weaken any observed effect. There have also been suggestions in the literature of ways to modify the capture rate, which may improve the viability of the magnetic dipole and spin-independent models, where the energy transport for small interaction strengths is obstructed by an unsaturated capture rate. For example, self-interaction can lead to considerable enhancement~\citep{frandsen10,catena16}, as dark matter particles could be captured by nuclei as well as dark matter already present in the Sun.
	
	There is a strong dependence of the capture rate on the velocity distribution of the dark matter halo~\citep{gould87b,choi14}. The properties of the halo are one of the largest uncertainties in the measurement. A halo-independent calculation is particularly interesting for magnetic dipole and anapole dark matter, given that refs.~\citep{delnobile13,delnobile14} suggested that both models could explain, albeit with some tension, both positive and negative direct detection experiments under the assumption of a modified halo. 
	
	
\section{Conclusions}	
	
	We have developed the formalism required for incorporating dark matter with an electromagnetic dipole moment into the Sun. Dark matter with electric, magnetic and anapole moments can occur where the spin couples to an electric field, magnetic field and electromagnetic current respectively. To incorporate each model into the Sun requires a description of the capture rate and energy transport. The effect can be parametrised in terms of the shape of the dark matter distribution within the Sun, and the energy carried by the dark matter. Simulations with the \texttt{DarkStec} code showed that it is possible to alleviate problems with the sound-speed profile, small frequency separations and convective zone radius by introducing electromagnetic dipole dark matter, but at the risk of creating conflict with the neutrino fluxes. Magnetic dipole and anapole dark matter both improve the total likelihood fit to the solar observables, but the required dipole moments are excluded by direct detection experiments~\citep{masso09,delnobile14,barger11,ho13,fitzpatrick10,gresham14,delnobile12,pospelov00,sigurdson04,banks10,heo11,cho10,fortin12}. Overall, it seems unlikely that dipole moment dark matter 
	can solve the discrepancies between solar models and helioseismology alone, at least in the regime where evaporation can be neglected. 
	
\section{Acknowledgements}

	The work of B.G., S.R. and A.G.W. is supported by the ARC Centre of Excellence for Particle Physics at the Terascale (CoEPP) (CE110001104) and the Centre for the Subatomic Structure of Matter (CSSM). P.S. is supported by STFC (ST/K00414X/1 and ST/N000838/1). A.S. is supported by 2014SGR-1458 (Generalitat de Catalunya), ESP2014-56003-R and ESP2015-66134-R (MINECO). M.W. is supported by the Australian Research Council Future Fellowship FT140100244.
	}{

	}
	

	\IfFileExists{JHEP.bst}{
		\bibliographystyle{JHEP}
	}{
		\bibliographystyle{references/JHEP}
	}

	\IfFileExists{references.bib}{
		\bibliography{references}
	}{
		\bibliography{references/references}

\providecommand{\href}[2]{#2}\begingroup\raggedright\begin{thebibliography}{100}

\bibitem{bernabei08}
R.~Bernabei, P.~Belli, F.~Cappella, R.~Cerulli, C.~J. Dai, A.~d'Angelo et~al.,
  \emph{First results from {DAMA}/{LIBRA} and the combined results with
  {DAMA}/{NaI}},
  \href{http://dx.doi.org/10.1140/epjc/s10052-008-0662-y}{\emph{Eur. Phys. J.
  C.} {\bf 56} (2008) 333}, [\href{http://arxiv.org/abs/arXiv:0804.2741}{{\tt
  arXiv:0804.2741}}].

\bibitem{cogent11}
{CoGeNT Collaboration}, \emph{Results from a search for light-mass dark matter
  with a $p$-type point contact germanium detector},
  \href{http://dx.doi.org/10.1086/508162}{\emph{Phys. Rev. Lett.} {\bf 106}
  (2011) 131301}, [\href{http://arxiv.org/abs/arXiv:1002.4703}{{\tt
  arXiv:1002.4703}}].

\bibitem{angloher16}
G.~Angloher, A.~Bento, C.~Bucci, L.~Canonica, X.~Defay, A.~Erb et~al.,
  \emph{Results on light dark matter particles with a low-threshold
  {CRESST}-{II} dark matter search},
  \href{http://dx.doi.org/10.1140/epjc/s10052-016-3877-3}{\emph{Eur. Phys. J.
  C.} {\bf 76} (2016) 25}, [\href{http://arxiv.org/abs/arXiv:1509.01515}{{\tt
  arXiv:1509.01515}}].

\bibitem{cdms13}
{CDMS Collaboration}, \emph{Silicon detector dark matter results from the final
  exposure of {CDMS} {II}},
  \href{http://dx.doi.org/10.1103/PhysRevLett.111.251301}{\emph{Phys. Rev.
  Lett.} {\bf 111} (2013) 251301},
  [\href{http://arxiv.org/abs/arXiv:1304.4279}{{\tt arXiv:1304.4279}}].

\bibitem{xenon11a}
{XENON10 Collaboration}, \emph{Search for light dark matter in {XENON10} data},
  \href{http://dx.doi.org/10.1103/PhysRevLett.107.051301}{\emph{Phys. Rev.
  Lett.} {\bf 107} (2011) 051301},
  [\href{http://arxiv.org/abs/arXiv:1207.5988}{{\tt arXiv:1207.5988}}].

\bibitem{xenon12}
{XENON100 Collaboration}, \emph{Dark matter results from 255 live days of
  {XENON100} data},
  \href{http://dx.doi.org/10.1103/PhysRevLett.109.181301}{\emph{Phys. Rev.
  Lett.} {\bf 109} (2012) 181301},
  [\href{http://arxiv.org/abs/arXiv:1207.5988}{{\tt arXiv:1207.5988}}].

\bibitem{coupp12}
{COUPP Collaboration}, \emph{First dark matter search results from a 4-kg
  {C$\mathrm{F}_3$I} bubble chamber operated in a deep underground site},
  \href{http://dx.doi.org/10.1103/PhysRevD.86.052001}{\emph{Phys. Rev. D} {\bf
  86} (2012) 052001}, [\href{http://arxiv.org/abs/arXiv:1204.3094}{{\tt
  arXiv:1204.3094}}].

\bibitem{simple12}
{SIMPLE Collaboration}, \emph{Final analysis and results of the phase {II}
  {SIMPLE} dark matter search},
  \href{http://dx.doi.org/10.1103/PhysRevLett.108.201302}{\emph{Phys. Rev.
  Lett.} {\bf 108} (2012) 201302},
  [\href{http://arxiv.org/abs/arXiv:1106.3014}{{\tt arXiv:1106.3014}}].

\bibitem{lux16}
{LUX Collaboration}, \emph{Results from a search for dark matter in {LUX} with
  332 live days of exposure},
  \href{http://arxiv.org/abs/arXiv:1608.07648}{{\tt arXiv:1608.07648}}.

\bibitem{supercdms16}
{SuperCDMS Collaboration}, \emph{New results from the search for low-mass
  weakly interacting massive particles with the {CDMS} low ionization threshold
  experiment},
  \href{http://dx.doi.org/10.1103/PhysRevLett.116.071301}{\emph{Phys. Rev.
  Lett.} {\bf 116} (2016) 071301},
  [\href{http://arxiv.org/abs/arXiv:1509.02448}{{\tt arXiv:1509.02448}}].

\bibitem{pandaxii16}
{PandaX-II Collaboration}, \emph{Dark matter results from the first 98.7-day
  data of {P}anda{X}-{II} experiment},
  \href{http://arxiv.org/abs/arXiv:1607.07400v3}{{\tt arXiv:1607.07400v3}}.

\bibitem{turckchieze12a}
S.~Turck-Chi\`{e}ze and I.~Lopes, \emph{Solar-stellar astrophysics and dark
  matter}, \href{http://dx.doi.org/10.1088/1674-4527/12/8/011}{\emph{Res.
  Astron. Astrophys.} {\bf 12} (2012) 1107}.

\bibitem{steigman78}
G.~Steigman, C.~L. Sarazin, H.~Quintana and J.~Faulkner, \emph{Dynamical
  interactions and astrophysical effects of stable heavy neutrinos},
  \href{http://dx.doi.org/10.1086/112290}{\emph{Astron. J.} {\bf 83} (1978)
  1050}.

\bibitem{faulkner85}
J.~Faulkner and R.~L. Gilliland, \emph{Weakly interacting, massive particles
  and the solar neutrino flux},
  \href{http://dx.doi.org/10.1086/163766}{\emph{Astrophys. J.} {\bf 299} (1985)
  994}.

\bibitem{spergel85}
D.~N. Spergel and W.~H. Press, \emph{Effect of hypothetical, weakly
  interacting, massive particles on energy transport in the solar interior},
  \href{http://dx.doi.org/10.1086/163336}{\emph{Astrophys. J.} {\bf 294} (1985)
  663}.

\bibitem{krauss85}
L.~M. Krauss, K.~Freese, D.~N. Spergel and W.~H. Press, \emph{Cold dark matter
  candidates and the solar neutrino problem},
  \href{http://dx.doi.org/10.1086/163767}{\emph{Astrophys. J.} {\bf 299} (1985)
  1001}.

\bibitem{krauss86}
L.~M. Krauss, M.~Srednicki and F.~Wilczek, \emph{{S}olar {S}ystem constraints
  and signatures for dark-matter candidates},
  \href{http://dx.doi.org/10.1103/PhysRevD.33.2079}{\emph{Phys. Rev. D} {\bf
  33} (1986) 2079}.

\bibitem{griest87}
K.~Griest and D.~Seckel, \emph{Cosmic asymmetry, neutrinos and the sun},
  \href{http://dx.doi.org/10.1016/0550-3213(87)90293-8}{\emph{Nucl. Phys. B}
  {\bf 283} (1987) 681}.

\bibitem{raby87}
S.~Raby and G.~B. West, \emph{A simple solution to the solar neutrino and
  missing mass problems},
  \href{http://dx.doi.org/10.1016/0550-3213(87)90671-7}{\emph{Nucl. Phys. B}
  {\bf 292} (1987) 793}.

\bibitem{raby88}
S.~Raby and G.~B. West, \emph{A fourth generation neutrino with a standard
  {H}iggs scalar solves both the solar neutrino and dark matter problems},
  \href{http://dx.doi.org/0.1016/0370-2693(88)90851-9}{\emph{Phys. Lett. B}
  {\bf 202} (1988) 47}.

\bibitem{giudice89}
G.~F. Giudice and E.~Roulet, \emph{A supersymmetric solution to the solar
  neutrino and dark matter problems},
  \href{http://dx.doi.org/10.1016/0370-2693(89)90396-1}{\emph{Phys. Lett. B}
  {\bf 219} (1989) 309}.

\bibitem{roulet89}
E.~Roulet and G.~Gelmini, \emph{Cosmions, cosmic asymmetry and underground
  detection}, \href{http://dx.doi.org/10.1016/0550-3213(89)90506-3}{\emph{Nucl.
  Phys. B} {\bf 325} (1989) 733}.

\bibitem{davis68}
R.~Davis, Jnr, D.~S. Harmer and K.~C. Hoffman, \emph{Search for neutrinos from
  the sun}, \href{http://dx.doi.org/10.1103/PhysRevLett.20.1205}{\emph{Phys.
  Rev. Lett.} {\bf 20} (1968) 1205}.

\bibitem{bahcall68}
J.~N. Bahcall and G.~Shaviv, \emph{Solar models and neutrino fluxes},
  \href{http://dx.doi.org/10.1086/149641}{\emph{Astrophys. J.} {\bf 153} (1968)
  113}.

\bibitem{gould87a}
A.~Gould, \emph{Weakly interacting massive particle distribution in and
  evaporation from the sun},
  \href{http://dx.doi.org/10.1086/165652}{\emph{Astrophys. J.} {\bf 321} (1987)
  560}.

\bibitem{gould87b}
A.~Gould, \emph{Resonant enhancements in weakly interacting massive particle
  capture by the {E}arth},
  \href{http://dx.doi.org/10.1086/165653}{\emph{Astrophys. J.} {\bf 321} (1987)
  571}.

\bibitem{gould90a}
A.~Gould and G.~Raffelt, \emph{Thermal conduction by massive particles},
  \href{http://dx.doi.org/10.1086/168568}{\emph{Astrophys. J.} {\bf 352} (1990)
  654}.

\bibitem{gould90b}
A.~Gould, \emph{Evaporation of {WIMP}s with arbitary cross sections},
  \href{http://dx.doi.org/10.1086/168840}{\emph{Astrophys. J.} {\bf 356} (1990)
  302}.

\bibitem{gould90c}
A.~Gould and G.~Raffelt, \emph{Cosmion energy transfer in stars: the {K}nusden
  limit}, \href{http://dx.doi.org/10.1086/168569}{\emph{Astrophys. J.} {\bf
  352} (1990) 699}.

\bibitem{littleton72}
J.~E. Littleton, H.~M. {van Horn} and H.~L. Helfer, \emph{Process of energy
  transport by longitudinal waves and the problem of solar neutrinos},
  \href{http://dx.doi.org/10.1086/151455}{\emph{Astrophys. J.} {\bf 173} (1972)
  677}.

\bibitem{superkamiokande98}
{Super-Kamiokande Collaboration}, \emph{Evidence for oscillation in atmospheric
  neutrinos}, \href{http://dx.doi.org/10.1103/PhysRevLett.81.1562}{\emph{Phys.
  Rev. Lett.} {\bf 81} (1998) 1562},
  [\href{http://arxiv.org/abs/arXiv:hep-ex/9807003}{{\tt
  arXiv:hep-ex/9807003}}].

\bibitem{sno01}
{SNO Collaboration}, \emph{Measurement of the rate of $\nu_e+d\rightarrow
  p+p+e^-$ interactions produced by $^8${B} solar neutrinos at the {S}udbury
  {N}eutrino {O}bservaory},
  \href{http://dx.doi.org/10.1103/PhysRevLett.87.071301}{\emph{Phys. Rev.
  Lett.} {\bf 87} (2001) 071301},
  [\href{http://arxiv.org/abs/arXiv:nucl-ex/0106015}{{\tt
  arXiv:nucl-ex/0106015}}].

\bibitem{gough96b}
D.~O. Gough, A.~G. Kosovichev, J.~Toomre, E.~Anderson, H.~M. Antia, S.~Basu
  et~al., \emph{The seismic structure of the sun},
  \href{http://dx.doi.org/10.1126/science.272.5266.1296}{\emph{Science} {\bf
  272} (1996) 1296}.

\bibitem{grevesse98}
N.~Grevesse and A.~J. Sauval, \emph{Standard solar composition},
  \href{http://dx.doi.org/10.1023/A:1005161325181}{\emph{Space Sci. Rev.} {\bf
  85} (1998) 161}.

\bibitem{bertello00a}
L.~Bertello, C.~J. Henney, R.~K. Ulrich, F.~Varadi, A.~G. Kosovichev, P.~H.
  Scherrer et~al., \emph{Comparison of frequencies and rotational splittings of
  solar acoustic modes of low angular degree from simultaneous {MDI} and {GOLF}
  observations}, \href{http://dx.doi.org/10.1086/308864}{\emph{Astrophys. J.}
  {\bf 535} (2000) 1066}.

\bibitem{bertello00b}
L.~Bertello, F.~Varadi and R.~K. Ulrich, \emph{Identification of solar acoustic
  modes of low angular degree and low radial order},
  \href{http://dx.doi.org/10.1086/312775}{\emph{Astrophys. J.} {\bf 537} (2000)
  L143}.

\bibitem{garcia01}
R.~A. Garc\'{\i}a, C.~R\'{e}gulo, S.~Turck-Chi\`{e}ze, L.~Bertello, A.~G.
  Kosovichev, A.~S. Brun et~al., \emph{Low-degree low-order solar p-modes as
  seen by {GOLF} on board {SOHO}},
  \href{http://dx.doi.org/10.1023/A:1010344721148}{\emph{Sol. Phys.} {\bf 200}
  (2001) 361}.

\bibitem{allendeprieto01}
C.~Allende~Prieto, D.~L. Lambert and M.~Asplund, \emph{The `forbidden'
  abundance of oxygen in the sun},
  \href{http://dx.doi.org/10.1086/322874}{\emph{Astrophys. J.} {\bf 556} (2001)
  L63}, [\href{http://arxiv.org/abs/arXiv:astro-ph/0106360}{{\tt
  arXiv:astro-ph/0106360}}].

\bibitem{allendeprieto02}
C.~Allende~Prieto, D.~L. Lambert and M.~Asplund, \emph{A reappraisal of the
  solar photospheric {C}/{O} ratio},
  \href{http://dx.doi.org/10.1086/342095}{\emph{Astrophys. J.} {\bf 573} (2002)
  L137}, [\href{http://arxiv.org/abs/arXiv:astro-ph/0206089}{{\tt
  arXiv:astro-ph/0206089}}].

\bibitem{asplund04}
M.~Asplund, N.~Grevesse, A.~J. Sauval, C.~Allende~Prieto and D.~Kiselman,
  \emph{Line formulation in solar granulation. {IV}. {OI} and {OH} lines and
  the photospheric {O} abundance},
  \href{http://dx.doi.org/10.1051/0004-6361:20034328}{\emph{Astron. Astrophys.}
  {\bf 417} (2004) 751},
  [\href{http://arxiv.org/abs/arXiv:astro-ph/0312290}{{\tt
  arXiv:astro-ph/0312290}}].

\bibitem{asplund05b}
M.~Asplund, N.~Grevesse, A.~J. Sauval, C.~Allende~Prieto and R.~Blomme,
  \emph{Line formation in solar granulation. {IV}. [{C} {I}], {C} {I}, {CH} and
  $\mathrm{C}_2$ lines in the photospheric {C} abundance},
  \href{http://dx.doi.org/10.1051/0004-6361:20041951}{\emph{Astron. Astrophys.}
  {\bf 431} (2005) 693},
  [\href{http://arxiv.org/abs/arXiv:astro-ph/0410681}{{\tt
  arXiv:astro-ph/0410681}}].

\bibitem{scott06}
P.~C. Scott, M.~Asplund, N.~Grevesse and A.~J. Sauval, \emph{Line formulation
  in solar granulation. {VII}. {CO} lines and the solar {C} and {O} isotopic
  abundances},
  \href{http://dx.doi.org/10.1051/0004-6361:20064986}{\emph{Astron. Astrophys.}
  {\bf 456} (2006) 675},
  [\href{http://arxiv.org/abs/arXiv:astro-ph/0605116}{{\tt
  arXiv:astro-ph/0605116}}].

\bibitem{melendez08}
J.~Mel\'{e}ndez and M.~Asplund, \emph{Another forbidden solar oxygen abundance:
  the [{O I}] 5577 a line},
  \href{http://dx.doi.org/10.1051/0004-6361:200810347}{\emph{Astron.
  Astrophys.} {\bf 490} (2008) 817},
  [\href{http://arxiv.org/abs/arXiv:0808.2796}{{\tt arXiv:0808.2796}}].

\bibitem{scott09c}
P.~Scott, M.~Asplund, N.~Grevesse and A.~J. Sauval, \emph{On the solar nickel
  and oxygen abundances},
  \href{http://dx.doi.org/10.1088/0004-637X/691/2/L119}{\emph{Astrophys. J.}
  {\bf 691} (2009) L119}, [\href{http://arxiv.org/abs/arXiv:0811.0815}{{\tt
  arXiv:0811.0815}}].

\bibitem{asplund09}
M.~Asplund, N.~Grevesse, A.~Jacques~Sauval and P.~Scott, \emph{The chemical
  composition of the sun},
  \href{http://dx.doi.org/10.1146/annurev.astro.46.060407.145222}{\emph{Annu.
  Rev. Astron. Astrophys.} {\bf 47} (2009) 481},
  [\href{http://arxiv.org/abs/arXiv:0909.0948}{{\tt arXiv:0909.0948}}].

\bibitem{scott15a}
P.~Scott, N.~Grevesse, M.~Asplund, A.~J. Sauval, K.~Lind, Y.~Takeda et~al.,
  \emph{The elemental composition of the sun. {I}. the intermediate mass
  elements {Na} to {Ca}},
  \href{http://dx.doi.org/10.1051/0004-6361/201424109}{\emph{Astron.
  Astrophys.} {\bf 573} (2015) A25},
  [\href{http://arxiv.org/abs/arXiv:1405.0279}{{\tt arXiv:1405.0279}}].

\bibitem{scott15b}
P.~Scott, M.~Asplund, N.~Grevesse, M.~Bergemann and A.~J. Sauval, \emph{The
  elemental composition of the sun. {II}. the iron group elements {Sc} to
  {Ni}}, \href{http://dx.doi.org/10.1051/0004-6361/201424110}{\emph{Astron.
  Astrophys.} {\bf 573} (2015) A26},
  [\href{http://arxiv.org/abs/arXiv:1405.0287}{{\tt arXiv:1405.0287}}].

\bibitem{grevesse15}
N.~Grevesse, P.~Scott, M.~Asplund and A.~J. Sauval, \emph{The elemental
  composition of the sun. {III}. the heavy elements {Cu} to {Th}},
  \href{http://dx.doi.org/0.1051/0004-6361/201424111}{\emph{Astron. Astrophys.}
  {\bf 573} (2015) A27}, [\href{http://arxiv.org/abs/arXiv:1405.0288}{{\tt
  arXiv:1405.0288}}].

\bibitem{basu04}
S.~Basu and H.~M. Antia, \emph{Constraining solar abundances using
  helioseismology}, \href{http://dx.doi.org/10.1086/421110}{\emph{Astrophys.
  J.} {\bf 606} (2004) L85},
  [\href{http://arxiv.org/abs/arXiv:astro-ph/0403485}{{\tt
  arXiv:astro-ph/0403485}}].

\bibitem{bahcall05b}
J.~N. Bahcall, S.~Basu, M.~Pinsonneualt and A.~M. Serenelli,
  \emph{Helioseismological implications of recent solar abundance
  determinations}, \href{http://dx.doi.org/10.1086/426070}{\emph{Astrophys. J.}
  {\bf 618} (2005) 1049},
  [\href{http://arxiv.org/abs/arXiv:astro-ph/0407060}{{\tt
  arXiv:astro-ph/0407060}}].

\bibitem{bahcall06}
J.~N. Bahcall, A.~M. Serenelli and S.~Basu, \emph{10,000 standard solar models:
  a {M}onte {C}arlo simulation},
  \href{http://dx.doi.org/10.1086/504043}{\emph{Astrophys. J. Supp. Ser.} {\bf
  165} (2006) 400}, [\href{http://arxiv.org/abs/arXiv:astro-ph/0511337}{{\tt
  arXiv:astro-ph/0511337}}].

\bibitem{yang07}
W.~M. Yang and S.~L. Bi, \emph{Solar models with revised abundances and
  opacities}, \href{http://dx.doi.org/10.1086/513694}{\emph{Astrophys. J.} {\bf
  658} (2007) L67}, [\href{http://arxiv.org/abs/arXiv:0805.3644}{{\tt
  arXiv:0805.3644}}].

\bibitem{basu08}
S.~Basu and H.~M. Antia, \emph{Helioseismology and solar abundances},
  \href{http://dx.doi.org/10.1016/j.physrep.2007.12.002}{\emph{Phys. Rep.} {\bf
  457} (2008) 217}, [\href{http://arxiv.org/abs/arXiv:0711.4590}{{\tt
  arXiv:0711.4590}}].

\bibitem{asplund05a}
M.~Asplund, N.~Grevesse and A.~J. Sauval, \emph{The solar chemical
  composition},  in \emph{Cosmic Abundances as Records of Stellar Evolution}
  (I.~Barnes, Thomas~G and F.~N. Bash, eds.), vol.~336, p.~25, 2005.

\bibitem{serenelli09}
A.~M. Serenelli, S.~Basu, J.~W. Ferguson and M.~Asplund, \emph{New solar
  composition: the problem with solar models revisited},
  \href{http://dx.doi.org/10.1088/0004-637X/705/2/L123}{\emph{Astrophys. J.}
  {\bf 705} (2009) L123}, [\href{http://arxiv.org/abs/arXiv:0909.2668}{{\tt
  arXiv:0909.2668}}].

\bibitem{serenelli11}
A.~M. Serenelli, W.~C. Haxton and C.~{Pe\~{n}a-Garay}, \emph{Solar models with
  accretion. {I}. application to the solar abundance problem},
  \href{http://dx.doi.org/10.1088/0004-637X/743/1/24}{\emph{Astrophys. J.} {\bf
  743} (2011) 24}, [\href{http://arxiv.org/abs/arXiv:1104.1639}{{\tt
  arXiv:1104.1639}}].

\bibitem{serenelli16}
A.~Serenelli, P.~Scott, F.~L. Villante, A.~C. Vincent, M.~Asplund, S.~Basu
  et~al., \emph{Implications of solar wind measurements for solar models and
  composition}, \href{http://dx.doi.org/10.1093/mnras/stw1927}{\emph{Mon. Not.
  R. Astron. Soc.} {\bf 463} (2016) 2},
  [\href{http://arxiv.org/abs/arXiv:1604.05318}{{\tt arXiv:1604.05318}}].

\bibitem{serenelli10}
A.~M. Serenelli and S.~Basu, \emph{Determining the initial helium abundance of
  the sun},
  \href{http://dx.doi.org/10.1088/0004-637X/719/1/865}{\emph{Astrophys. J.}
  {\bf 719} (2010) 865}, [\href{http://arxiv.org/abs/arXiv:1006.0244}{{\tt
  arXiv:1006.0244}}].

\bibitem{bottino02}
A.~Bottino, G.~Fiorentini, N.~Fornengo, B.~Ricci, S.~Copel and F.~L. Villante,
  \emph{Does solar physics provide constraints to weakly interacting massive
  particles}, \href{http://dx.doi.org/10.1103/PhysRevD.66.053005}{\emph{Phys.
  Rev. D} {\bf 66} (2002) 053005},
  [\href{http://arxiv.org/abs/arXiv:hep-ph/0206211}{{\tt
  arXiv:hep-ph/0206211}}].

\bibitem{lopes02b}
I.~P. Lopes, G.~Bertone and J.~Silk, \emph{Solar seismic model as a new
  constraint on supersymmetric dark matter},
  \href{http://dx.doi.org/10.1046/j.1365-8711.2002.05835.x}{\emph{Mon. Not. R.
  Astron. Soc.} {\bf 337} (2002) 1179},
  [\href{http://arxiv.org/abs/arXiv:astro-ph/0205066}{{\tt
  arXiv:astro-ph/0205066}}].

\bibitem{lopes02a}
I.~P. Lopes, J.~Silk and S.~H. Hansen, \emph{Helioseismology as a new
  constraint on supersymmetric dark matter},
  \href{http://dx.doi.org/10.1046/j.1365-8711.2002.05238.x}{\emph{Mon. Not. R.
  Astron. Soc.} {\bf 331} (2002) 361},
  [\href{http://arxiv.org/abs/arXiv:astro-ph/0111530}{{\tt
  arXiv:astro-ph/0111530}}].

\bibitem{cumberbatch10}
D.~T. Cumberbatch, J.~A. Guzik, J.~Silk, L.~S. Watson and S.~M. West,
  \emph{Light {WIMP}s in the sun: {C}onstraints from helioseismology},
  \href{http://dx.doi.org/10.1103/PhysRevD.82.103503}{\emph{Phys. Rev. D} {\bf
  82} (2010) 103503}, [\href{http://arxiv.org/abs/arXiv:1005.5102}{{\tt
  arXiv:1005.5102}}].

\bibitem{taoso10}
M.~Taoso, F.~Iocco, G.~Meynet, G.~Bertone and P.~Eggenberger, \emph{Effect of
  low mass dark matter particles on the sun},
  \href{http://dx.doi.org/10.1103/PhysRevD.82.083509}{\emph{Phys. Rev. D} {\bf
  82} (2010) 083509}, [\href{http://arxiv.org/abs/arXiv:1005.5711}{{\tt
  arXiv:1005.5711}}].

\bibitem{vincent14}
A.~C. Vincent and P.~Scott, \emph{Thermal conduction by dark matter with
  velocity and momentum-dependent cross sections},
  \href{http://dx.doi.org/10.1088/1475-7516/2014/04/019}{\emph{J. Cosmol.
  Astropart. Phys.} {\bf 2014} (2014) 019},
  [\href{http://arxiv.org/abs/arXiv:1311.2074}{{\tt arXiv:1311.2074}}].

\bibitem{vincent15a}
A.~C. Vincent, P.~Scott and A.~Serenelli, \emph{Possible indication of
  momentum-dependent asymmetric dark matter in the sun},
  \href{http://dx.doi.org/10.1103/PhysRevLett.114.081302}{\emph{Phys. Rev.
  Lett.} {\bf 114} (2015) 081302},
  [\href{http://arxiv.org/abs/arXiv:1411.6626}{{\tt arXiv:1411.6626}}].

\bibitem{vincent15b}
A.~C. Vincent, A.~Serenelli and P.~Scott, \emph{Generalised form factor dark
  matter in the sun},
  \href{http://dx.doi.org/10.1088/1475-7516/2015/08/040}{\emph{J. Cosmol.
  Astropart. Phys.} {\bf 2015} (2015) 040},
  [\href{http://arxiv.org/abs/arXiv:1504.04378}{{\tt arXiv:1504.04378}}].

\bibitem{vincent16}
A.~C. Vincent, P.~Scott and A.~Serenelli, \emph{Updated constraints on velocity
  and momentum-dependent asymetric dark matter},
  \href{http://dx.doi.org/10.1088/1475-7516/2016/11/007}{\emph{J. Cosmology.
  Astropart. Phys.} {\bf 2016} (2016) 007},
  [\href{http://arxiv.org/abs/arXiv:1605.06502}{{\tt arXiv:1605.06502}}].

\bibitem{pospelov00}
M.~Pospelov and T.~{ter Veldhuis}, \emph{Direct and indirect limits on the
  electromagnetic form factors of {WIMP}s},
  \href{http://dx.doi.org/10.1093/mnras/274.3.964}{\emph{Phys. Lett. B} {\bf
  480} (2000) 181}, [\href{http://arxiv.org/abs/arXiv:hep-ph/0003010}{{\tt
  arXiv:hep-ph/0003010}}].

\bibitem{sigurdson04}
K.~Sigurdson, M.~Doran, A.~Kurylov, R.~R. Caldwell and M.~Kamionkowsi,
  \emph{Dark-matter electric and magnetic dipole moments},
  \href{http://dx.doi.org/10.1103/PhysRevD.70.083501}{\emph{Phys. Rev. D} {\bf
  70} (2004) 083501}, [\href{http://arxiv.org/abs/arXiv:astro-ph/0406355}{{\tt
  arXiv:astro-ph/0406355}}].

\bibitem{masso09}
E.~Mass\'{o}, S.~Mohanty and S.~Rao, \emph{Dipolar dark matter}, {\emph{Phys.
  Rev. D} {\bf 80} (2009) 036009},
  [\href{http://arxiv.org/abs/arXiv:0906.1979}{{\tt arXiv:0906.1979}}].

\bibitem{banks10}
T.~Banks, J.-F. Fortin and S.~Thomas, \emph{Direct detection of dark matter
  electromagnetic dipole moments},
  \href{http://arxiv.org/abs/arXiv:1007.5515}{{\tt arXiv:1007.5515}}.

\bibitem{cho10}
W.~S. Cho, J.-H. Huh, I.-W. Kim, J.~E. Kim and B.~Kyae, \emph{Constraining
  {WIMP} magnetic moment from {CDMS} {II} experiment},
  \href{http://dx.doi.org/10.1016/j.physletb.2010.02.081}{\emph{Phys. Lett. B}
  {\bf 687} (2010) 6}, [\href{http://arxiv.org/abs/arXiv:1001.0579}{{\tt
  arXiv:1001.0579}}].

\bibitem{fitzpatrick10}
A.~L. Fitzpatrick and K.~M. Zurek, \emph{Dark moments and the {DAMA}-{CoGeNT}
  puzzle}, \href{http://dx.doi.org/10.1103/PhysRevD.82.075004}{\emph{Phys. Rev.
  D} {\bf 82} (2010) 075004}, [\href{http://arxiv.org/abs/arXiv:1007.5325}{{\tt
  arXiv:1007.5325}}].

\bibitem{barger11}
V.~Barger, W.-Y. Keung and D.~Marfatia, \emph{Electromagnetic properties of
  dark matter: dipole moments and charge form factor},
  \href{http://dx.doi.org/10.1016/j.physletb.2010.12.008}{\emph{Phys. Lett. B}
  {\bf 696} (2011) 74}, [\href{http://arxiv.org/abs/arXiv:1007.4345}{{\tt
  arXiv:1007.4345}}].

\bibitem{heo11}
J.~H. Heo, \emph{Electric dipole moment of {D}irac fermionic dark matter},
  \href{http://dx.doi.org/10.1016/j.physletb.2011.06.088}{\emph{Phys. Lett. B}
  {\bf 702} (2011) 205}, [\href{http://arxiv.org/abs/arXiv:0902.2643}{{\tt
  arXiv:0902.2643}}].

\bibitem{barger12}
V.~Barger, W.-Y. Keung, D.~Marfatia and P.-Y. Tseng, \emph{Dipole moment dark
  matter at the {LHC}},
  \href{http://dx.doi.org/10.1016/j.physletb.2012.09.036}{\emph{Phys. Lett. B}
  {\bf 717} (2012) 219}, [\href{http://arxiv.org/abs/arXiv:1206.0640}{{\tt
  arXiv:1206.0640}}].

\bibitem{delnobile12}
E.~{Del Nobile}, C.~Kouvaris, P.~Panci, F.~Sannino and J.~Virkaj\"{a}rvi,
  \emph{Light magnetic dark matter in direct detection searches},
  \href{http://dx.doi.org/10.1088/1475-7516/2012/08/010}{\emph{J. Cosmol.
  Astropart. Phys.} {\bf 2012} (2012) 010},
  [\href{http://arxiv.org/abs/arXiv:1203.6652}{{\tt arXiv:1203.6652}}].

\bibitem{fortin12}
J.-F. Fortin and T.~M.~P. Tait, \emph{Collider constraints on
  dipole-interacting dark matter},
  \href{http://dx.doi.org/10.1103/PhysRevD.85.063506}{\emph{Phys. Rev. D} {\bf
  85} (2012) 063506}, [\href{http://arxiv.org/abs/arXiv:1103.3289}{{\tt
  arXiv:1103.3289}}].

\bibitem{ho13}
C.~M. Ho and R.~J. Scherrer, \emph{Anapole dark matter},
  \href{http://dx.doi.org/10.1016/j.physletb.2013.04.039}{\emph{Phys. Lett. B}
  {\bf 722} (2013) 341}, [\href{http://arxiv.org/abs/arXiv:1211.0503}{{\tt
  arXiv:1211.0503}}].

\bibitem{delnobile14}
E.~{Del Nobile}, G.~B. Gelmini, P.~Gondolo and J.-H. Huh, \emph{Direct
  detection of light anapole and magnetic dipole {DM}},
  \href{http://dx.doi.org/10.1088/1475-7516/2014/06/002}{\emph{J. Cosmol.
  Astropart. Phys.} {\bf 2014} (2014) 002},
  [\href{http://arxiv.org/abs/arXiv:1401.4508}{{\tt arXiv:1401.4508}}].

\bibitem{cabralrosetti14}
L.~G. Cabral-Rosetti, M.~Mondrag\'{o}n and E.~Reyes~P\'{e}rez, \emph{Toroidal
  dipole moment of the {LSP} in the {cMSSM}},  in \emph{{PASCOS} 2012 -- 18th
  {I}nternational {S}ymposium on {P}article {S}trings and {C}osmology, Journal
  of Physics: Conference Series}, vol.~485, p.~012019, 2014.
\newblock \href{http://arxiv.org/abs/arXiv:1206.5052}{{\tt arXiv:1206.5052}}.
\newblock \href{http://dx.doi.org/10.1086/307932}{DOI}.

\bibitem{gao14}
Y.~Gao, C.~M. Ho and R.~J. Scherrer, \emph{Anapole dark matter at the {LHC}},
  \href{http://dx.doi.org/10.1103/PhysRevD.89.045006}{\emph{Phys. Rev. D} {\bf
  89} (2014) 045006}, [\href{http://arxiv.org/abs/arXiv:1311.5630}{{\tt
  arXiv:1311.5630}}].

\bibitem{gresham14}
M.~I. Gresham and K.~M. Zurek, \emph{Light dark matter anomalies after {LUX}},
  \href{http://dx.doi.org/10.1103/PhysRevD.89.016017}{\emph{Phys. Rev. D} {\bf
  89} (2014) 016017}, [\href{http://arxiv.org/abs/arXiv:1311.2082}{{\tt
  arXiv:1311.2082}}].

\bibitem{cabralrosetti16}
L.~G. Cabral-Rosetti, M.~Mondrag\'{o}n and E.~Reyes~P\'{e}rez, \emph{Anapole
  moment of the lightest neutralino in the c{MSSM}},
  \href{http://dx.doi.org/10.1016/j.nuclphysb.2016.03.025}{\emph{Nuclear
  Physics B} {\bf 907} (2016) 1},
  [\href{http://arxiv.org/abs/arXiv:1504.01213}{{\tt arXiv:1504.01213}}].

\bibitem{mohanty15}
S.~Mohanty and S.~Rao, \emph{Detecting dipolar dark matter in beam dump
  experiments},  \href{http://arxiv.org/abs/arXiv:1506.06462}{{\tt
  arXiv:1506.06462}}.

\bibitem{lopes14}
I.~Lopes, K.~Kadota and J.~Silk, \emph{Constraint on light dipole dark matter
  from helioseismology},
  \href{http://dx.doi.org/10.1088/2041-8205/780/2/L15}{\emph{Astrophys. J.
  Letters} {\bf 780} (2014) 15},
  [\href{http://arxiv.org/abs/arXiv:1310.0673}{{\tt arXiv:1310.0673}}].

\bibitem{helm56}
R.~H. Helm, \emph{Inelastic and elastic scattering of 187-{M}e{V} elecrons from
  selected even-even nuclei},
  \href{http://dx.doi.org/10.1103/PhysRev.104.1466}{\emph{Phys. Rev.} {\bf 104}
  (1956) 1466}.

\bibitem{catena15}
R.~Catena and B.~Schwabe, \emph{Form factors for dark matter capture by the sun
  in effective theories},
  \href{http://dx.doi.org/10.1088/1475-7516/2015/04/042}{\emph{J. Cosmol.
  Astropart. Phys.} {\bf 2015} (2015) 042},
  [\href{http://arxiv.org/abs/arXiv:1501.03729}{{\tt arXiv:1501.03729}}].

\bibitem{lux14}
{LUX Collaboration}, \emph{First results from the {LUX} dark matter experiment
  at the {S}anford {U}nderground {R}esearch {F}acility},
  \href{http://dx.doi.org/10.1103/PhysRevLett.112.091303}{\emph{Phys. Rev.
  Lett.} {\bf 112} (2014) 091303},
  [\href{http://arxiv.org/abs/arXiv:1310.8214}{{\tt arXiv:1310.8214}}].

\bibitem{kaplan09}
D.~E. Kaplan, M.~A. Luty and K.~M. Zurek, \emph{Asymmetric dark matter},
  \href{http://dx.doi.org/10.1103/PhysRevD.79.115016}{\emph{Phys. Rev. D} {\bf
  79} (2009) 115016}, [\href{http://arxiv.org/abs/arXiv:0901.4117}{{\tt
  arXiv:0901.4117}}].

\bibitem{petraki13}
K.~Petraki and R.~R. Volkas, \emph{Review of asymmetric dark matter},
  \href{http://dx.doi.org/10.1142/S0217751X13300287}{\emph{Int. J. Mod. Phys.
  A} {\bf 28} (2013) 1330028},
  [\href{http://arxiv.org/abs/arXiv:1305.4939}{{\tt arXiv:1305.4939}}].

\bibitem{blennow15}
M.~Blennow and S.~Clementz, \emph{Asymmetric capture of {D}irac dark matter by
  the {S}un}, \href{http://dx.doi.org/10.1088/1475-7516/2015/08/036}{\emph{J.
  Cosmol. Astropart. Phys.} {\bf 2015} (2015) 036},
  [\href{http://arxiv.org/abs/arXiv:1504.05813}{{\tt arXiv:1504.05813}}].

\bibitem{busoni16}
{G. Busoni, A. De Simone, P. Scott and A. Vincent, \textit{in preparation}
  (2017)}.

\bibitem{scott09b}
P.~Scott, M.~Fairbairn and J.~Edsj\"{o}, \emph{Dark stars at the galactic
  centre - the main sequence},
  \href{http://dx.doi.org/10.1111/j.1365-2966.2008.14282.x}{\emph{Mon. Not. R.
  Astron. Soc.} {\bf 394} (2009) 82},
  [\href{http://arxiv.org/abs/arXiv:0809.1871}{{\tt arXiv:0809.1871}}].

\bibitem{krstic99}
P.~S. Krsti\'{c} and D.~R. Schultz, \emph{Consistent definitions for, and
  relationships among, cross sections for elastic scattering of hydrogen ions,
  atoms, and molecules},
  \href{http://dx.doi.org/http://dx.doi.org/10.1103/PhysRevA.60.2118}{\emph{Phys.
  Rev. A} {\bf 60} (1999) 2118}.

\bibitem{tulin13}
S.~Tulin, H.-B. Yu and K.~M. Zurek, \emph{Beyond collisionless dark matter:
  Particle physics dynamics for dark matter halo structure},
  \href{http://dx.doi.org/10.1103/PhysRevD.87.115007}{\emph{Phys. Rev. D} {\bf
  87} (2013) 115007}, [\href{http://arxiv.org/abs/arXiv:1302.3898}{{\tt
  arXiv:1302.3898}}].

\bibitem{stone05}
N.~J. Stone, \emph{Table of nuclear magnetic dipole and electric quadrupole
  moments}, \href{http://dx.doi.org/10.1016/j.adt.2005.04.001}{\emph{At. Data
  Nucl. Data Tables} {\bf 90} (2005) 75}.

\bibitem{kippenhahn67}
R.~Kippenhahn, A.~Weigert and E.~Hofmeister, \emph{Methods for calculating
  stellar evolution}, {\emph{Methods of Computational Physics} {\bf 7} (1967)
  129}.

\bibitem{thomas67}
H.-C. Thomas, \emph{Sternentwicklung {VIII}. {D}er {H}elium-{F}lash bei einem
  {S}tern von 1.3 {S}onnenmassen}, {\emph{{Z}eitschrit f\"{u}r {A}strophysik}
  {\bf 67} (1967) 420}.

\bibitem{weiss89}
A.~Weiss, \emph{The progenitor of {SN} 1987{A}: {U}ncertain evolution of a
  {$20~M_\odot$} star},
  \href{http://dx.doi.org/10.1086/167302}{\emph{Astrophys. J.} {\bf 339} (1989)
  365}.

\bibitem{wagenhuber94}
J.~Wagenhuber and A.~Weiss, \emph{Numerical methods for {AGB} evolution},
  {\emph{Astron. Astrophys.} {\bf 286} (1994) 121}.

\bibitem{weiss00}
A.~Weiss and H.~Schlattl, \emph{Age-luminosity relations for low-mass
  metal-poor stars}, \href{http://dx.doi.org/10.1051/aas:2000223}{\emph{Astron.
  Astrophys. Suppl. Ser.} {\bf 144} (2000) 487}.

\bibitem{weiss08}
A.~Weiss and H.~Schlattl, \emph{{GARSTEC} - the {G}arching {S}tellar
  {E}volution code},
  \href{http://dx.doi.org/10.1007/s10509-007-9606-5}{\emph{Astrophys. Space
  Sci.} {\bf 316} (2008) 99}.

\bibitem{eggleton71}
P.~P. Eggleton, \emph{The evolution of low mass stars},
  \href{http://dx.doi.org/10.1093/mnras/151.3.351}{\emph{Mon. Not. R. Astron.
  Soc.} {\bf 151} (1971) 351}.

\bibitem{eggleton72}
P.~P. Eggleton, \emph{Composition changes during stellar evolution},
  \href{http://dx.doi.org/10.1093/mnras/156.3.361}{\emph{Mon. Not. R. Astron.
  Soc.} {\bf 156} (1972) 361}.

\bibitem{pols95}
O.~R. Pols, C.~A. Tout, P.~P. Eggleton and Z.~Han, \emph{Approximate input
  physics for stellar modelling},
  \href{http://dx.doi.org/10.1093/mnras/274.3.964}{\emph{Mon. Not. R. Astron.
  Soc.} {\bf 274} (1995) 964},
  [\href{http://arxiv.org/abs/arXiv:astro-ph/9504025}{{\tt
  arXiv:astro-ph/9504025}}].

\bibitem{paxton04}
B.~Paxton, \emph{{EZ} to evolve {ZAMS} stars: {A} program derived from
  {E}ggleton's stellar evolution code},
  \href{http://dx.doi.org/10.1086/422345}{\emph{Publ. Astron. Soc. Pac.} {\bf
  116} (2004) 699}, [\href{http://arxiv.org/abs/arXiv:astro-ph/0405130}{{\tt
  arXiv:astro-ph/0405130}}].

\bibitem{gondolo04}
P.~Gondolo, J.~Edsj\"{o}, P.~Ullio, L.~Bergstr\"{o}m, M.~Schelke and E.~A.
  Baltz, \emph{{D}ark{SUSY}: computing supersymmetric dark matter properties
  numerically}, \href{http://dx.doi.org/10.1088/1475-7516/2004/07/008}{\emph{J.
  Cosmol. Astropart. Phys.} {\bf 2004} (2004) 008},
  [\href{http://arxiv.org/abs/arXiv:astro-ph/0406204}{{\tt
  arXiv:astro-ph/0406204}}].

\bibitem{fairbairn08}
M.~Fairbairn, P.~Scott and J.~Edsj\"{o}, \emph{The zero age main sequence of
  {WIMP} burners},
  \href{http://dx.doi.org/10.1103/PhysRevD.77.047301}{\emph{Phys. Rev. D} {\bf
  77} (2008) 047301}, [\href{http://arxiv.org/abs/arXiv:0710.3396}{{\tt
  arXiv:0710.3396}}].

\bibitem{scott08b}
P.~Scott, M.~Fairbairn and J.~Edsj\"{o}, \emph{Impacts of {WIMP} dark matter
  upon stellar evolution: main-sequence stars},  in \emph{Proceedings of the
  Identification of Dark matter Conference}, p.~1, 2008.
\newblock \href{http://arxiv.org/abs/arXiv:0810.5560}{{\tt arXiv:0810.5560}}.

\bibitem{scott08a}
P.~C. Scott, J.~Edsj\'{o} and M.~Fairbairn, \emph{Low mass stellar evolution
  with {WIMP} capture and annihilation},  in \emph{Dark Matter in Astrophysics
  and Particle Physics}, p.~387, 2008.
\newblock \href{http://arxiv.org/abs/arXiv:0711.0991}{{\tt arXiv:0711.0991}}.

\bibitem{scott09a}
P.~Scott, J.~Edjs\"{o} and M.~Fairbairn, \emph{The {D}ark{S}tars code: {A}
  publicly available dark stellar evolution package},  in \emph{Dark Matter in
  Astrophysics and Particle Physics}, p.~320, 2009.
\newblock \href{http://arxiv.org/abs/arXiv:0904.2395}{{\tt arXiv:0904.2395}}.
\newblock \href{http://dx.doi.org/10.1142/9789814293792_0024}{DOI}.

\bibitem{antonelli13}
V.~Antonelli, L.~Miramonti, C.~{Pe\~{n}a-Garay} and A.~Serenelli, \emph{Solar
  neutrinos}, \href{http://dx.doi.org/10.1155/2013/351926}{\emph{Adv. High
  Energy Phys.} {\bf 2013} (2013) 351926},
  [\href{http://arxiv.org/abs/arXiv:1208.1356}{{\tt arXiv:1208.1356}}].

\bibitem{bergstrom16}
J.~Bergst\"{o}m, M.~C. Gonzalez-Garcia, M.~Maltoni, C.~P. {n}a Garay, A.~M.
  Serenelli and N.~Song, \emph{Updated determination of the solar neutrino
  fluxes from solar neutrino data},
  \href{http://dx.doi.org/10.1007/JHEP03(2016)132}{\emph{J. High Energy Phys.}
  {\bf 3} (2016) 132}, [\href{http://arxiv.org/abs/arXiv:1601.00972}{{\tt
  arXiv:1601.00972}}].

\bibitem{castellani93}
V.~Castellani, S.~Degl'Innocenti and G.~Fiorentini, \emph{The $pp$ reaction in
  the sun and solar neutrinos},
  \href{http://dx.doi.org/10.1016/0370-2693(93)90045-J}{\emph{Phys. Lett. B}
  {\bf 303} (1993) 68}, [\href{http://arxiv.org/abs/arXiv:1212.2985}{{\tt
  arXiv:1212.2985}}].

\bibitem{castellani94}
V.~Castellani, S.~Degl'Innocenti, G.~Fiorentini, M.~Lissia and B.~Ricci,
  \emph{Neutrinos from the sun: {E}xperimental results confronted with solar
  models}, \href{http://dx.doi.org/10.1103/PhysRevD.50.4749}{\emph{Phys. Rev.
  D} {\bf 50} (1994) 4749},
  [\href{http://arxiv.org/abs/arXiv:astro-ph/9405064}{{\tt
  arXiv:astro-ph/9405064}}].

\bibitem{bahcall04}
J.~N. Bahcall and M.~H. Pinsonneault, \emph{What do we (not) know theoretically
  about solar neutrino fluxes?},
  \href{http://dx.doi.org/10.1103/PhysRevLett.92.121301}{\emph{Phys. Rev.
  Lett.} {\bf 92} (2004) 121301},
  [\href{http://arxiv.org/abs/arXiv:astro-ph/0402114}{{\tt
  arXiv:astro-ph/0402114}}].

\bibitem{bahcall05a}
J.~N. Bahcall, A.~M. Serenelli and S.~Basu, \emph{New solar opacities,
  abundances, helioseismology, and neutrino fluxes},
  \href{http://dx.doi.org/10.1086/428929}{\emph{Astrophys. J.} {\bf 621} (2005)
  L85}, [\href{http://arxiv.org/abs/arXiv:astro-ph/0412440}{{\tt
  arXiv:astro-ph/0412440}}].

\bibitem{Socas07}
H.~{Socas-Navarro} and A.~A. {Norton}, \emph{The solar oxygen crisis: Probably
  not the last word}, {\emph{Astrophys. J.} {\bf 660} (2007) L153}.

\bibitem{Koesterke08}
L.~{Koesterke}, C.~{Allende Prieto} and D.~L. {Lambert}, \emph{Center-to-limb
  variation of solar three-dimensional hydrodynamical simulations},
  \href{http://dx.doi.org/10.1086/587471}{\emph{Astrophys. J.} {\bf 680} (2008)
  764--773}, [\href{http://arxiv.org/abs/arXiv:0802.2177}{{\tt
  arXiv:0802.2177}}].

\bibitem{Pereira09}
T.~M.~D. {Pereira}, M.~{Asplund} and D.~{Kiselman}, \emph{Oxygen lines in solar
  granulation. {II}. {C}entre-to-limb variation, {NLTE} line formation, blends,
  and the solar oxygen abundance},
  \href{http://dx.doi.org/10.1051/0004-6361/200912840}{\emph{Astron.
  Astrophys.} {\bf 508} (2009) 1403},
  [\href{http://arxiv.org/abs/arXiv:0909.2310}{{\tt arXiv:0909.2310}}].

\bibitem{Lind11}
K.~{Lind}, M.~{Asplund}, P.~S. {Barklem} and A.~K. {Belyaev}, \emph{Non-{LTE}
  calculations for neutral {Na} in late-type stars using improved atomic data},
  \href{http://dx.doi.org/10.1051/0004-6361/201016095}{\emph{Astron.
  Astrophys.} {\bf 528} (2011) A103},
  [\href{http://arxiv.org/abs/arXiv:1102.2160}{{\tt arXiv:1102.2160}}].

\bibitem{Bergemann12}
M.~{Bergemann}, K.~{Lind}, R.~{Collet}, Z.~{Magic} and M.~{Asplund},
  \emph{Non-{LTE} line formation of fe in late-type stars - {I}. {S}tandard
  stars with {1D} and $\langle${3D}$\rangle$ model atmospheres},
  \href{http://dx.doi.org/10.1111/j.1365-2966.2012.21687.x}{\emph{Mon. Not. R.
  Astron. Soc} {\bf 427} (2012) 27},
  [\href{http://arxiv.org/abs/arXiv:1207.2455}{{\tt arXiv:1207.2455}}].

\bibitem{Mashonkina12}
L.~{Mashonkina}, A.~{Ryabtsev} and A.~{Frebel}, \emph{Non-{LTE} effects on the
  lead and thorium abundance determinations for cool stars},
  \href{http://dx.doi.org/10.1051/0004-6361/201218790}{\emph{Astron.
  Astrophys.} {\bf 540} (2012) A98},
  [\href{http://arxiv.org/abs/arXiv:1202.2630}{{\tt arXiv:1202.2630}}].

\bibitem{caffau11}
E.~{Caffau}, H.-G. {Ludwig}, M.~{Steffen}, B.~{Freytag} and P.~{Bonifacio},
  \emph{Solar chemical abundances determined with a {CO5BOLD} {3D} model
  atmosphere}, \href{http://dx.doi.org/10.1007/s11207-010-9541-4}{\emph{Sol.
  Phys.} {\bf 268} (2011) 255},
  [\href{http://arxiv.org/abs/arXiv:1003.1190}{{\tt arXiv:1003.1190}}].

\bibitem{vSZ16}
R.~{von Steiger} and T.~H. {Zurbuchen}, \emph{Solar metallicity derived from in
  situ solar wind composition},
  \href{http://dx.doi.org/10.3847/0004-637X/816/1/13}{\emph{Astrophys. J.} {\bf
  816} (2016) 13}.

\bibitem{basu09}
S.~Basu, W.~J. Chaplin, Y.~Elsworth, R.~New and A.~M. Serenelli, \emph{Fresh
  insights on the structure of the solar core},
  \href{http://dx.doi.org/10.1088/0004-637X/699/2/1403}{\emph{Astrophys. J.}
  {\bf 699} (2009) 1403}, [\href{http://arxiv.org/abs/arXiv:0905.0651}{{\tt
  arXiv:0905.0651}}].

\bibitem{deglinnoccenti97}
S.~Degl'Innoccenti, W.~A. Dziembowski, G.~Fiorentini and B.~Ricci,
  \emph{Helioseismology and standard solar models},
  \href{http://dx.doi.org/10.1016/S0927-6505(97)00004-2}{\emph{Astropart.
  Phys.} {\bf 3} (1997) 77}.

\bibitem{basu07}
S.~Basu, W.~J. Chaplin, Y.~Elsworth, R.~New, A.~M. Serenelli and G.~A. Verner,
  \emph{Solar abundances and helioseismology: {F}ine-structure spacings and
  separation ratios of low-degree $p$-modes},
  \href{http://dx.doi.org/10.1086/509820}{\emph{Astrophys. J.} {\bf 655} (2007)
  660}, [\href{http://arxiv.org/abs/arXiv:astro-ph/0610052}{{\tt
  arXiv:astro-ph/0610052}}].

\bibitem{roxburgh03}
I.~W. Roxburgh and S.~V. Vorontsov, \emph{The ratio of small to large
  separations of acoustic oscillations as a diagnostic of the interior of
  solar-like stars},
  \href{http://dx.doi.org/10.1051/0004-6361:20031318}{\emph{Astron. Astrophys.}
  {\bf 411} (2003) 215}.

\bibitem{chaplin07}
W.~J. Chaplin, A.~M. Serenelli, S.~Basu, Y.~Elsworth, R.~New and G.~A. Verner,
  \emph{Solar heavy element abundance: constraints from frequency separation
  ratios of low-degree $p$ modes},
  \href{http://dx.doi.org/10.1086/522578}{\emph{Astrophys. J.} {\bf 670} (2007)
  872}, [\href{http://arxiv.org/abs/arXiv:0705.3154}{{\tt arXiv:0705.3154}}].

\bibitem{spiegel92}
E.~A. Spiegel and J.-P. Zahn, \emph{The solar tachocline}, {\emph{Astron.
  Astrophys.} {\bf 265} (1992) 106}.

\bibitem{christensendalsgaard91}
J.~Christensen-Dalsgaard, D.~O. Gough and M.~J. Thompson, \emph{The depth of
  the solar convection zone},
  \href{http://dx.doi.org/10.1086/170441}{\emph{Astrophys. J.} {\bf 378} (1991)
  413}.

\bibitem{basu97}
S.~Basu and H.~M. Antia, \emph{Seismic measurement of the depth of the solar
  convection zone}, \href{http://dx.doi.org/10.1093/mnras/287.1.189}{\emph{Mon.
  Not. R. Astron. Soc.} {\bf 287} (1997) 189}.

\bibitem{basu98}
S.~Basu, \emph{Effects of errors in the solar radius on helioseismic
  inferences},
  \href{http://dx.doi.org/10.1046/j.1365-8711.1998.01690.x}{\emph{Mon. Not. R.
  Astron. Soc.} {\bf 298} (1998) 719},
  [\href{http://arxiv.org/abs/arXiv:astro-ph/9712133}{{\tt
  arXiv:astro-ph/9712133}}].

\bibitem{frandsen10}
M.~T. Frandsen and S.~Sarkar, \emph{Asymmetric dark matter and the {S}un},
  \href{http://dx.doi.org/10.1103/PhysRevLett.105.011301}{\emph{Phys. Rev.
  Lett.} {\bf 105} (2010) 011301},
  [\href{http://arxiv.org/abs/arXiv:1003.4505}{{\tt arXiv:1003.4505}}].

\bibitem{catena16}
R.~Catena and A.~Widmark, \emph{{WIMP} capture by the {S}un in the effective
  theory of dark matter self-interactions},
  \href{http://arxiv.org/abs/arXiv:1609.04825}{{\tt arXiv:1609.04825}}.

\bibitem{choi14}
K.~Choi, C.~Rott and Y.~Itow, \emph{Impact of the dark matter velocity
  distribution on capture rates in the sun},
  \href{http://dx.doi.org/10.1088/1475-7516/2014/05/049}{\emph{J. Cosmol.
  Astropart. Phys.} {\bf 2014} (2014) 049},
  [\href{http://arxiv.org/abs/arXiv:1312.0273}{{\tt arXiv:1312.0273}}].

\bibitem{delnobile13}
E.~{Del Nobile}, G.~Gelmini, P.~Gondolo and J.-H. Huh, \emph{Generalized halo
  independent comparison of direct dark matter detection data},
  \href{http://dx.doi.org/10.1088/1475-7516/2013/10/048}{\emph{J. Cosmol.
  Astropart. Phys.} {\bf 2013} (2013) 048},
  [\href{http://arxiv.org/abs/arXiv:1306.5273}{{\tt arXiv:1306.5273}}].

\end{thebibliography}\endgroup
	}


\end{document}